\documentclass[journal]{IEEEtran}
\usepackage{cite}
\usepackage{amsmath,amssymb,amsfonts}
\usepackage{algorithmic}
\usepackage{graphicx}
\usepackage{textcomp}
\def\BibTeX{{\rm B\kern-.05em{\sc i\kern-.025em b}\kern-.08em
    T\kern-.1667em\lower.7ex\hbox{E}\kern-.125emX}}
\usepackage{amssymb}
\usepackage{amsmath}
\usepackage{graphicx}
\usepackage{bbm}
\usepackage{booktabs}
\usepackage{multirow}
\usepackage{multicol}
\usepackage[flushleft]{threeparttable}
\usepackage{caption}
\usepackage{subfig}
\usepackage{stackengine}
\usepackage{makecell}
\usepackage{balance}
\usepackage{fancyhdr}

\usepackage[square,sort,comma,numbers]{natbib}

\begin{document}
\title{MulViMotion: Shape-aware 3D Myocardial Motion Tracking from Multi-View Cardiac MRI}

\author{Qingjie Meng,
        Chen Qin,
        Wenjia Bai,
        Tianrui Liu,
        Antonio de Marvao, \\
        Declan P O'Regan, 
        and Daniel Rueckert,~\IEEEmembership{Fellow, ~IEEE}
\thanks{This work is supported by the British Heart Foundation (RG/19/6/34387, RE/18/4/34215); Medical Research Council (MC-A658-5QEB0); National Institute for Health Research (NIHR) Imperial College Biomedical Research Centre; Wellcome Trust IEH [102431]. This research has been conducted using the UK Biobank resource under Application 40616}
\thanks{Q. Meng, W. Bai, T. Liu and D. Rueckert are with the Biomedical Image Analysis Group, Department of Computing, Imperial College London, SW7 2AZ, UK, (e-mail: q.meng16$\vert$w.bai$\vert$t.liu15$\vert$drueckert@imperial.ac.uk).}
\thanks{Corresponding author: Qingjie Meng}
\thanks{W. Bai is also with Department of Brain Sciences, Imperial College London. D. Rueckert is also with Faculty of Informatics and Medicine, Technical University Munich, 85748, Germany.}
\thanks{C. Qin is with the Institute for Digital Communications, School of Engineering, University of Edinburgh, EH9 9JL, UK. (e-mail: Chen.Qin@ed.ac.uk)}
\thanks{D. P. O'Regan and A. de Marvao are with the MRC London Institute of Medical Sciences, Imperial College London, W12 0HS, UK. (e-mail: declan.oregan$\vert$antonio.de-marvao@imperial.ac.uk). D. P. O'Regan and D. Rueckert are joint senior authors.}
}

\markboth{ACCEPTED BY IEEE TRANSACTIONS ON MEDICAL IMAGING}%
{Meng \MakeLowercase{\textit{et al.}}: 3D motion estimation from multi-view cine CMR}

\maketitle

\begin{abstract}
Recovering the 3D motion of the heart from cine cardiac magnetic resonance (CMR) imaging enables the assessment of regional myocardial function and is important for understanding and analyzing cardiovascular disease. However, 3D cardiac motion estimation is challenging because the acquired cine CMR images are usually 2D slices which limit the accurate estimation of through-plane motion.
To address this problem, we propose a novel multi-view motion estimation network (MulViMotion), which integrates 2D cine CMR images acquired in short-axis and long-axis planes to learn a consistent 3D motion field of the heart. In the proposed method, a hybrid 2D/3D network is built to generate dense 3D motion fields by learning fused representations from multi-view images. To ensure that the motion estimation is consistent in 3D, a shape regularization module is introduced during training, where shape information from multi-view images is exploited to provide weak supervision to 3D motion estimation.
We extensively evaluate the proposed method on 2D cine CMR images from 580 subjects of the UK Biobank study for 3D motion tracking of the left ventricular myocardium. Experimental results show that the proposed method quantitatively and qualitatively outperforms competing methods.
\end{abstract}


\section{Introduction}
\label{sec:introduction}
\IEEEPARstart{T}{he} motion of the beating heart is a rhythmic pattern of non-linear trajectories regulated by the circulatory system and cardiac neuroautonomic control~\cite{Bello2019, Stefanovska1999, Ivanov1998}. Estimating cardiac motion is an important step for the exploration of cardiac function and the diagnosis of cardiovascular diseases~\cite{Shen2005, Bello2019, Reindl2019}. In particular, left ventricular (LV) myocardial motion tracking enables spatially and temporally localized assessment of LV function~\cite{Puyol2019}. This is helpful for the early and accurate detection of LV dysfunction and myocardial diseases~\cite{Ibrahim2011, Claus2015}.

\begin{figure}
    \centering
    \hspace{-1cm}
    \begin{tabular}{c}
         \raisebox{1\height}{\rotatebox[origin=c]{90}{\makecell{~\scalebox{0.8}{\textbf{End-diastole}}}}} \\
         \raisebox{0.7\height}{\rotatebox[origin=c]{90}{\makecell{~\scalebox{0.8}{\textbf{End-systole}}}}}
    \end{tabular}
    \hspace{-0.8cm}
    \setcounter{subfigure}{0}
    \subfloat[]{
    \begin{tabular}{c}
         \includegraphics[height=2cm]{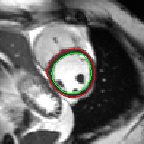}  \\
         \includegraphics[height=2cm]{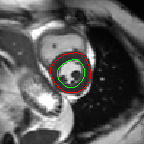} 
    \end{tabular}
    }
    \hspace{-0.8cm}
    \setcounter{subfigure}{1}
    \subfloat[]{
    \begin{tabular}{c}
         \includegraphics[height=2cm]{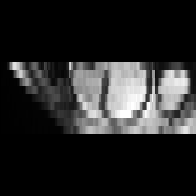}  \\
         \includegraphics[height=2cm]{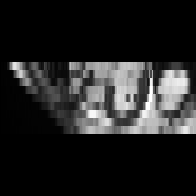} 
    \end{tabular}
    }
    \hspace{-0.8cm}
    \setcounter{subfigure}{2}
    \subfloat[]{
    \begin{tabular}{c}
         \includegraphics[height=2cm]{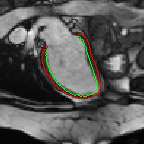}  \\
         \includegraphics[height=2cm]{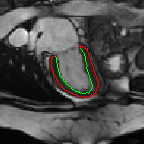} 
    \end{tabular}
    }
    \hspace{-0.8cm}
    \setcounter{subfigure}{3}
    \subfloat[]{
    \begin{tabular}{c}
         \includegraphics[height=2cm]{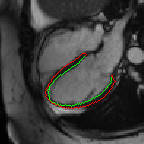}  \\
         \includegraphics[height=2cm]{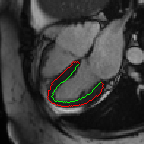} 
    \end{tabular}
    }
    \hspace{-0.9cm}
    \caption[Examples of 2D cine CMR scans of a healthy subject. Cine CMR scans are acquired from short-axis (SAX) view and two long-axis (LAX) views. The SAX view contains a stack of 2D images while each LAX view contains a single 2D image. (a) $XY$-plane of the SAX stack. (b) $XZ$-plane of the SAX stack. (c) LAX 2-chamber (2CH) view. (d) LAX 4-chamber (4CH) view. Red and green contours show the epicardium and endocardium, respectively. The area between these contours is the myocardium of the left ventricle. We show the end-diastolic (ED) frame (top row) and the end-systolic (ES) frame (bottom row) of the cine CMR image sequence.]{Examples of 2D cine CMR scans of a healthy subject. Cine CMR scans are acquired from short-axis (SAX) view and two long-axis (LAX) views. The SAX view contains a stack of 2D images while each LAX view contains a single 2D image. (a) $XY$-plane of the SAX stack. (b) $XZ$-plane of the SAX stack. (c) LAX 2-chamber (2CH) view. (d) LAX 4-chamber (4CH) view. Red and green contours\footnotemark ~show the epicardium and endocardium, respectively. The area between these contours is the myocardium of the left ventricle. We show the end-diastolic (ED) frame (top row) and the end-systolic (ES) frame (bottom row) of the cine CMR image sequence.}
    \label{dataintro}
\end{figure}
\footnotetext{The contours are generated based on~\cite{Duan2019} and a manual quality control. Detailed information is shown in Sec.~\ref{exp_setups}.}

Cine cardiac magnetic resonance (CMR) imaging supports motion analysis by acquiring sequences of 2D images in different views. Each image sequence covers the complete cardiac cycle containing end-diastolic (ED) and end-systolic (ES) phases~\cite{Ginat2011}. 
Two types of anatomical views are identified, including (1) short-axis (SAX) view and (2) long-axis (LAX) view such as 2-chamber (2CH) view and 4-chamber (4CH) view (Fig.~\ref{dataintro}). The SAX sequences typically contain a stack of 2D slices sampling from base to apex in each frame (\emph{e.g.}, 9-12 slices). The LAX sequences contain a single 2D slice that is approximately orthogonal to the SAX plane in each frame. These acquired images have high temporal resolution, high signal-to-noise ratio as well as high contrast between the blood pool and myocardium. With these properties, cine CMR imaging has been utilized in recent works for 2D myocardial motion estimation, \emph{e.g.},~\cite{Qin2018, ZhengQ2019, Qin2020, Bai2020, Yu2020}.

2D myocardial motion estimation only considers motion in either the SAX plane or LAX plane and does not provide complete 3D motion information for the heart. This may lead to inaccurate assessment of cardiac function. Therefore, 3D motion estimation that recovers myocardial deformation in the $X$, $Y$ and $Z$ directions is important. However, estimating 3D motion fields from cine CMR images remains challenging because (1) SAX stacks have much lower through-plane resolution (typically 8 mm slice thickness) than in-plane resolution (typically 1.8 x 1.8 mm), (2) image quality can be negatively affected by slice misalignment in SAX stacks as only one or two slices are acquired during a single breath-hold, and (3) high-resolution 2CH and 4CH view images are too spatially sparse to estimate 3D motion fields on their own.

In this work, we take full advantage of both SAX and LAX (2CH and 4CH) view images, and propose a multi-view motion estimation network for 3D myocardial motion tracking from cine CMR images. In the proposed method, a hybrid 2D/3D network is developed for 3D motion estimation. This hybrid network learns combined representations from multi-view images to estimate a 3D motion field from the ED frame to any $t$-th frame in the cardiac cycle.
To guarantee an accurate motion estimation, especially along the longitudinal direction (\emph{i.e.}, the $Z$ direction), a shape regularization module is introduced to leverage anatomical shape information for motion estimation during training. This module encourages the estimated 3D motion field to correctly transform the 3D shape of the myocardial wall from the ED frame to the $t$-th frame. Here anatomical shape is represented by edge maps that show the contour of the cardiac anatomy.
During inference, the hybrid network generates a sequence of 3D motion fields between paired frames (ED and $t$-th frames), which represents the myocardial motion across the cardiac cycle.
The main contributions of this paper are summarized as follows:
\begin{itemize}
    \item We develop a solution to a challenging cardiac motion tracking problem: learning 3D motion fields from a set of 2D SAX and LAX cine CMR images. We propose an end-to-end trainable multi-view motion estimation network (MulViMotion) for 3D myocardial motion tracking. 
    
    \item The proposed method enables accurate 3D motion tracking by combining multi-view images using both latent information and shape information: (1) the representations of multi-view images are combined in the latent space for the generation of 3D motion fields; (2) the complementary shape information from multi-view images is exploited in a shape regularization module to provide explicit constraint on the estimated 3D motion fields.
 
    \item The proposed method is trained in a weakly supervised manner which only requires sparsely annotated data in different 2D SAX and LAX views and requires no ground truth 3D motion fields. The 2D edge maps from the corresponding SAX and LAX planes provide weak supervision to the estimated 3D edge maps for guiding 3D motion estimation in the shape regularization module.
  
    \item We perform extensive evaluations for the proposed method on 580 subjects from the UK Biobank study. We further present qualitative analysis on the CMR images with severe slice misalignment and we explore the applicability of our method for wall thickening measurement.

\end{itemize}

\section{Related work}

\subsubsection{Conventional motion estimation methods} 
A common method for quantifying cardiac motion is to track noninvasive markers. CMR myocardial tagging provides tissue markers (stripe-like darker tags) in myocardium which can deform with myocardial motion~\cite{Zerhouni1988}. By tracking the deformation of markers, dense displacement fields can be retrieved in the imaging plane. Harmonic phase (HARP) technique is the most representative approach for motion tracking in tagged images~\cite{Osman2000, ChenT2010, LiuX2012}.
Several other methods have been proposed to compute dense displacement fields from dynamic myocardial contours or surfaces using geometrical and biomechanical modeling~\cite{Wang2001, Papademetris2001}. For example, Papademetris et al.~\cite{Papademetris2001} proposed a Bayesian estimation framework for myocardial motion tracking from 3D echocardiography. In addition, image registration has been applied to cardiac motion estimation in previous works. Craene et al.~\cite{Craene2012} introduced continuous spatio-temporal B-spline kernels for computing a 4D velocity field, which enforced temporal consistency in motion recovery. Rueckert et al.~\cite{Rueckert1999} proposed a free form deformation (FFD) method for general non-rigid image registration. This method has been used for cardiac motion estimation in many recent works, \emph{e.g.},~\cite{Chandrashekara2003, Shen2005, Shi2012, Tobon2013, Puyol2018, Bello2019, Puyol2019, Bai2020}. Thirion~\cite{Thirion1998} built a demons algorithm which utilizes diffusing models for image matching and further cardiac motion tracking. Based on this work, Vercauteren et al.~\cite{Vercauteren2007} adapted demons algorithm to provide non-parametric diffeomorphic transformation and McLeod et al.~\cite{McLeod2011} introduced an elastic-like regularizer to improve the incompressibility of deformation recovery.

\begin{figure*}[tpb]
 \centering
 \includegraphics[width=\textwidth, trim=1.5cm 7.5cm 9.5cm 0.9cm, clip]{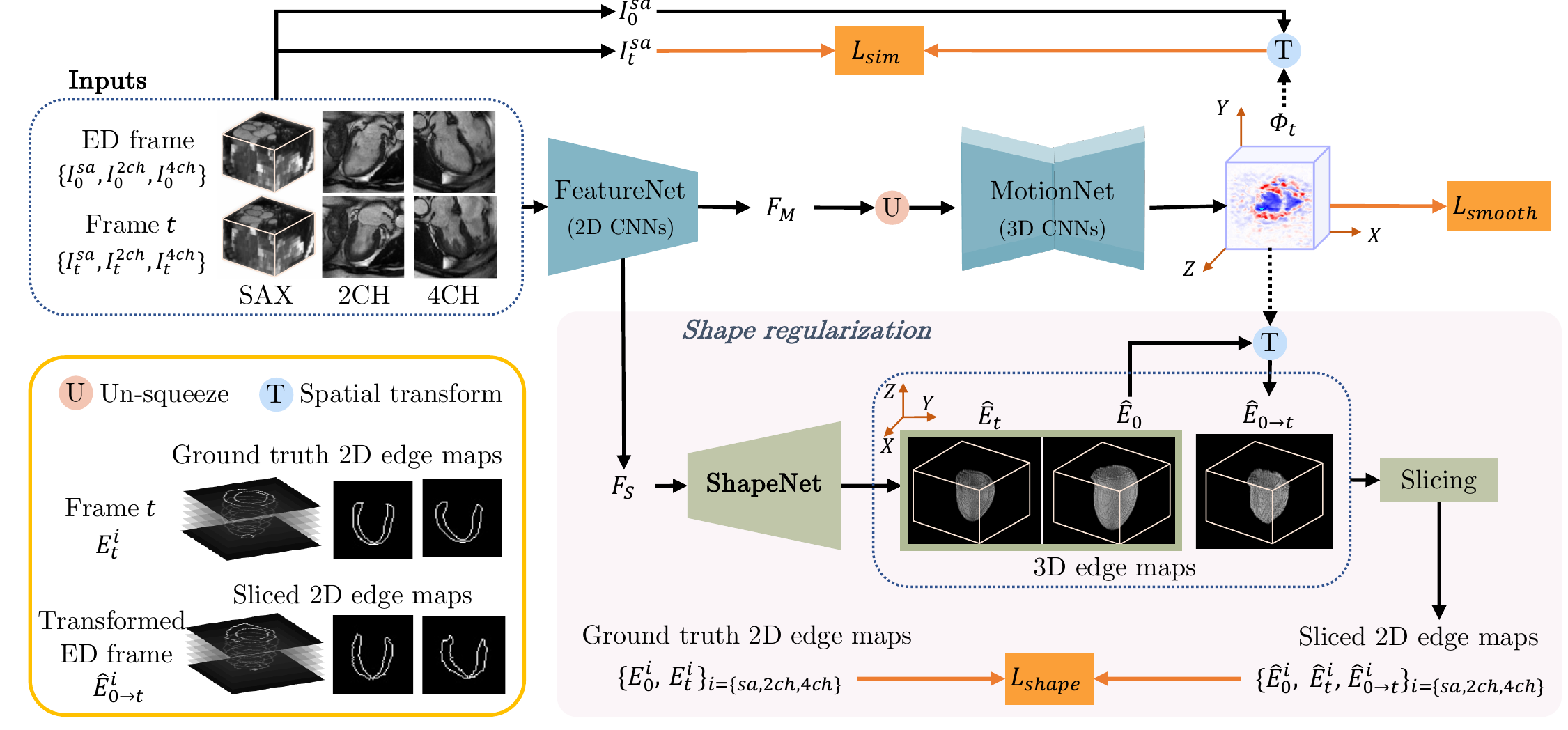}
 \caption{An overview of MulViMotion. We use a hybrid 2D/3D network to estimate a 3D motion field $\Phi_t$ from the input multi-view images. In the hybrid network, FeatureNet learns multi-view motion feature $F_M$ and multi-view shape feature $F_S$ from the input, followed by MotionNet which generates $\Phi_t$ based on $F_M$. A shape regularization module leverages anatomical shape information for 3D motion estimation. It encourages the predicted 3D edge maps of the myocardial wall $\hat{E}_0$/$\hat{E}_t$ (predicted from $F_S$ using ShapeNet) and the warped 3D edge map $\hat{E}_{0\to t}$ (warped from ED frame to the $t$-th frame by $\Phi_t$) to be consistent with the ground truth 2D edge maps defined on multi-view images. Shape regularization is only used during training.
 }
 \label{methed_outline}
\end{figure*}

\subsubsection{Deep learning-based motion estimation methods}
In recent years, deep convolutional neural networks (CNNs) have been successfully applied to medical image analysis, which has inspired the exploration of deep learning-based cardiac motion estimation approaches.
Qin et al.~\cite{Qin2018} proposed a multi-task framework for joint estimation of segmentation and motion. This multi-task framework contains a shared feature encoder which enables a weakly-supervised segmentation. 
Zheng et al.~\cite{ZhengQ2019} proposed a method for cardiac pathology classification based on cardiac motion. Their method utilizes a modified U-Net~\cite{Ronneberger2015} to generate flow maps between ED frame and any other frame.
For cardiac motion tracking in multiple datasets, Yu et al.~\cite{Yu2020} considered the distribution mismatch problem and proposed a meta-learning-based online model adaption framework. Different from these methods which estimate motion in cine CMR, Ye et al.~\cite{Ye2021} proposed a deep learning model for tagged image motion tracking. In their work, the motion field between any two consecutive frames is first computed, followed by estimating the Lagrangian motion field between ED frame and any other frame. Most of these existing deep learning-based methods aim at 2D motion tracking by only using SAX stacks. In contrast, our method focuses on 3D motion tracking by fully combining multiple anatomical views (\emph{i.e.}, SAX, 2CH and 4CH), which is able to estimate both in-plane and through-plane myocardial motion.

\subsubsection{Multi-view based cardiac analysis}
Different anatomical scan views usually contain complementary information and the combined multiple views can be more descriptive than a single view. Chen et al.~\cite{ChenC2019} utilized both SAX and LAX views for 2D cardiac segmentation, where the features of multi-view images are combined in the bottleneck of 2D U-Net. Puyol-Antón et al.~\cite{Puyol2018} introduced a framework that separately uses multi-view images for myocardial strain analysis. In their method, the SAX view is used for radial and circumferential strain estimation while the LAX view is used for longitudinal strain estimation. Abdelkhalek et al.~\cite{Abdelkhalek2020} proposed a 3D myocardial strain estimation framework, where the point clouds from SAX and LAX views are aligned for surface reconstruction. Attar et al.~\cite{Attar2019} proposed a framework for 3D cardiac shape prediction, in which the features of multi-view images are concatenated in CNNs to predict the 3D shape parameters. In this work, we focus on using multi-view images for 3D motion estimation. Compared to most of these existing works which only combine the features of multi-view images in the latent space (\emph{e.g.},~\cite{ChenC2019, Attar2019}), our method additionally combines complementary shape information from multiple views to predict anatomically plausible 3D edge map of myocardial wall on different time frames, which provides guidance for 3D motion estimation.


\section{Method}

Our goal is to estimate 3D motion fields of the LV myocardium from multi-view 2D cine CMR images.
We formulate our task as follows: Let $\mathbf{I}^{SA}={\{I_t^{sa}\in \mathbb{R}^{H\times W\times D}|0\leqslant t\leqslant T-1\}}$ be a SAX sequence which contains stacks of 2D images ($D$ slices) and $\mathbf{I}^{LA}={\{I_t^{2ch}\in \mathbb{R}^{H\times W},I_t^{4ch}\in \mathbb{R}^{H\times W}|0\leqslant t\leqslant T-1\}}$ be LAX sequences which contain 2D images in the 2CH and 4CH views. $H$ and $W$ are the height and width of each image and $T$ is the number of frames. We want to train a network to estimate a 3D motion field $\Phi_t\in \mathbb{R}^{H\times W\times D  \times 3}$ by using the multi-view images of the ED frame ($\{I_0^{sa}, I_0^{2ch}, I_0^{4ch}\}$) and of any $t$-th frame ($\{I_t^{sa}, I_t^{2ch}, I_t^{4ch}\}$). $\Phi_t$ describes the motion of the LV myocardium from ED frame to the $t$-th frame. For each voxel in $\Phi_t$, we estimate its displacement in the $X$, $Y$, $Z$ directions.

\begin{figure*}[pthb]
 \centering
 \includegraphics[width=\textwidth, trim=1.7cm 11.7cm 10.2cm 0.8cm, clip]{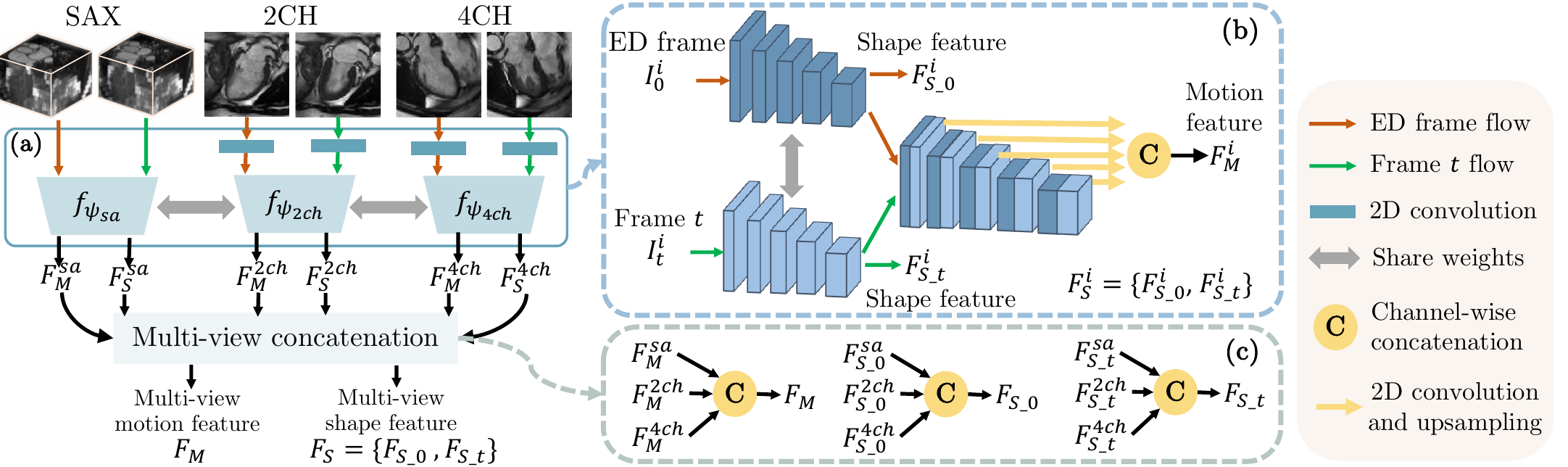}
 \caption{An overview of FeatureNet. FeatureNet takes multi-view images as input and extracts multi-view motion feature $F_M$ and multi-view shape feature $F_S$. Panel (a) describes multi-scale feature fusion. Panel (b) shows the 2D encoder $f_{\psi_{i}}$, where $i=\{sa, 2ch, 4ch\}$ refers to SAX, 2CH and 4CH views. Panel (c) describes the combination of multi-view features.
 }
 \label{FeatureNet_outline}
\end{figure*}

To solve this task, we propose MulViMotion that estimates 3D motion fields from multi-view images with shape regularization. The schematic architecture of our method is shown in Fig.~\ref{methed_outline}. A hybrid 2D/3D network that contains FeatureNet (2D CNNs) and MotionNet (3D CNNs) is used to predict $\Phi_t$ from the input multi-view images. FeatureNet learns multi-view multi-scale features and is used to extract multi-view motion feature $F_M$ and multi-view shape feature $F_S$ from the input. MotionNet generates $\Phi_t$ based on $F_M$. 
A shape regularization module is used to leverage anatomical shape information for 3D motion estimation during training. In this module, 3D edge maps of the myocardial wall are predicted from $F_S$ using ShapeNet and warped from ED frame to the $t$-th frame by $\Phi_t$. The sparse ground truth 2D edge maps derived from the multi-view images provide weak supervision to the predicted and warped 3D edge maps, and thus encourage an accurate estimation of $\Phi_t$, especially in the $Z$ direction. Here, a slicing step is used to extract corresponding multi-view planes from the 3D edge maps in order to compare 3D edge maps with 2D ground truth. During inference, a 3D motion field is directly generated from the input multi-view images by the hybrid network, without using shape regularization.


\subsection{3D motion estimation}
\label{movEst}

\subsubsection{Multi-view multi-scale feature extraction (FeatureNet)}
\label{mvmvfeaExt}
The first step of 3D motion estimation is to extract internal representations from the input 2D multi-view images $\{I_{j}^{{sa}}, I_j^{2ch}, I_j^{4ch}|j=\{0,t\}\}$.
We build FeatureNet to simultaneously learn motion and shape feature from the input because the motion and shape of the myocardial wall are closely related and can provide complementary information to each other~\cite{Cheng2017, Qin2018, Ta2020}.
FeatureNet consists of (1) multi-scale feature fusion and (2) multi-view concatenation (see Fig.~\ref{FeatureNet_outline}).

In the multi-scale feature fusion (Fig.~\ref{FeatureNet_outline} (a)), the input multi-view images are unified to $D$-channel 2D feature maps by applying 2D convolution on 2CH and 4CH view images. Then three 2D encoders $\{f_{\psi_i}| i=\{{sa}, {2ch}, {4ch}\}\}$ are built to extract motion and shape features from each anatomical view,
\begin{equation}\label{feanet_encoder}
\{F_M^{i}, F_S^{i}\}=f_{\psi_i}(I_0^{i}, I_{t}^{i}), \quad i=\{{{sa}}, {2ch}, {4ch}\}.
\end{equation}
Here, $i$ represents anatomical views and $\psi_i$ refers to the network parameters of $f_{\psi_i}$. $F_M^{i}$ and $F_S^{i}$ are the learned motion feature and shape feature, respectively. 
As these encoders aim to extract the same type of information (\emph{i.e.}, shape and motion information), the three encoders share weights to learn representations that are useful and related to different views. 

In each encoder, representations at different scales are fully exploited for feature extraction. $\{f_{\psi_i}| i=\{{sa}, {2ch}, {4ch}\}\}$ consists of (1) a Siamese network that extracts features from both ED frame and $t$-th frame, and (2) feature-fusion layers that concatenate multi-scale features from pairs of frames (Fig.~\ref{FeatureNet_outline} (b)). 
From the Siamese network, the last feature maps of the two streams are used as shape feature of the ED frame ($F_{S\_0}^{i}$) and the $t$-th frame ($F_{S\_t}^{i}$), respectively,  and $F_S^{i}=\{F_{S\_0}^{i}, F_{S\_t}^{i}\}$. All features across different scales from both streams are combined by feature-fusion layers to generate motion feature $F_M^{i}$. In detail, these multi-scale features are upsampled to the original resolution by a convolution and upsampling operation and then combined using a concatenation layer.

With the obtained $\{F_M^{i}, F_S^{i}|i=\{sa,2ch,4ch\}\}$, a multi-view concatenation generates the multi-view motion feature $F_M$ and the multi-view shape feature $F_S$ via channel-wise concatenation $\textrm{C}(\cdot,\cdot,\cdot)$ (see Fig.~\ref{FeatureNet_outline} (c)),
\begin{equation}\label{FeatureNet_cat}
F_M=\textrm{C}(F_M^{sa},F_M^{2ch},F_M^{4ch}), F_{S\_j}=\textrm{C}(F_{S\_j}^{sa}, F_{S\_j}^{2ch}, F_{S\_j}^{4ch}).
\end{equation}
Here $j=\{0,t\}$ and $F_S=\{F_{S\_0}, F_{S\_t}\}$.

The FeatureNet model is composed of 2D CNNs which learns 2D features from the multi-view images and inter-slice correlation from SAX stacks. The obtained $F_M$ is first unified to $D$-channels using 2D convolution and then is used to predict $\Phi_t$ in the next step. The obtained $F_S$ is used for shape regularization in Sec.~\ref{shape_regular}.

\subsubsection{Motion estimation (MotionNet)}
In this step, we introduce MotionNet to predict the 3D motion field $\Phi_t$ by learning 3D representations from the multi-view motion feature $F_M$. MotionNet is built with a 3D encoder-decoder architecture. $\Phi_t$ is predicted by MotionNet with
\begin{equation}\label{motionNet}
\Phi_t = g_\theta (U(F_M)),
\end{equation}
where $g_\theta$ represents MotionNet and $\theta$ refers to the network parameters of $g_\theta$. The function $U(\cdot)$ denotes an un-squeeze operation which changes $F_M$ from a stack of 2D feature maps to a 3D feature map by adding an extra dimension.

\subsubsection{Spatial transform (Warping)}
\label{warpping}
Inspired by the successful application of spatial transformer networks~\cite{Jaderberg2015, Caballero2017}, the SAX stack of the ED frame ($I_0^{sa}$) can be transformed to the $t$-th frame using the motion field $\Phi_t$. For voxel with location $p$ in the transformed SAX stack ($I_{0\to t}^{sa}$), we compute the corresponding location $p'$ in $I_0^{sa}$ by $p'=p+\Phi_t(p)$. As image values are only defined at discrete locations, the value at $p$ in $I_{0\to t}^{sa}$ is computed from $p'$ in $I_0^{sa}$ using trillinear interpolation\footnote{This is implemented by Pytorch function grid\_sample().}.

\subsubsection{Motion loss}
\label{motionEstLoss}
As true dense motion fields of paired frames are usually unavailable in real practice, we propose an unsupervised motion loss $\mathcal{L}_{mov}$ to evaluate the 3D motion estimation model using only the input SAX stack ($I_t^{sa}$) and the generated 3D motion field ($\Phi_t$). 
$\mathcal{L}_{mov}$ consists of two components: (1) an image similarity loss $\mathcal{L}_{sim}$ that penalizes appearance difference between $I_t^{sa}$ and $I_{0\to t}^{sa}$, and (2) a local smoothness loss $\mathcal{L}_{smooth}$ that penalizes the gradients of $\Phi_t$,
\begin{equation}\label{motionloss}
\mathcal{L}_{mov} = \mathcal{L}_{sim} + \lambda \mathcal{L}_{smooth}.
\end{equation}
Here $\lambda$ is a hyper-parameter, $\mathcal{L}_{sim}$ is defined by voxel-wise mean squared error and $\mathcal{L}_{smooth}$ is the Huber loss used in~\cite{Caballero2017, Qin2018} which encourages a smooth $\Phi_t$, \begin{equation}\label{simliarityloss}
\mathcal{L}_{sim} = \frac{1}{N}\sum_{i=1}^{N}(I_t^{sa}(p_i)-I_{0\to t}^{sa}(p_i))^2,
\end{equation}
\begin{equation}\label{smoothloss}
\begin{split}
&\mathcal{L}_{smooth} = \sqrt{\epsilon+\sum_{i=1}^{N}\|\triangledown\Phi_t(p_i)\|^2},\\
&\triangledown\Phi_t(p_i)=(\frac{\partial\Phi_t(p_i)}{\partial x}, \frac{\partial\Phi_t(p_i)}{\partial y}, \frac{\partial\Phi_t(p_i)}{\partial z}).
\end{split}
\end{equation}
Here $\frac{\partial\Phi_t(p_i)}{\partial x}\approx \Phi_t(p_{i_x}+1, p_{i_y}, p_{i_z})-\Phi_t(p_{i_x}, p_{i_y}, p_{i_z})$ and we use the same approximation to $\frac{\partial\Phi_t(p_i)}{\partial y}$ and $\frac{\partial\Phi_t(p_i)}{\partial z}$. Same to~\cite{Caballero2017,Qin2018}, $\epsilon$ is set to 0.01. 
In Eq.~\ref{simliarityloss} and Eq.~\ref{smoothloss}, $p_i$ is the $i$th voxel and $N$ denotes the number of voxels. 

Note that $\mathcal{L}_{sim}$ is only applied to SAX stacks because 2D images in 2CH and 4CH views typically consist of only one slice and can not be directly warped by a 3D motion field. 

\subsection{Shape regularization}
\label{shape_regular}
The motion loss ($\mathcal{L}_{mov}$) on its own is not sufficient to guarantee motion estimation in the $Z$ direction due to the low through-plane resolution in SAX stacks.
To address this problem, we introduce a shape regularization module which ensures the 3D edge map of the myocardial wall is correct before and after $\Phi_t$ warping, and thus enables an accurate estimation of $\Phi_t$. Here, the ground truth 2D edge maps derived from the multi-view images provide weak supervision to the predicted and warped 3D edge maps.

\subsubsection{Shape estimation (ShapeNet)}
\begin{figure}[tb]
 \centering
 \includegraphics[width=\textwidth, trim=1.4cm 15.7cm 13.5cm 0.7cm, clip]{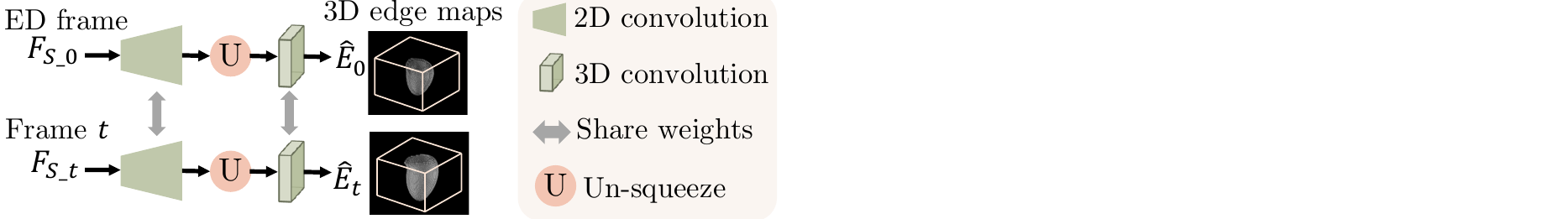}
 \caption{An overview of ShapeNet. ShapeNet predicts the 3D edge maps of the LV myocardial wall in ED frame and the $t$-th frame from the corresponding shape features $F_{S\_0}$ and $F_{S\_t}$. 
 }
 \label{shapenet_outline}
\end{figure}

ShapeNet is built to generate the 3D edge map of the myocardial wall in the ED frame ($\hat{E}_0$) and the $t$-th frame ($\hat{E}_t$) from $F_S=\{F_{S\_0}, F_{S\_t}\}$,
\begin{equation}\label{edgeMap}
\hat{E}_0 = h_1 (F_{S\_0}), \hat{E}_t = h_2 (F_{S\_t}).
\end{equation}
Here $h_1$ and $h_2$ are the two branches in ShapeNet which contain shared 2D decoders and 3D convolutional layers in order to learn 3D edge maps from 2D features for all frames (Fig.~\ref{shapenet_outline}). The dimension of $\hat{E}_0$ and $\hat{E}_t$ are $H\times W\times D$.
With the spatial transform in Sec.~\ref{warpping}, $\hat{E}_0$ is warped to the $t$-th frame by $\Phi_t$, which generates the transformed 3D edge map $\hat{E}_{0\to t}$.
Then $\hat{E}_0$, $\hat{E}_t$ and $\hat{E}_{0\to t}$ are weakly supervised by ground truth 2D edge maps.

\subsubsection{Slicing}
\label{slicerSection}
To compare the 3D edge maps with 2D ground truth, we use 3D masks $\{M^{sa}, M^{2ch}, M^{4ch}\}$ to extract SAX, 2CH and 4CH view planes from $\hat{E}_0$, $\hat{E}_t$ and $\hat{E}_{0\to t}$ with
\begin{equation}\label{slicer}
\hat{E}_0^i = M^i\odot \hat{E}_0, \hat{E}_t^i = M^i\odot \hat{E}_t, \hat{E}_{0\to t}^i = M^i\odot \hat{E}_{0\to t},
\end{equation}
where $i=\{sa, 2ch, 4ch\}$ represents anatomical views and $\odot$ refers to element-wise multiplication. These 3D masks describe the locations of multi-view images in SAX stacks and are generated based on the input during image preprocessing. 


\subsubsection{Shape loss}
\label{l_shape}
The sliced 2D edge maps $\{\hat{E}_0^i, \hat{E}_t^i, \hat{E}_{0\to t}^i|i=\{sa, 2ch, 4ch\}\}$ are compared to 2D ground truth $\{E_0^{i}, E_t^{i}|i=\{sa,2ch,4ch\}\}$ by a shape loss $\mathcal{L}_{shape}$,
\begin{equation}\label{shapeLoss}
\mathcal{L}_{shape}=\mathcal{L}_0^S+\mathcal{L}_t^S+\mathcal{L}_{0\to t}^S.
\end{equation}
For each component in $\mathcal{L}_{shape}$, we utilize cross-entropy loss ($\textrm{CE}(\cdot,\cdot)$) to measure the similarity of edge maps, \emph{e.g.},
\begin{equation}\label{ceLoss}
\mathcal{L}_0^S=\sum\nolimits_{i=\{sa,2ch,4ch\}}{CE}(\hat{E}_0^i, E_0^i).
\end{equation}
Same to Eq.~\ref{ceLoss}, $\mathcal{L}_{t}^S$ is computed by $\{\hat{E}_t^i, E_t^i\}$ and $\mathcal{L}_{0\to t}^S$ is computed by $\{\hat{E}_{0\to t}^i, E_t^i\}$.

\subsection{Optimization}
Our model is an end-to-end trainable framework and the overall objective is a linear combination of all loss functions 
\begin{equation}\label{Loss}
min\{\mathcal{L}_{sim}+\lambda\mathcal{L}_{smooth}+\beta\mathcal{L}_{shape}\}, 
\end{equation}
where $\lambda$ and $\beta$ are hyper-parameters chosen experimentally depending on the dataset. We use the Adam optimizer ($\text{learning rate}=10^{-4}$) to update the parameters of MulViMotion. Our model is implemented by Pytorch and is trained on a NVIDIA Tesla T4 GPU with 16 GB of memory.

\section{Experiments}
We demonstrate our method on the task of 3D myocardial motion tracking. We evaluate the proposed method using quantitative metrics such as Dice, Hausdorff distance, volume difference and Jacobian determinant. Geometric mesh is used to provide qualitative results with 3D visualization. We compared the proposed method with other state-of-the-art motion estimation methods and performed extensive ablation study. In addition, we show the effectiveness of the proposed method on the subjects with severe slice misalignment. We further explore the applicability of the proposed method for wall thickening measurement.
We show the key results in the main paper. More results (\emph{e.g.}, dynamic videos) are shown in the Appendix. 

\subsection{Experiment setups}
\label{exp_setups}
\subsubsection{Data}
We performed experiments on randomly selected 580 subjects from the UK Biobank study\footnote{Application number 40616, https://www.ukbiobank.ac.uk/}. All participants gave written informed consent~\cite{bycroft2018uk}. The participant characteristics are shown in Table~\ref{clinical_character}. The CMR images of all subjects are acquired by a 1.5 Tesla scanner (MAGNETOM Aera, Syngo Platform VD13A, Siemens Healthcare, Erlangen, Germany).
Each subject contains SAX, 2CH and 4CH view cine CMR sequences and each sequence contains 50 frames. More CMR acquisition details for UK Biobank study can be found in~\cite{Petersen2015}.
For image preprocessing, (1) SAX view images were resampled by linear interpolation from a spacing of $\sim 1.8\times 1.8\times 10mm$ to a spacing of $1.25\times 1.25\times 2mm$ while 2CH and 4CH view images were resampled from $\sim 1.8\times 1.8 mm$ to $1.25\times 1.25mm$, (2) by keeping the middle slice of the resampled SAX stacks in the center, zero-padding was used on top or bottom if necessary to reshape the resampled SAX stacks to $64$ slices, (3) to cover the whole LV as the ROI, based on the center of the LV in the middle slice, the resampled SAX stacks were cropped to a size of $128\times 128\times 64$ (note that we computed the center of the LV based on the LV myocardium segmentation of the middle slice of the SAX stack), (4) 2CH and 4CH view images were cropped to $128\times 128$ based on the center of the intersecting line between the middle slice of the cropped SAX stack and the 2CH/4CH view image, (5) each frame was independently normalized to zero mean and unitary standard deviation, and (6) 3D masks (Eq.~\ref{slicer}) were computed by a coordinate transformation using DICOM image header information of SAX, 2CH and 4CH view images. 
Note that 2D SAX slices used in the shape regularization module were unified to $9$ adjacent slices for all subjects, including the middle slice and $4$ upper and lower slices.
With this image preprocessing, the input SAX, 2CH and 4CH view images cover the whole LV in the center.

\begin{table}[tb]
\centering
\caption{Participant characteristics. Data are mean$\pm$standard deviation for continuous variables and number of participant for categorical variable.}
\label{clinical_character}
\begin{tabular}{lc}
\toprule[1.2pt]
Parameter                            & 
Value (Subject number is 580)                   \\
\midrule
Age (years)   &
64$\pm$8    \\
Sex (Female/Male)    &
325 / 255 \\
Ejection fraction ($\%$)   &
60$\pm$6    \\
Weight (kg)    &
74$\pm$15 \\
Height (cm)    &
169$\pm$9 \\
Body mass index (kg/m$^2$)    &
26$\pm$4 \\
Diastolic blood pressure (mm Hg)    &
79$\pm$10 \\
Systolic blood pressure (mm Hg)    &
138$\pm$19 \\
\bottomrule[1.2pt]
\end{tabular}
\end{table}

3D high-resolution segmentations of these subjects were automatically generated using the 4Dsegment tool~\cite{Duan2019} based on the resampled SAX stacks, followed by manual quality control. The obtained segmentations have been shown to be useful in clinical applications (\emph{e.g.},~\cite{Bello2019}), and thus we use them to generate ground truth 2D edge maps (Fig.~\ref{dataintro}) in this work. In detail, we utilize the obtained 3D masks to extract SAX, 2CH and 4CH view planes from these 3D segmentations and then use contour extraction
to obtain $\{E_0^{i}, E_t^{i}|i=\{sa,2ch,4ch\}\}$ used in Sec.~\ref{slicerSection}. Note that we use \textit{3D segmentation(s)} to refer to the 3D segmentations obtained by~\cite{Duan2019} in this section.

We split the dataset into 450/50/80 for train/validation/test and train MulViMotion for 300 epochs. 
The hyper-parameters in Eq.~\ref{Loss} are selected as $\lambda=0.005, \beta=5$.

\subsubsection{Evaluation metrics}
\label{metrics}
We use segmentations to provide quantitative evaluation to the estimated 3D motion fields. This is the same evaluation performed in other cardiac motion tracking literature~\cite{Qin2018, ZhengQ2019, Yu2020}. Here, 3D segmentations obtained by~\cite{Duan2019} are used in the evaluation metrics. The framework in~\cite{Duan2019} performs learning-based segmentation, followed by an atlas-based refinement step to ensure robustness towards potential imaging artifacts. The generated segmentations are anatomically meaningful and spatially consistent. As our work aims to estimate real 3D motion of the heart from the acquired CMR images, such segmentations that approximate the real shape of the heart can provide a reasonable evaluation. In specific, on test data, we estimate the 3D motion field $\Phi_{ES}$ from ED frame to ES frame, which shows large deformation. Then we warp the 3D segmentation of the ED frame ($S_{ED}$) to ES frame by $\Phi_{ES}$. 
Finally, we compared the transformed 3D segmentation ($S_{ED\to ES}$) with the ground truth 3D segmentation of the ES frame ($S_{ES}$) using following metrics. Note that the ES frame is identified as the frame with the least image intensity similarity to the ED frame. 

\textbf{Dice score} and \textbf{Hausdorff distance (HD)} are utilized to respectively quantify the volume overlap and contour distance between $S_{ES}$ and $S_{ED\to ES}$. A high value of Dice and a low value of HD represent an accurate 3D motion estimation. 

\textbf{Volume difference (VD)} is computed to evaluate the volume preservation, as incompressible motion is desired within the myocardium~\cite{McLeod2011, Shi2012, LiuX2012, Qin2020}. ${VD}=|V(S_{ED})-V(S_{ED\to ES})|/V(S_{ED})$, where $V(\cdot)$ computes the number of voxels in the segmentation volume. A low value of VD means a good volume preservation ability of $\Phi_{ES}$. 

The \textbf{Jacobian determinant} ${det(J_{\Phi_{ES}}})$ ($J_{\Phi_{ES}}=\triangledown\Phi_{ES}$) is employed to evaluate the local behavior of $\Phi_{ES}$: A negative Jacobian determinant ${det(J_{\Phi_{ES}}(p))<0}$ indicates that the motion field at position $p$ results in folding and leads to non-diffeomorphic transformations. Therefore, a low number of points with ${det(J_{\Phi_{ES}}(p))<0}$ corresponds to
an anatomically plausible deformation from ED frame to ES frame and thus indicates a good $\Phi_{ES}$. We count the percentage of voxels in the myocardial wall with ${det(J_{\Phi_{ES}}(p))<0}$ in the evaluation.

\subsubsection{Baseline methods} 
We compared the proposed method with three cardiac motion tracking methods, including two conventional methods and one deep learning method. The first baseline is a B-spline free form deformation (FFD) algorithm ~\cite{Rueckert1999} which has been used in many recent cardiac motion tracking works~\cite{Tobon2013, Puyol2018, Bello2019, Puyol2019, Bai2020}. We use the FFD approach implemented in the MIRTK toolkit\footnote{http://mirtk.github.io/}. The second baseline is a diffeomorphic  Demons (dDemons) algorithm~\cite{Vercauteren2007} which has been used in~\cite{Qin2020} for cardiac motion tracking. We use a SimpleITK software package as the dDemons implementation\footnote{https://github.com/InsightSoftwareConsortium/SimpleITK-Notebooks/blob/master/Python/66\_Registration\_Demons.ipynb}. In addition, the UNet architecture has been used in many recent works for image registration~\cite{Balakrishnan2019, XuZ2020, Ta2020}, and thus our third baseline is a deep learning method with 3D-UNet~\cite{ociek2016}. The input of 3D-UNet baseline is paired frames ($I_0^{sa}, I_t^{sa}$) and output is a 3D motion field. Eq.~\ref{motionloss} is used as the loss function for this baseline. We implemented 3D-UNet based on its online code\footnote{https://github.com/wolny/pytorch-3dunet}. 
For the baseline methods with hyper-parameters, we evaluated several sets of parameter values. The hyper-parameters that achieve the best Dice score on the validation set are selected. 

\subsection{3D myocardial motion tracking} 

\begin{figure}[ptb]
 \centering
 \subfloat[Warped 3D segmentations overlaid on multi-view images]{
 \begin{tabular}{@{\hspace{-1\tabcolsep}}c@{\hspace{0.3\tabcolsep}}c@{\hspace{0.3\tabcolsep}}c@{\hspace{0.3\tabcolsep}}c@{\hspace{0.3\tabcolsep}}c@{\hspace{0.3\tabcolsep}}c}
  \raisebox{1\height}{\rotatebox[origin=c]{90}{\makecell{~\scalebox{0.8}{\textbf{SAX}}}}}
  \includegraphics[height=1.6cm]{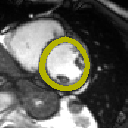} &
  \includegraphics[height=1.6cm]{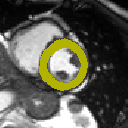} &
  \includegraphics[height=1.6cm]{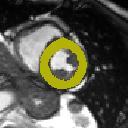} &
  \includegraphics[height=1.6cm]{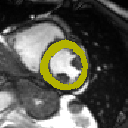} &
  \includegraphics[height=1.6cm]{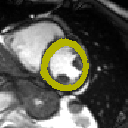} \\
  \raisebox{1.5\height}{\rotatebox[origin=c]{90}{\makecell{~\scalebox{0.8}{\textbf{2CH}}}}} 
  \includegraphics[height=1.6cm]{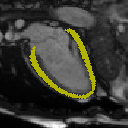} &
  \includegraphics[height=1.6cm]{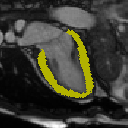} &
  \includegraphics[height=1.6cm]{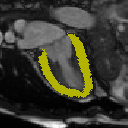} &
  \includegraphics[height=1.6cm]{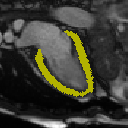} &
  \includegraphics[height=1.6cm]{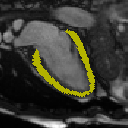} \\
  \raisebox{1.5\height}{\rotatebox[origin=c]{90}{\makecell{~\scalebox{0.8}{\textbf{4CH}}}}} 
  \includegraphics[height=1.6cm]{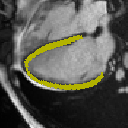} &
  \includegraphics[height=1.6cm]{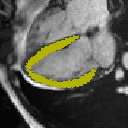} &
  \includegraphics[height=1.6cm]{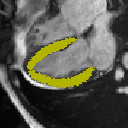} &
  \includegraphics[height=1.6cm]{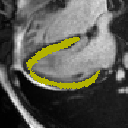} &
  \includegraphics[height=1.6cm]{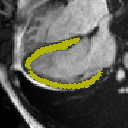} \\
  ~~~
  \raisebox{0.1\height}{\rotatebox[origin=c]{0}{\makecell{~\scalebox{0.8}{\textbf{t=0}}}}} &
  \raisebox{0.1\height}{\rotatebox[origin=c]{0}{\makecell{~\scalebox{0.8}{\textbf{t=10}}}}} &
  \raisebox{0.1\height}{\rotatebox[origin=c]{0}{\makecell{~\scalebox{0.8}{\textbf{t=20}}}}} &
  \raisebox{0.1\height}{\rotatebox[origin=c]{0}{\makecell{~\scalebox{0.8}{\textbf{t=30}}}}} &
  \raisebox{0.1\height}{\rotatebox[origin=c]{0}{\makecell{~\scalebox{0.8}{\textbf{t=40}}}}}
  \end{tabular}
  } \\
 \subfloat[The ground truth vs. the warped SAX stacks]{
 \begin{tabular}{@{\hspace{-1\tabcolsep}}c@{\hspace{0.3\tabcolsep}}c@{\hspace{0.3\tabcolsep}}c@{\hspace{0.3\tabcolsep}}c@{\hspace{0.3\tabcolsep}}c@{\hspace{0.3\tabcolsep}}c}
  \raisebox{2\height}{\rotatebox[origin=c]{90}{\makecell{~\scalebox{0.8}{\textbf{GT}}}}}
  \includegraphics[height=1.6cm]{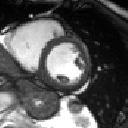} &
  \includegraphics[height=1.6cm]{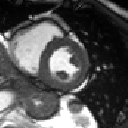} &
  \includegraphics[height=1.6cm]{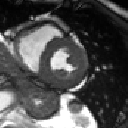} &
  \includegraphics[height=1.6cm]{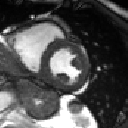} &
  \includegraphics[height=1.6cm]{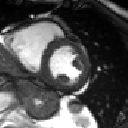} \\
  \raisebox{0.9\height}{\rotatebox[origin=c]{90}{\makecell{~\scalebox{0.8}{\textbf{Warped}}}}} 
  \includegraphics[height=1.6cm]{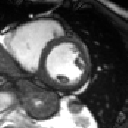} &
  \includegraphics[height=1.6cm]{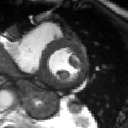} &
  \includegraphics[height=1.6cm]{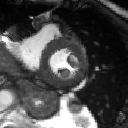} &
  \includegraphics[height=1.6cm]{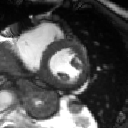} &
  \includegraphics[height=1.6cm]{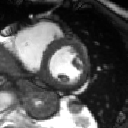} \\
  \raisebox{0.7\height}{\rotatebox[origin=c]{90}{\makecell{~\scalebox{0.8}{\textbf{GT$-$Warped}}}}} 
  \includegraphics[height=1.6cm, trim=0.2cm 0.2cm 0.2cm 0.2cm, clip]{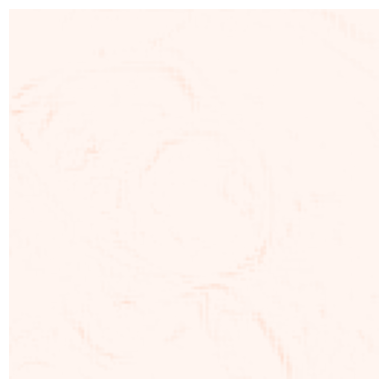} &
  \includegraphics[height=1.6cm, trim=0.2cm 0.2cm 0.2cm 0.2cm, clip]{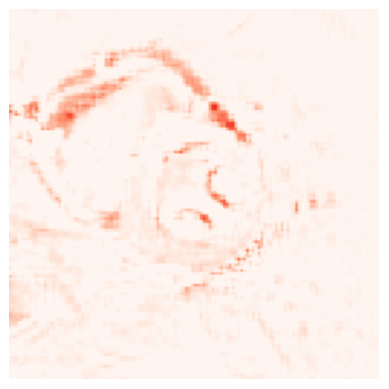} &
  \includegraphics[height=1.6cm, trim=0.2cm 0.2cm 0.2cm 0.2cm, clip]{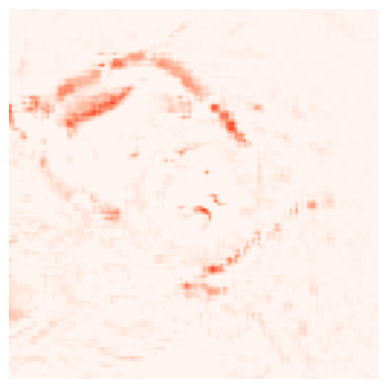} &
  \includegraphics[height=1.6cm, trim=0.2cm 0.2cm 0.2cm 0.2cm, clip]{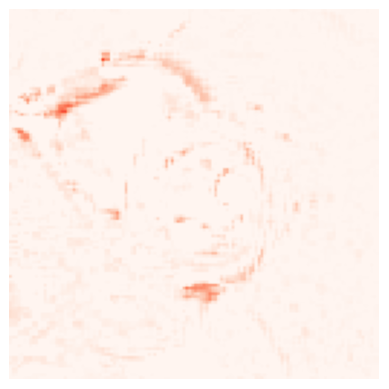} &
  \includegraphics[height=1.6cm, trim=0.2cm 0.2cm 0.2cm 0.2cm, clip]{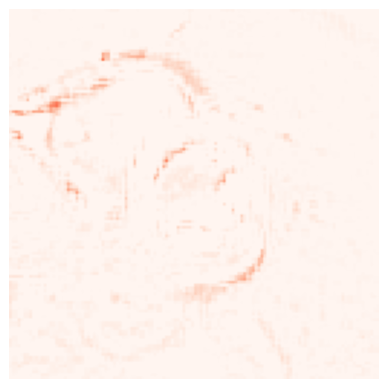} \\
  ~~~
  \raisebox{0.1\height}{\rotatebox[origin=c]{0}{\makecell{~\scalebox{0.8}{\textbf{t=0}}}}} &
  \raisebox{0.1\height}{\rotatebox[origin=c]{0}{\makecell{~\scalebox{0.8}{\textbf{t=10}}}}} &
  \raisebox{0.1\height}{\rotatebox[origin=c]{0}{\makecell{~\scalebox{0.8}{\textbf{t=20}}}}} &
  \raisebox{0.1\height}{\rotatebox[origin=c]{0}{\makecell{~\scalebox{0.8}{\textbf{t=30}}}}} &
  \raisebox{0.1\height}{\rotatebox[origin=c]{0}{\makecell{~\scalebox{0.8}{\textbf{t=40}}}}}
  \end{tabular}
  }
  \caption{Examples of motion tracking results. 3D motion fields generated by MulViMotion are used to warp 3D segmentations and SAX stacks from ED frame to the $t$-th frame. (a) The warped segmentations overlaid on SAX, 2CH and 4CH view images. (b) The ground truth (GT) and the warped SAX stacks as well as their difference maps (\emph{i.e.}, GT$-$Warped).}
  \label{warpsegimg_proposed}
\end{figure}
\subsubsection{Motion tracking performance} 
\begin{figure}[pt]
    \centering
    \begin{tabular}{@{\hspace{-1\tabcolsep}}c@{\hspace{0.5\tabcolsep}}c@{\hspace{-0.8\tabcolsep}}}
         \subfloat[A single test subject]{\includegraphics[height=3.2cm, trim=0.6cm 0.5cm 0.5cm 0.5cm, clip]{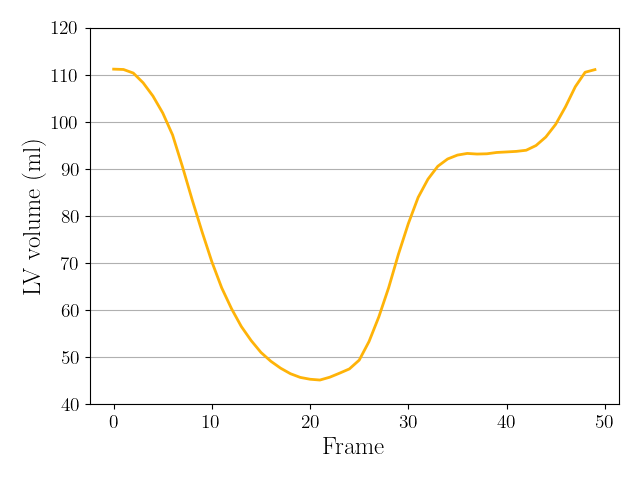}} &
         \subfloat[All test subjects]{\includegraphics[height=3.2cm, trim=0.6cm 0.5cm 0.5cm 0.5cm, clip]{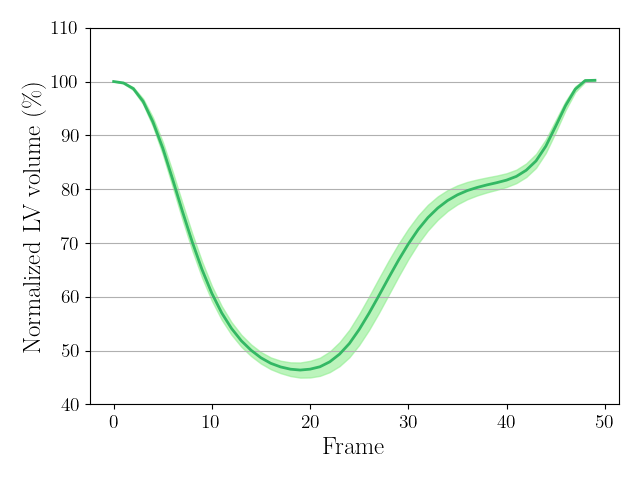}
         }
    \end{tabular}
    \caption{The results of LV volume across the cardiac cycle. (a) Results on a randomly selected test subject. (b) Results on all test subjects (mean values and confidence interval are presented). Note that, for each subject in (b), we normalized LV volume (dividing LV volume in all time frames by that in the ED frame) and show the average results of all test subjects.}
    \label{lv_vol_mass}
\end{figure}

For each test subject, MulViMotion is utilized to estimate 3D motion fields in the full cardiac cycle. With the obtained $\{\Phi_t|t=[0,49]\}$, we warp the 3D segmentation of ED frame ($t=0$) to the $t$-th frame. Fig.~\ref{warpsegimg_proposed} (a) shows that the estimated $\Phi_t$ enables the warped 3D segmentation to match the myocardial area in images from different anatomical views. In addition, we warp the SAX stack of the ED frame ($I_0^{sa}$) to the $t$-th frame. Fig.~\ref{warpsegimg_proposed} (b) shows the effectiveness of $\Phi_t$ by comparing the warped and the ground truth SAX view images. By utilizing the warped 3D segmentation, we further compute established clinical biomarkers.
Fig.~\ref{lv_vol_mass} demonstrates the curve of LV volume over time. The shape of the curve are consistent with reported results in the literature~\cite{Qin2018,Clough2019}.

\begin{table*}[htb]
\centering
\caption{Comparison of other cardiac motion tracking methods. $\uparrow$ indicates the higher value the better while $\downarrow$ indicates the lower value the better. Results are reported as ``mean (standard deviation)" for Dice, Hausdorff distance (HD), volume difference (VD) and negative Jacobian determinant ($det(J_{\Phi_{ES}})<0$). CPU and GPU runtimes are reported as the average inference time for a single subject. Best results in bold.}
\label{comparison_methods}
\resizebox{\textwidth}{!}{
\begin{tabular}{lccccccc}
\toprule[1.2pt]
Methods                            & 
Anatomical views                   &
Dice $\uparrow$                    &
HD (mm) $\downarrow$             &
VD ($\%$) $\downarrow$             &
$det(J_{\Phi_{ES}})<0$ ($\%$) $\downarrow$             &
Times CPU (s) $\downarrow$             &
Times GPU (s) $\downarrow$             \\
\midrule
FFD~\cite{Rueckert1999}                        &
SAX                                              &
0.7250 (0.0511)                                              &
20.1138 (5.1130)                                              &
14.45 (6.87)                                              &
11.94 (5.01)                                              &
15.91                                              &
--                                              \\
dDemons~\cite{Vercauteren2007}                            &
SAX                                  &
0.7219 (0.0422)                                              &
18.3945 (3.5650)                                             &
14.46 (6.38)                                              &
\textbf{0.13 (0.17)}                                             &
28.32                                            &
--                                              \\
3D-UNet~\cite{ociek2016}                                &
SAX                                             &
0.7382 (0.0293)                                         &
17.4785 (3.1030)                                               &
30.97 (9.89)                                             &
0.95 (1.05)                                              &
16.88                                            &
\textbf{1.09}                                             \\
\midrule
MulViMotion                                           &
SAX, 2CH, 4CH                                              &
\textbf{0.8200 (0.0348)}                                              &
\textbf{14.5937 (4.2449)}                                            &
\textbf{8.62 (4.85)}                                             &
0.93 (0.94)                                              &
\textbf{3.55}                                              &
1.15                                             \\
\bottomrule[1.2pt]
\end{tabular}
}
\end{table*}

\begin{figure*}[tb]
 \centering
 \begin{tabular}{@{\hspace{-0.5\tabcolsep}}c@{\hspace{0.3\tabcolsep}}c@{\hspace{0.3\tabcolsep}}c@{\hspace{0.3\tabcolsep}}c@{\hspace{0.3\tabcolsep}}c@{\hspace{0.3\tabcolsep}}c@{\hspace{0.3\tabcolsep}}c@{\hspace{0.3\tabcolsep}}c}
  \raisebox{1\height}{\rotatebox[origin=c]{90}{\makecell{~\scalebox{0.8}{\textbf{FFD~\cite{Rueckert1999}}}}}}
  \includegraphics[height=2cm, trim=13cm 2cm 14cm 3.5cm, clip]{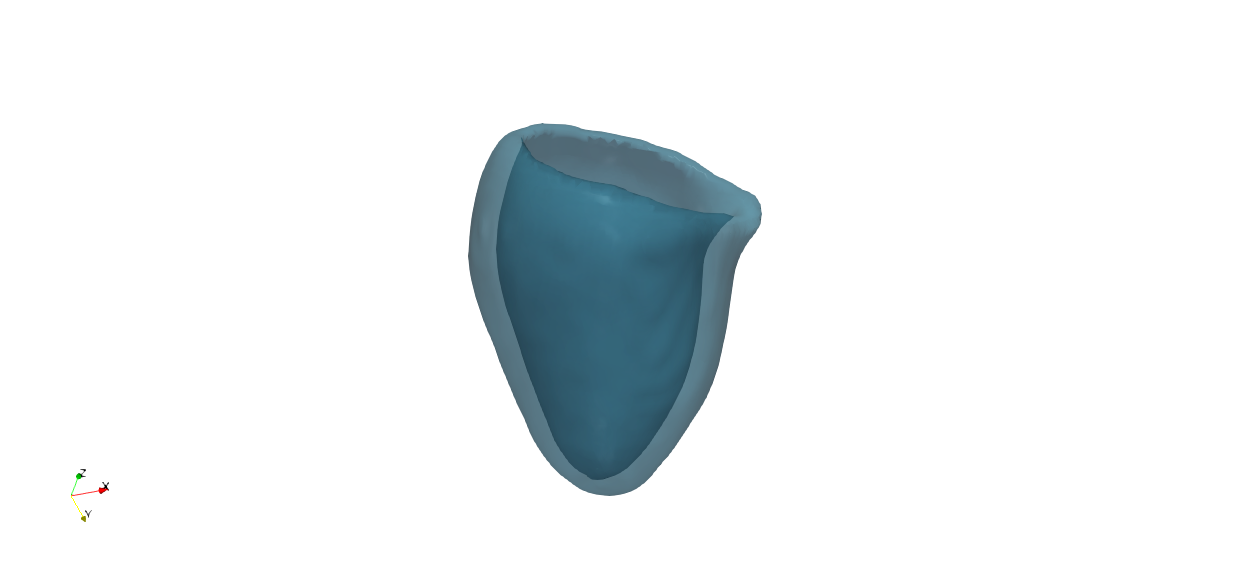} &
  \includegraphics[height=2cm, trim=14cm 2cm 14cm 3.5cm, clip]{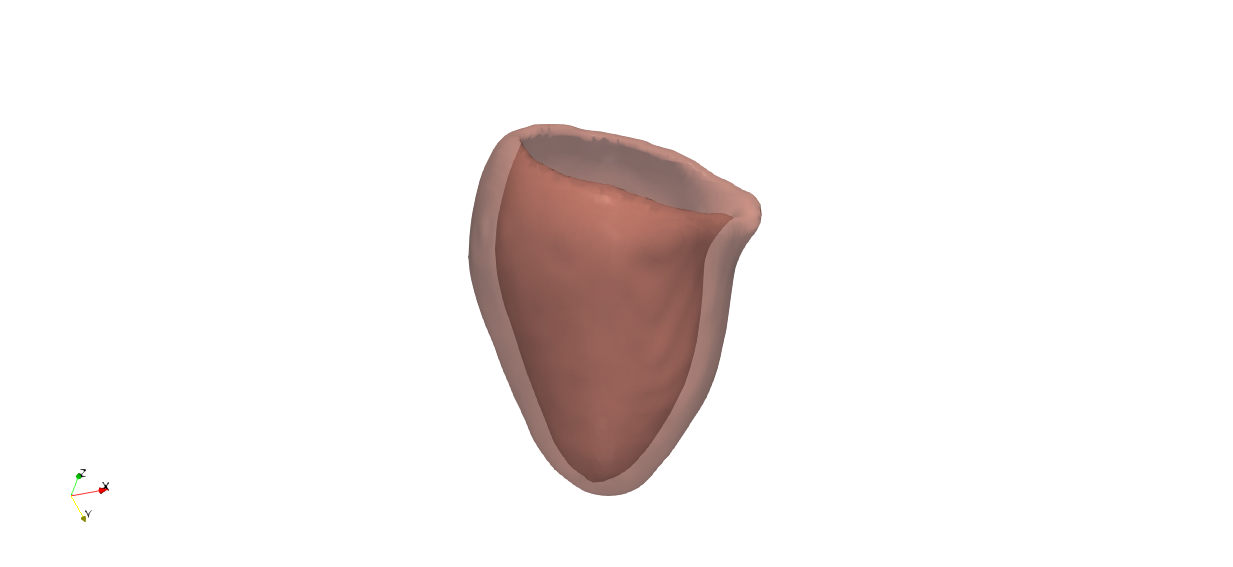} &
  \includegraphics[height=2cm, trim=14cm 2cm 14cm 3.5cm, clip]{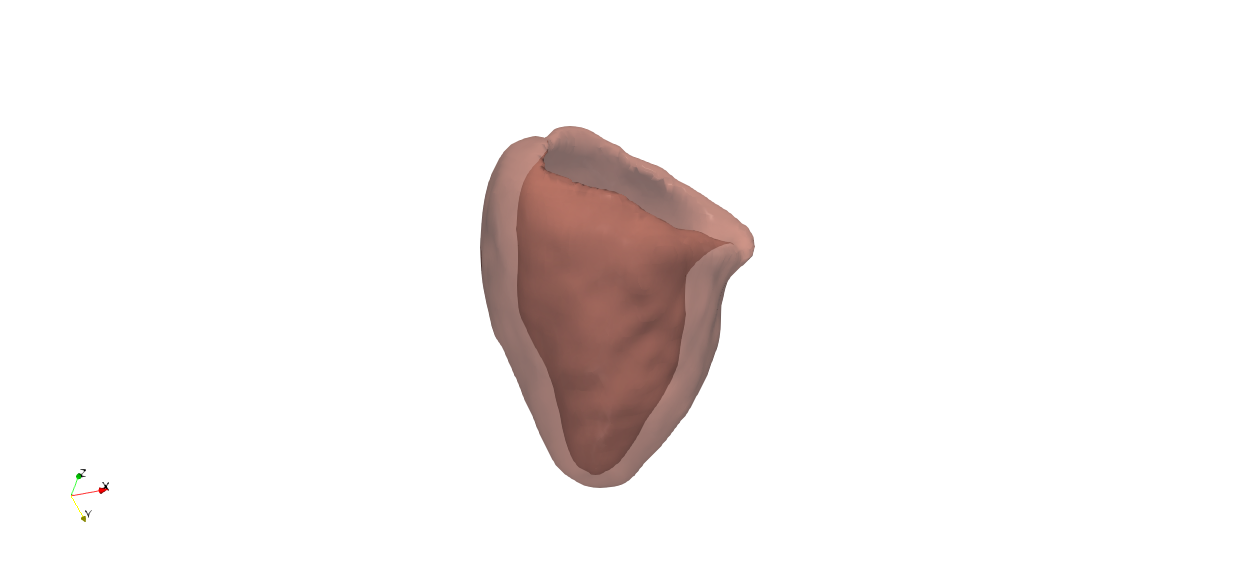} &
  \includegraphics[height=2cm, trim=14cm 2cm 14cm 3.5cm, clip]{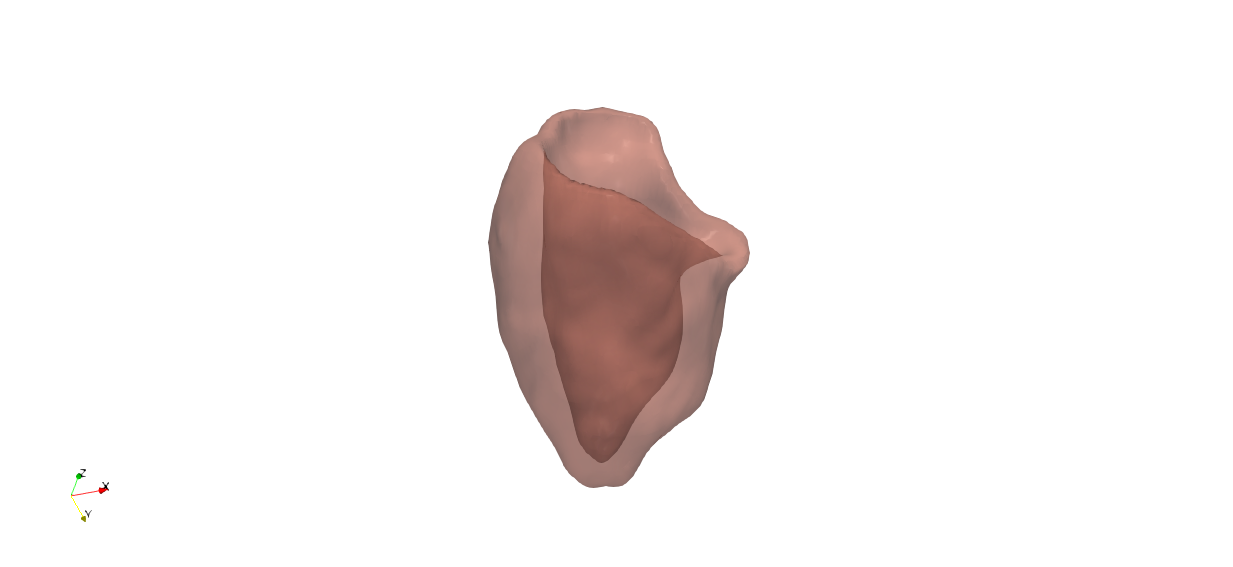} &
  \includegraphics[height=2cm, trim=14cm 2cm 14cm 3.5cm, clip]{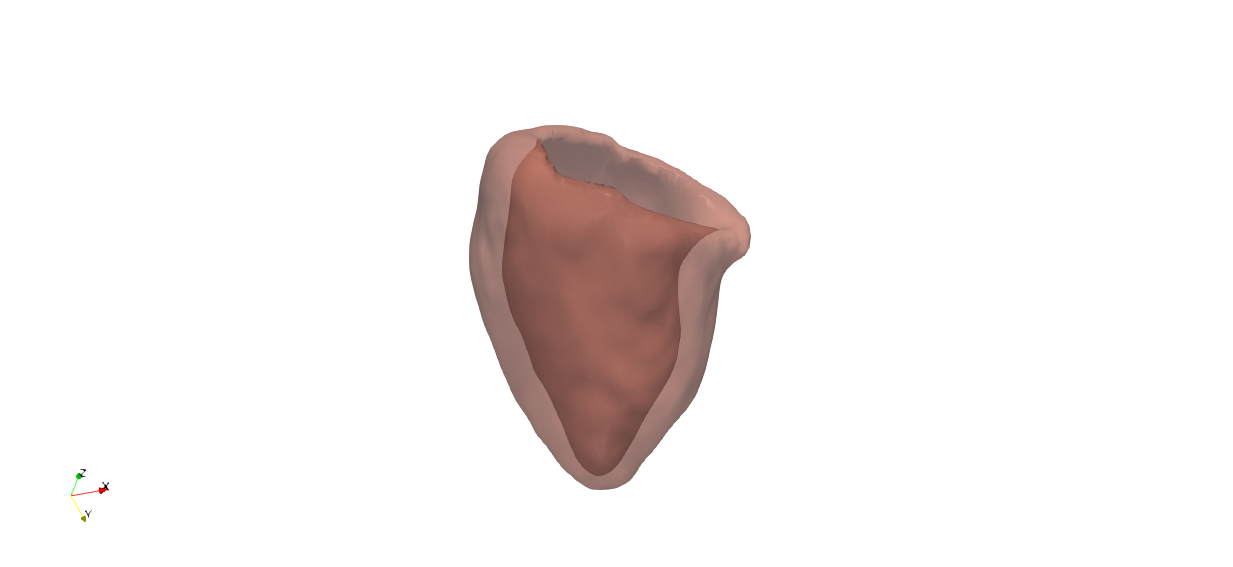} &
  \includegraphics[height=2cm, trim=14cm 2cm 14cm 3.5cm, clip]{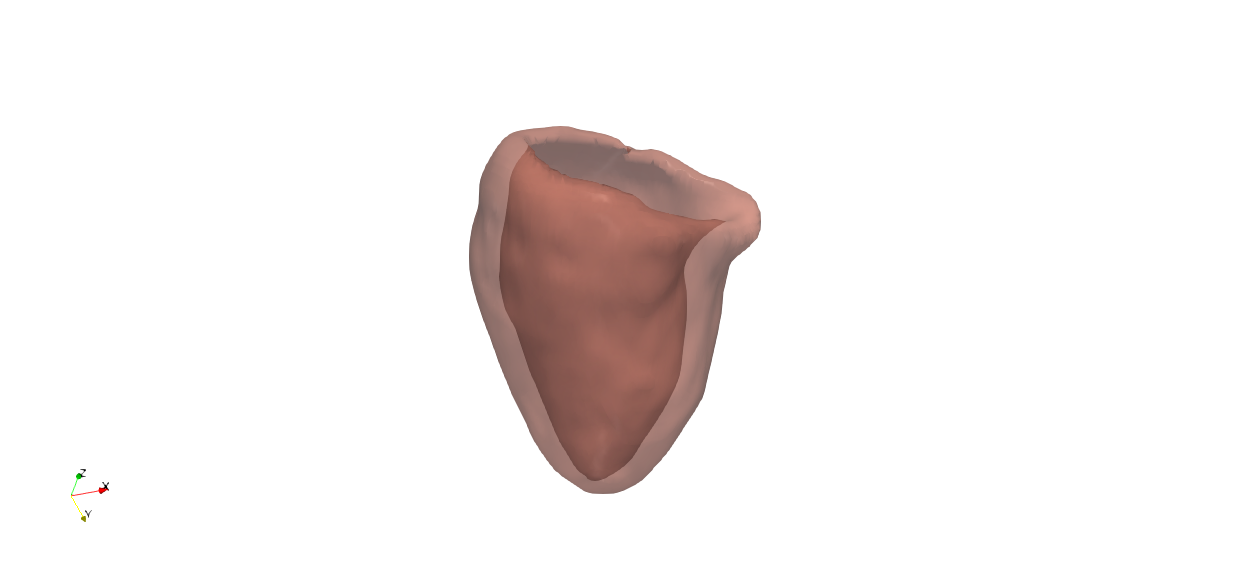} &
  \includegraphics[height=2cm, trim=14cm 2cm 14cm 3.5cm, clip]{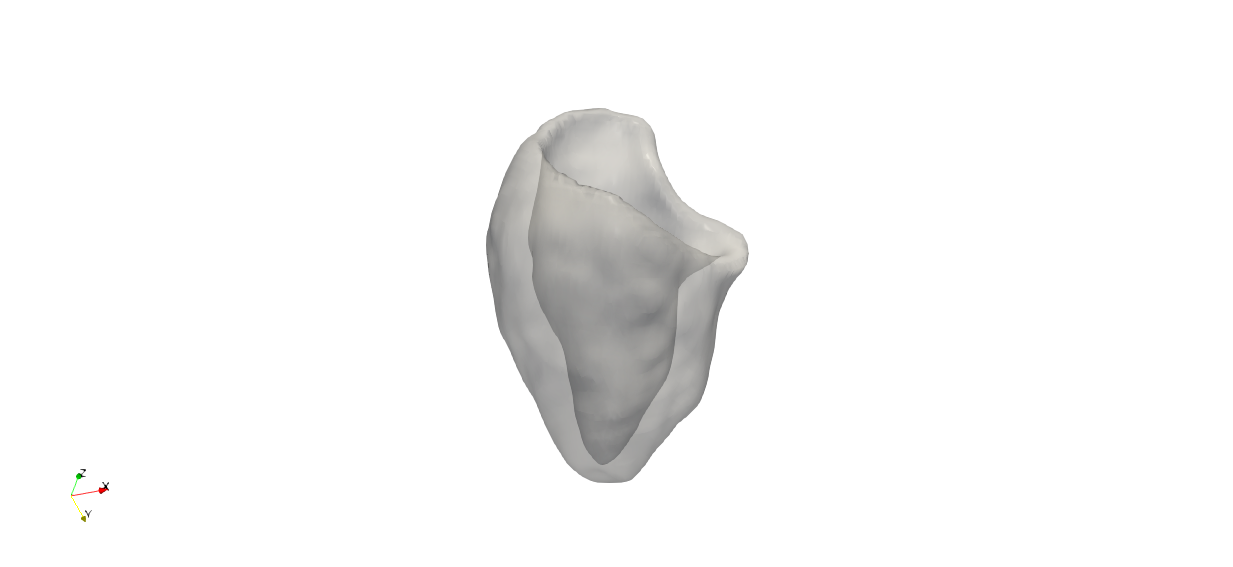} &
  \includegraphics[height=2cm, trim=14cm 2cm 14cm 3.5cm, clip]{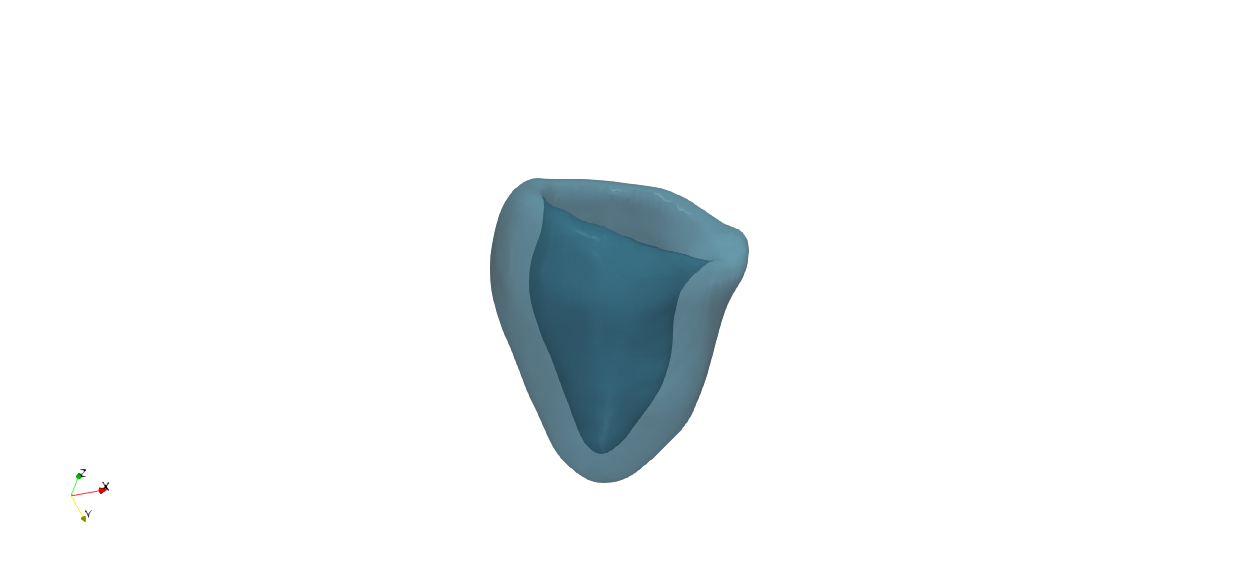} \\
  \raisebox{1\height}{\rotatebox[origin=c]{90}{\makecell{~\scalebox{0.8}{\textbf{dDemons~\cite{Vercauteren2007}}}}}}
  \includegraphics[height=2cm, trim=13cm 2cm 14cm 3.5cm, clip]{ed_gt.png} &
  \includegraphics[height=2cm, trim=14cm 2cm 14cm 3.5cm, clip]{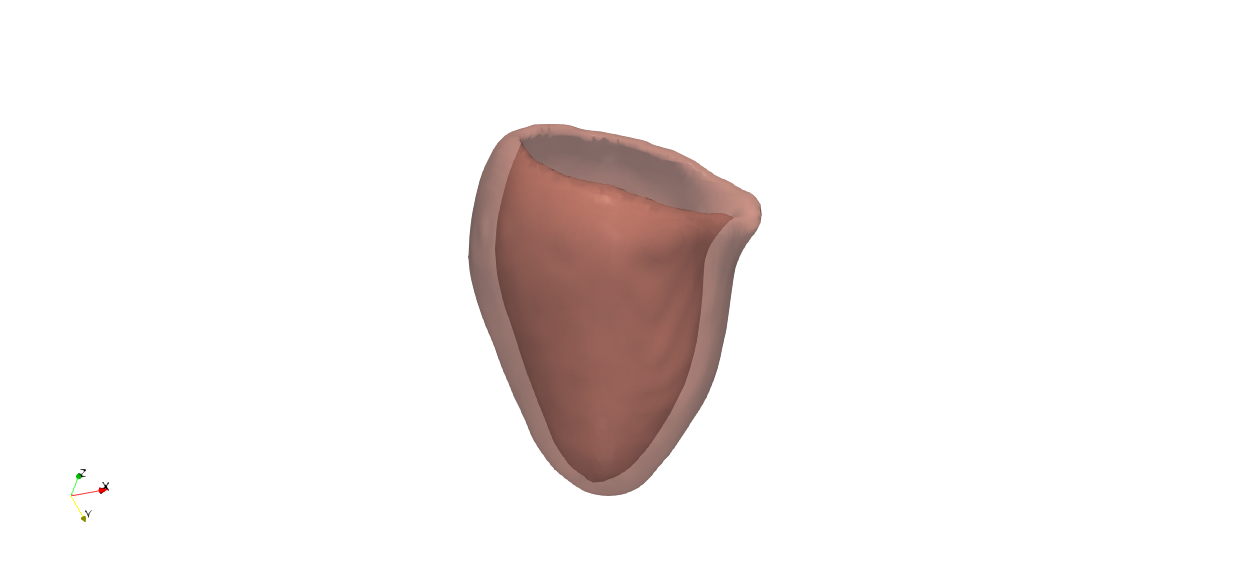} &
  \includegraphics[height=2cm, trim=14cm 2cm 14cm 3.5cm, clip]{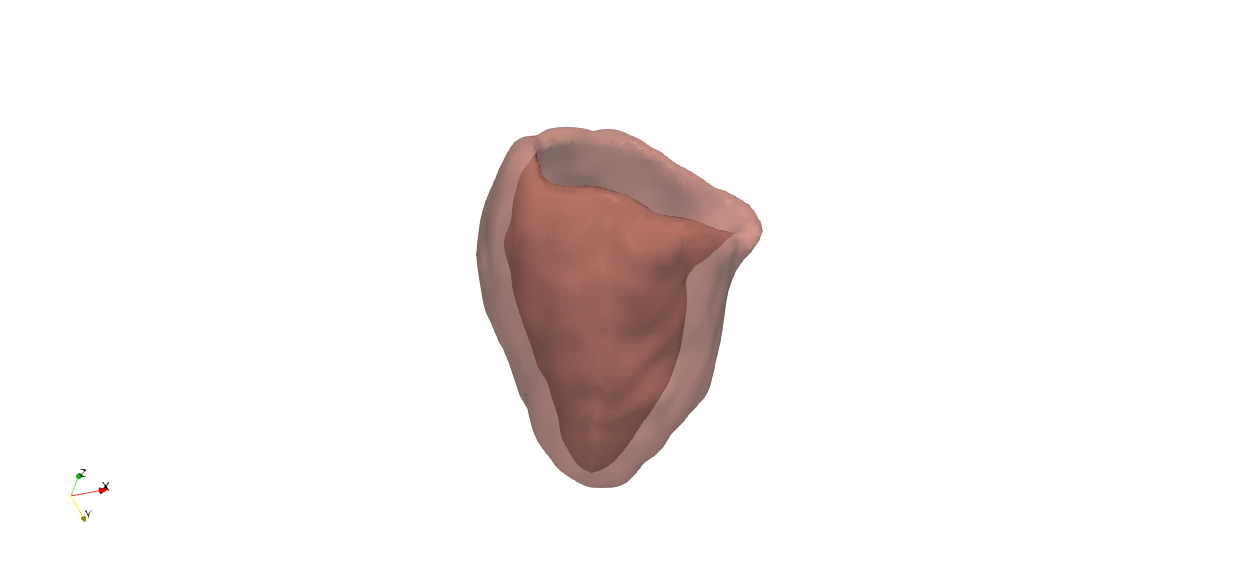} &
  \includegraphics[height=2cm, trim=14cm 2cm 14cm 3.5cm, clip]{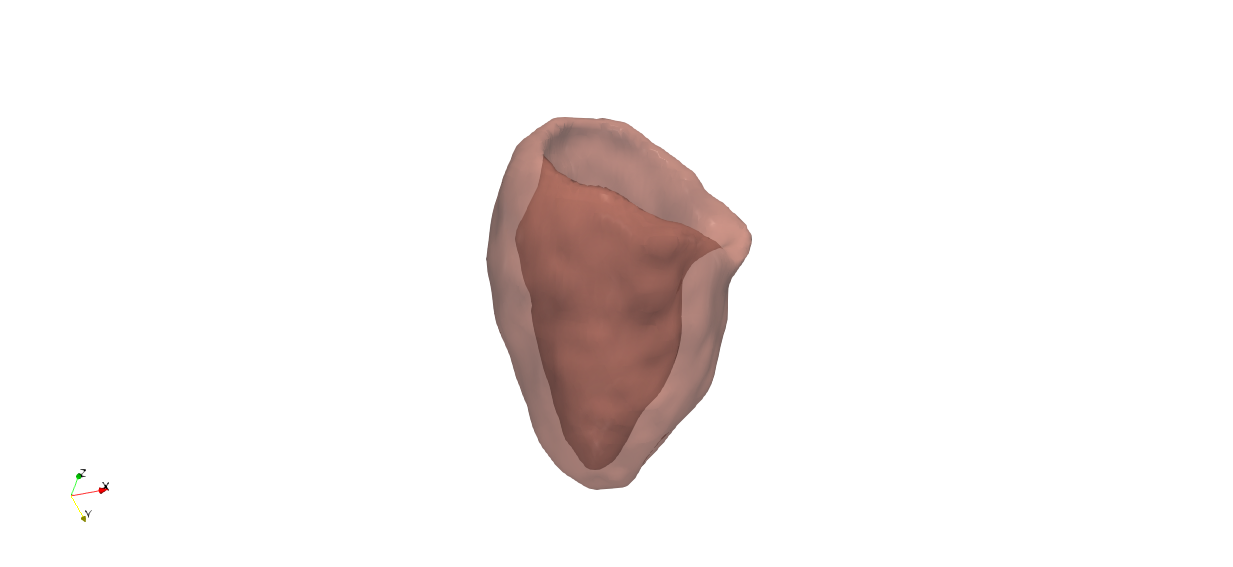} &
  \includegraphics[height=2cm, trim=14cm 2cm 14cm 3.5cm, clip]{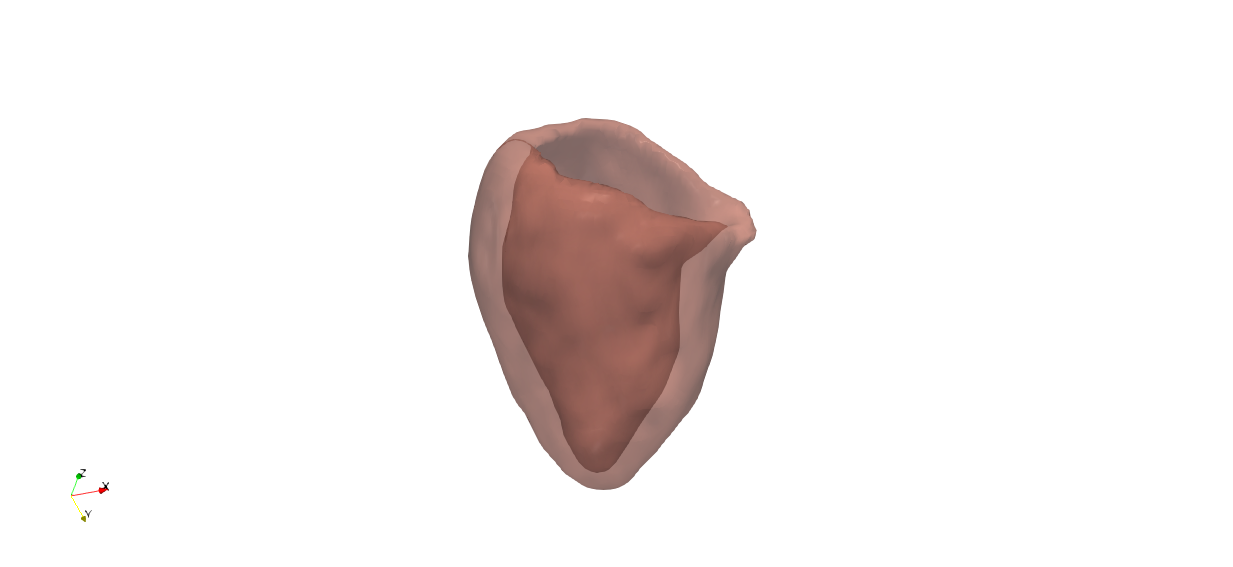} &
  \includegraphics[height=2cm, trim=14cm 2cm 14cm 3.5cm, clip]{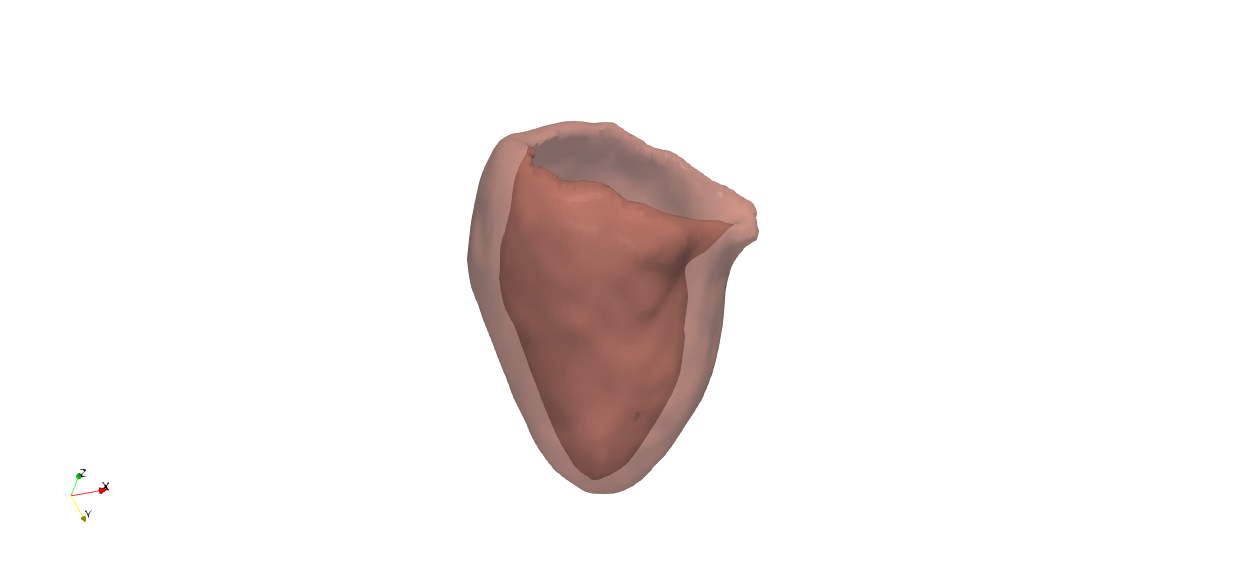} &
  \includegraphics[height=2cm, trim=14cm 2cm 14cm 3.5cm, clip]{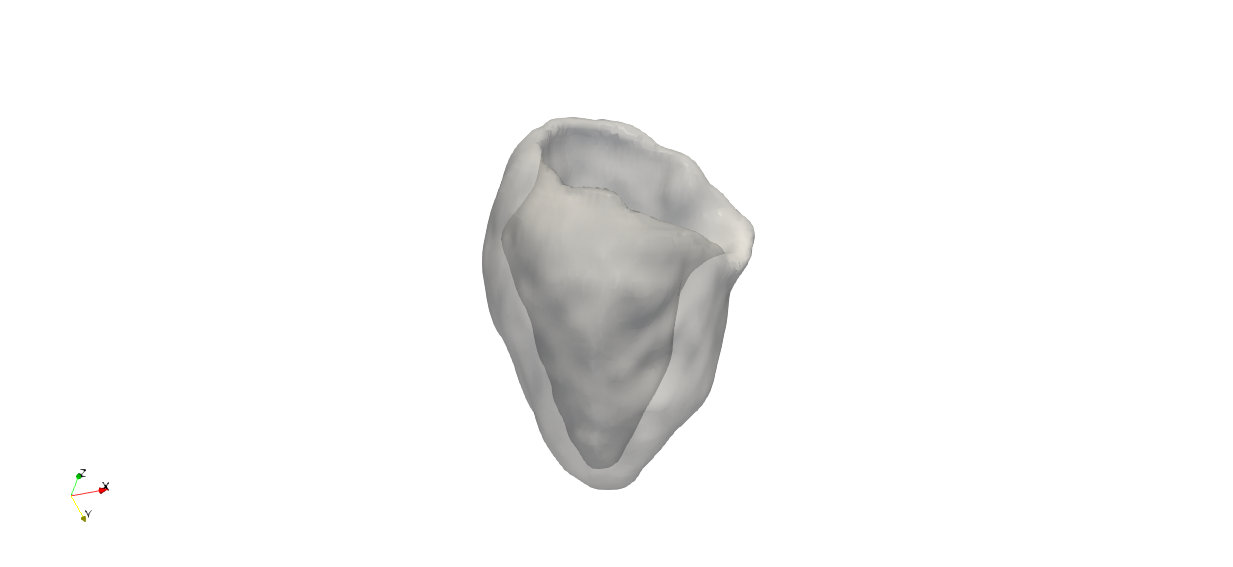} &
  \includegraphics[height=2cm, trim=14cm 2cm 14cm 3.5cm, clip]{es_gt.png} \\
  \raisebox{1\height}{\rotatebox[origin=c]{90}{\makecell{~\scalebox{0.8}{\textbf{3D-UNet~\cite{ociek2016}}}}}}
  \includegraphics[height=2cm, trim=13cm 2cm 14cm 3.5cm, clip]{ed_gt.png} &
  \includegraphics[height=2cm, trim=14cm 2cm 14cm 3.5cm, clip]{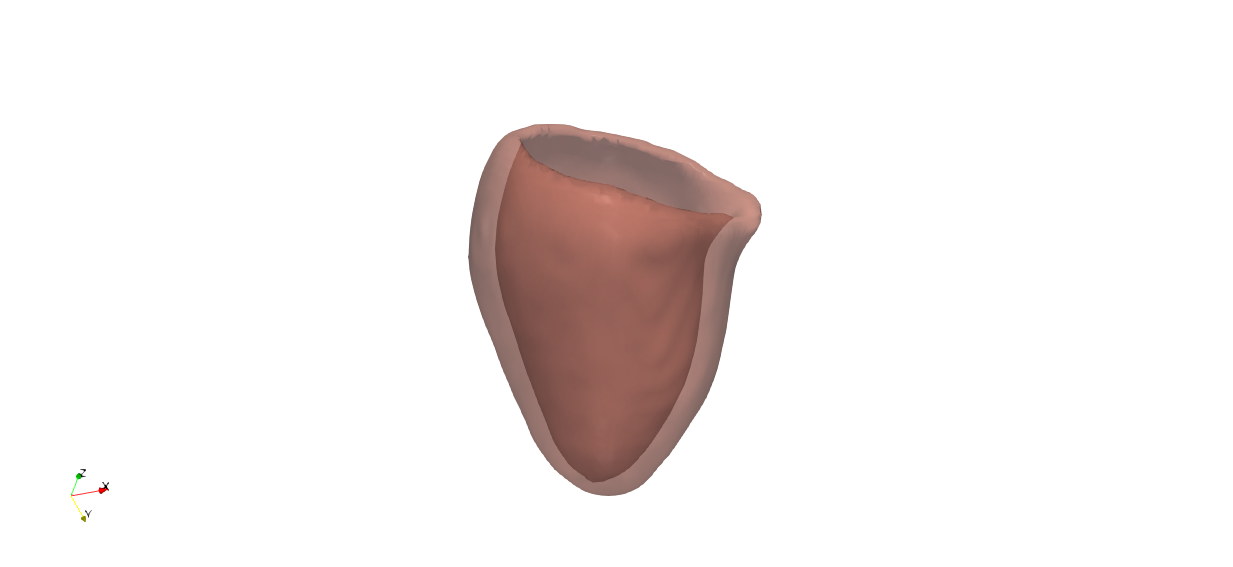} &
  \includegraphics[height=2cm, trim=14cm 2cm 14cm 3.5cm, clip]{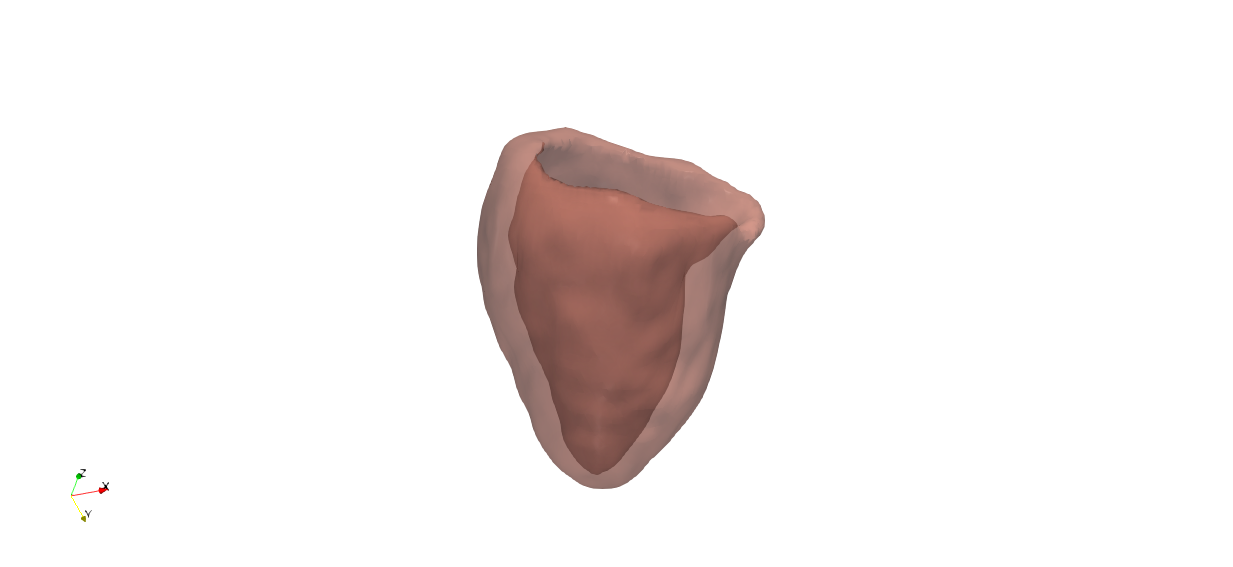} &
  \includegraphics[height=2cm, trim=14cm 2cm 14cm 3.5cm, clip]{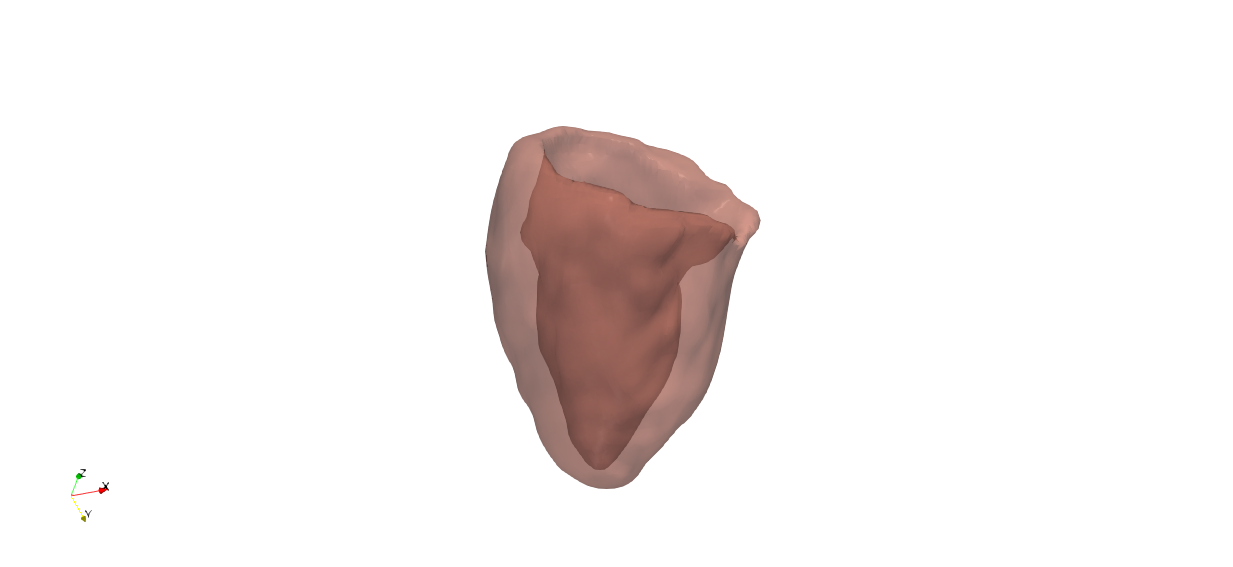} &
  \includegraphics[height=2cm, trim=14cm 2cm 14cm 3.5cm, clip]{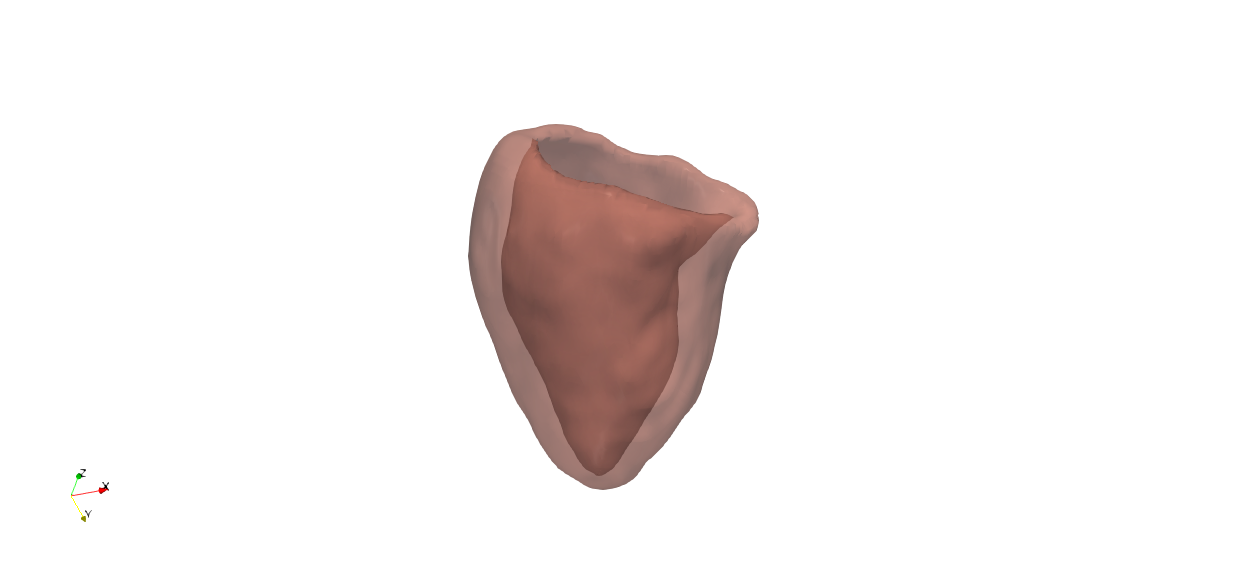} &
  \includegraphics[height=2cm, trim=14cm 2cm 14cm 3.5cm, clip]{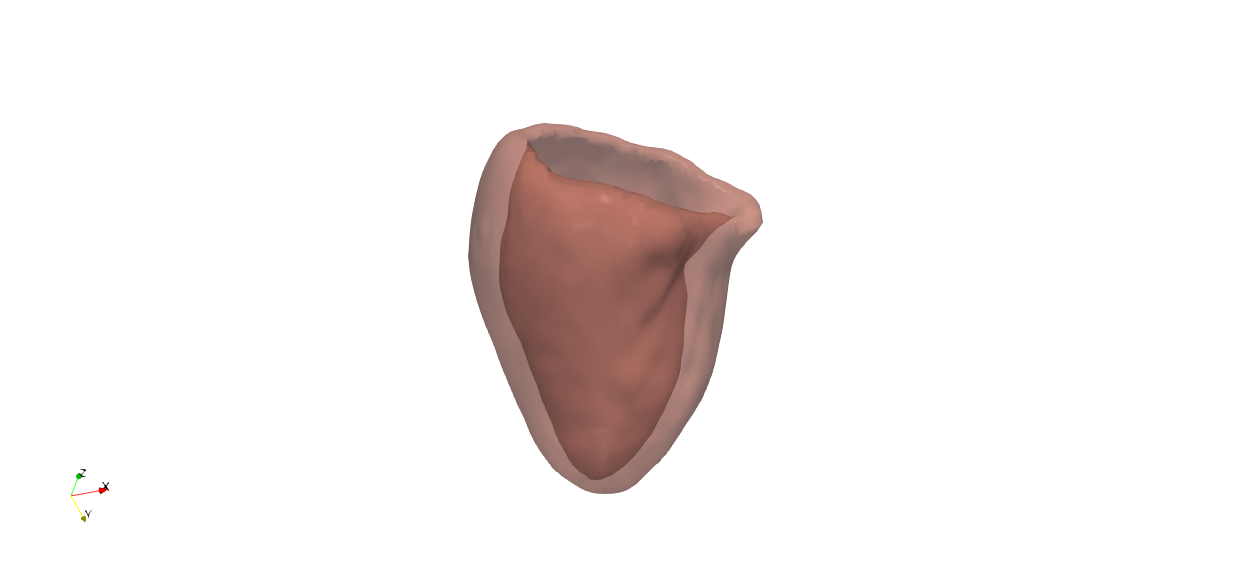} &
  \includegraphics[height=2cm, trim=14cm 2cm 14cm 3.5cm, clip]{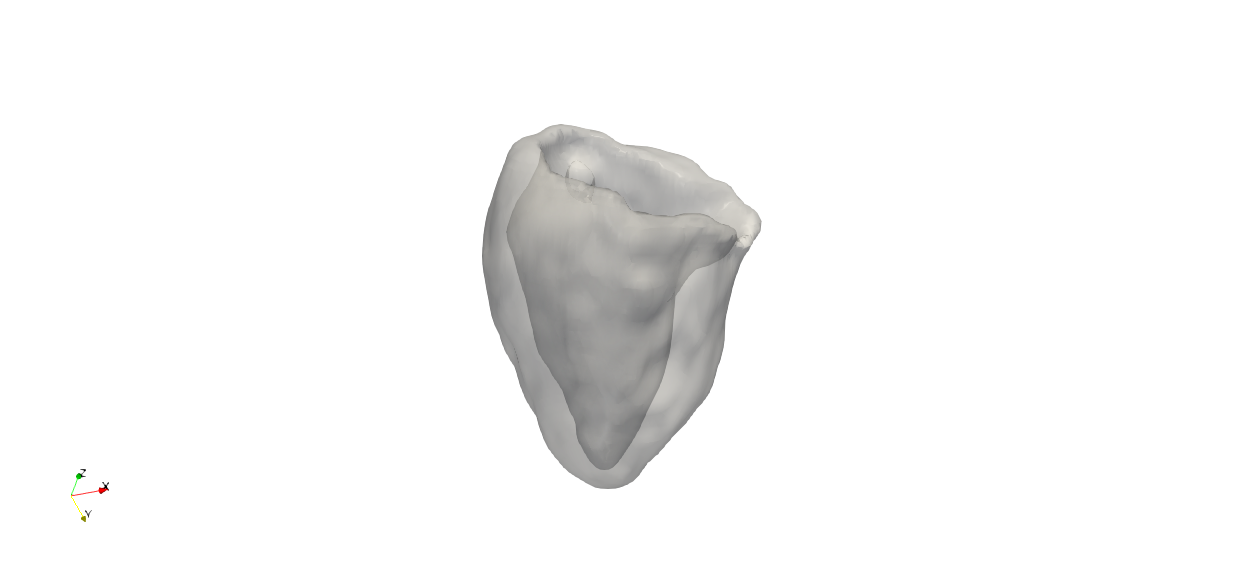} &
  \includegraphics[height=2cm, trim=14cm 2cm 14cm 3.5cm, clip]{es_gt.png} \\
  \raisebox{1\height}{\rotatebox[origin=c]{90}{\makecell{~\scalebox{0.8}{\textbf{MulViMotion}}}}}
  \includegraphics[height=2cm, trim=13cm 2cm 14cm 3.5cm, clip]{ed_gt.png} &
  \includegraphics[height=2cm, trim=14cm 2cm 14cm 3.5cm, clip]{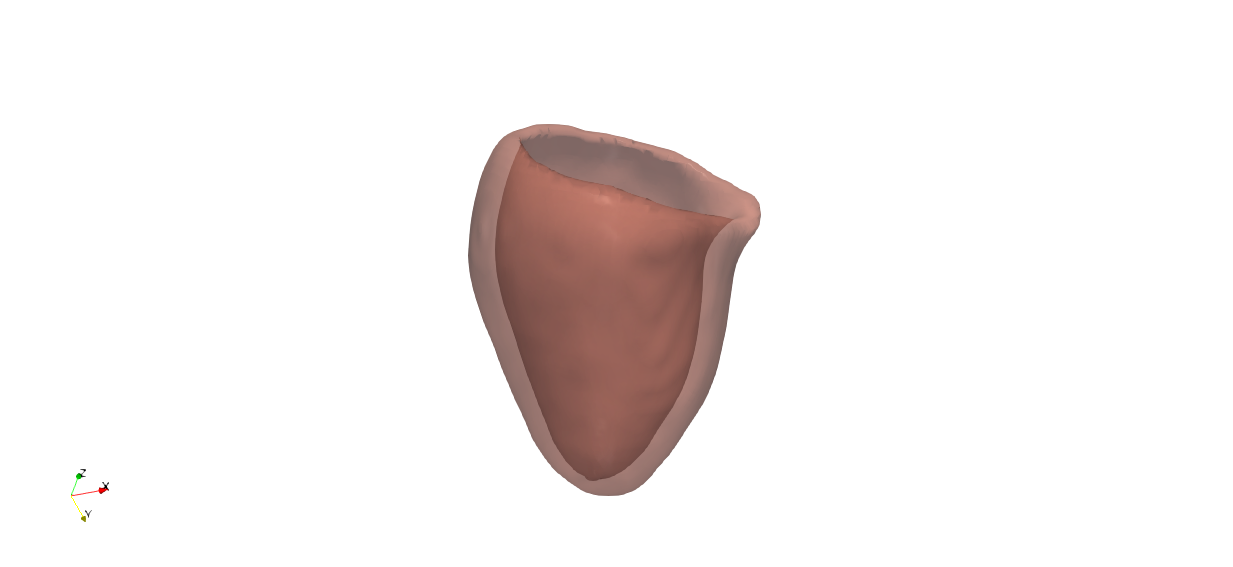} &
  \includegraphics[height=2cm, trim=14cm 2cm 14cm 3.5cm, clip]{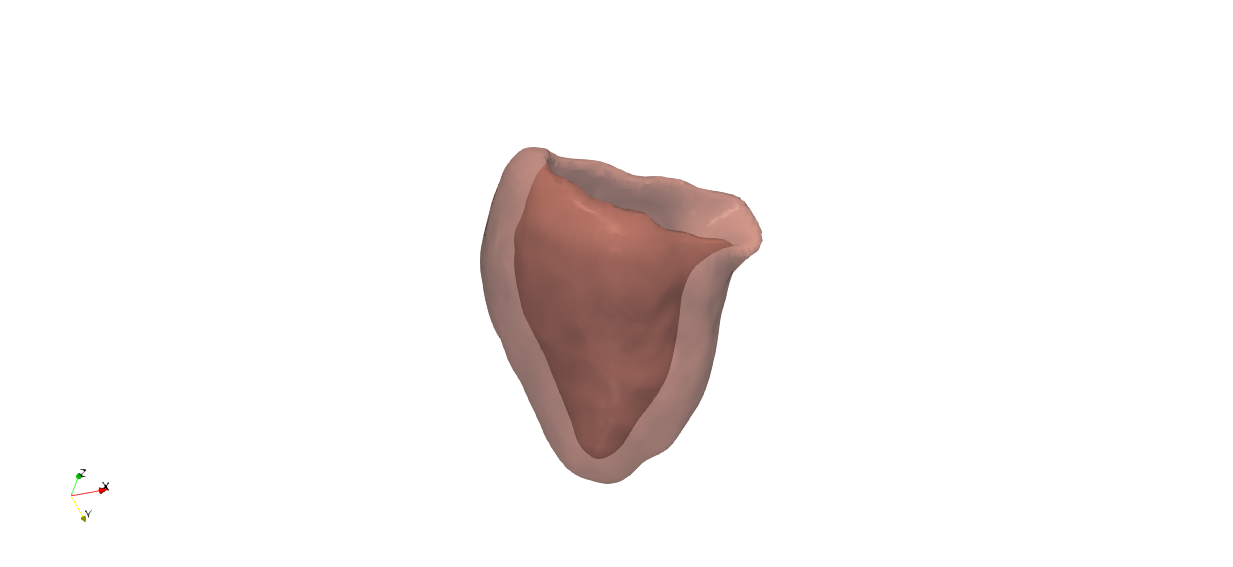} &
  \includegraphics[height=2cm, trim=14cm 2cm 14cm 3.5cm, clip]{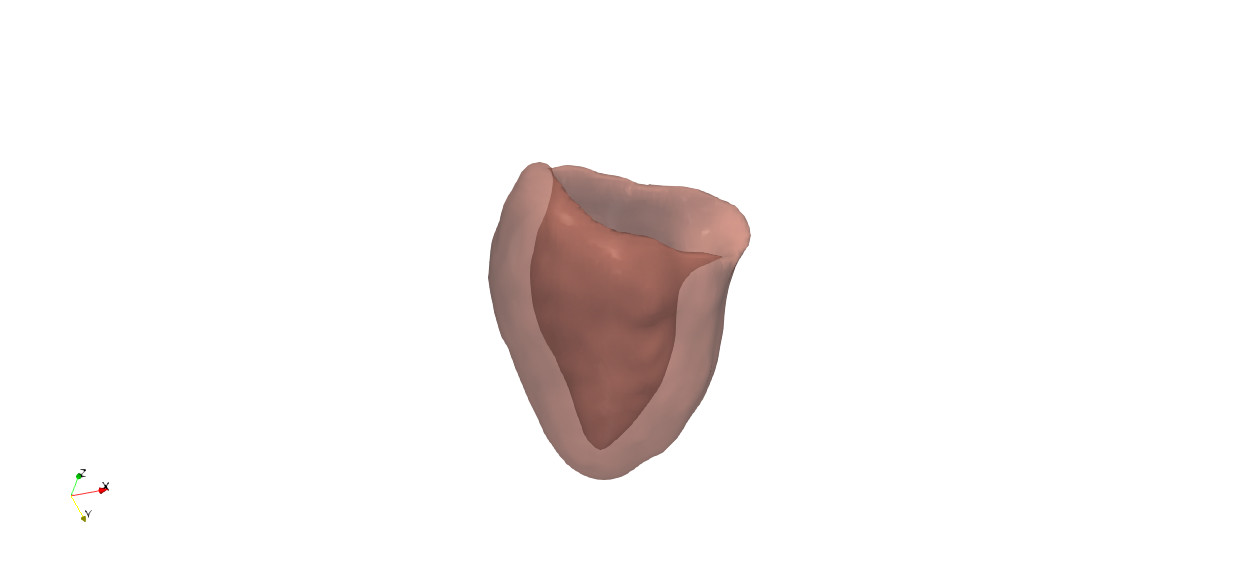} &
  \includegraphics[height=2cm, trim=14cm 2cm 14cm 3.5cm, clip]{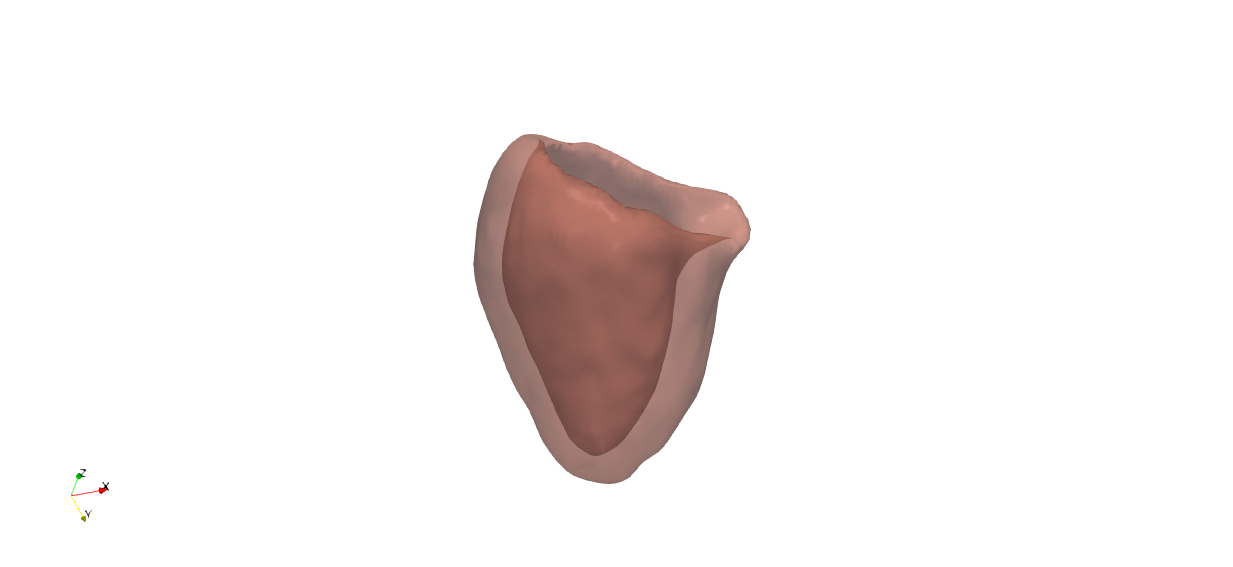} &
  \includegraphics[height=2cm, trim=14cm 2cm 14cm 3.5cm, clip]{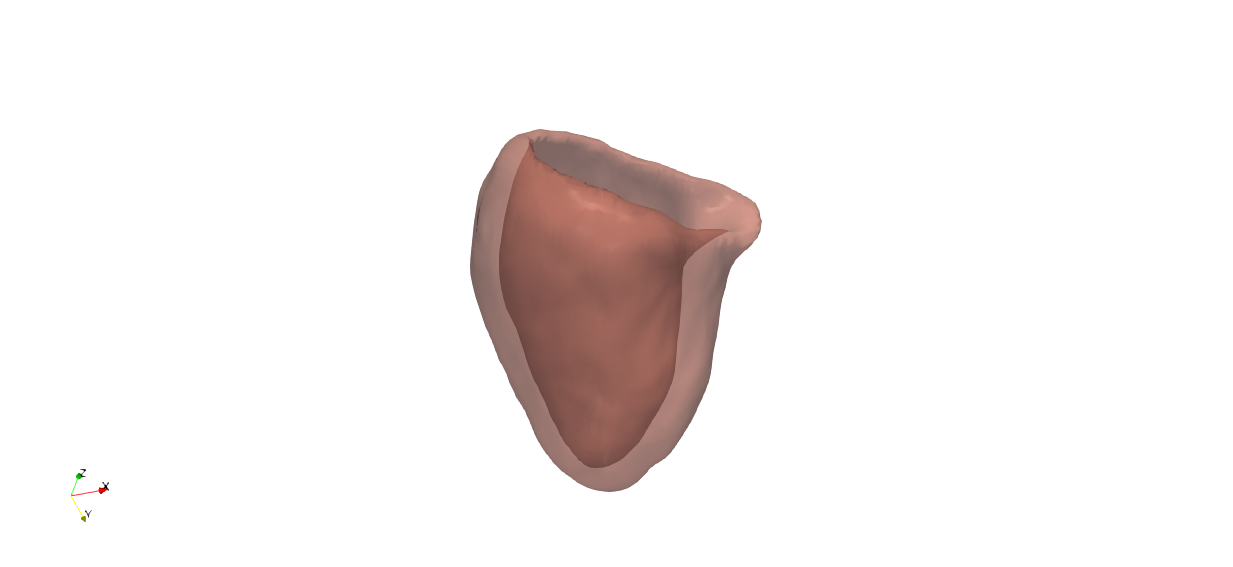} &
  \includegraphics[height=2cm, trim=14cm 2cm 14cm 3.5cm, clip]{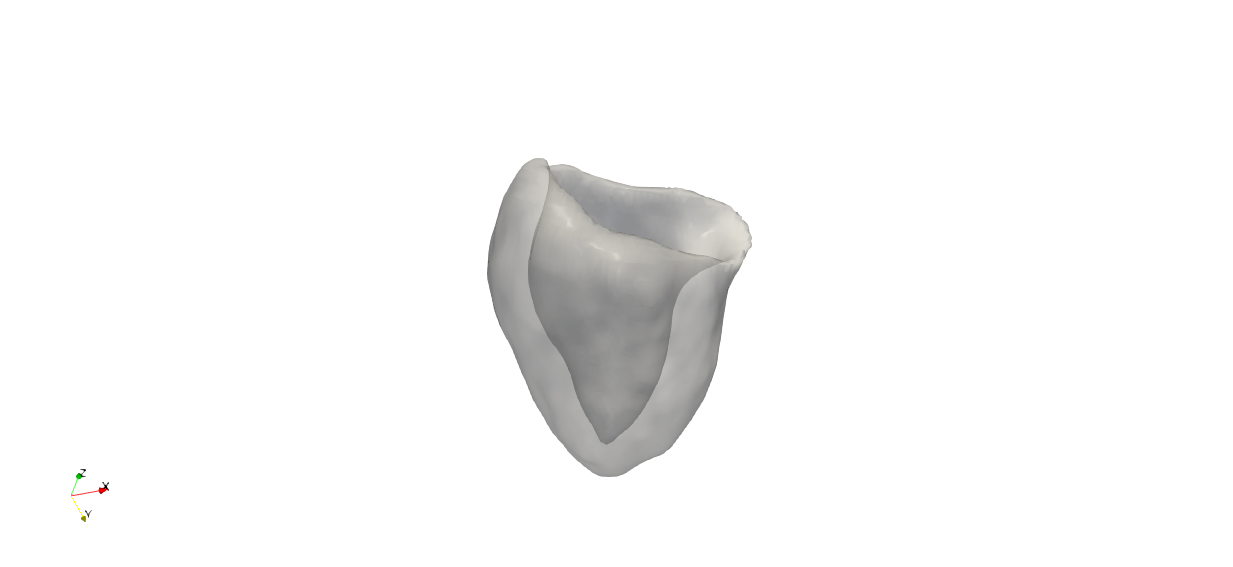} &
  \includegraphics[height=2cm, trim=14cm 2cm 14cm 3.5cm, clip]{es_gt.png} \\
  ~~~
  \raisebox{0.1\height}{\rotatebox[origin=c]{0}{\makecell{~\scalebox{0.8}{\textbf{ED frame (GT)}}}}} &
  \raisebox{0.1\height}{\rotatebox[origin=c]{0}{\makecell{~\scalebox{0.8}{\textbf{t=0}}}}} &
  \raisebox{0.1\height}{\rotatebox[origin=c]{0}{\makecell{~\scalebox{0.8}{\textbf{t=10}}}}} &
  \raisebox{0.1\height}{\rotatebox[origin=c]{0}{\makecell{~\scalebox{0.8}{\textbf{t=20}}}}} &
  \raisebox{0.1\height}{\rotatebox[origin=c]{0}{\makecell{~\scalebox{0.8}{\textbf{t=30}}}}} &
  \raisebox{0.1\height}{\rotatebox[origin=c]{0}{\makecell{~\scalebox{0.8}{\textbf{t=40}}}}} &
  \raisebox{0.1\height}{\rotatebox[origin=c]{0}{\makecell{~\scalebox{0.8}{\textbf{ES frame (warped)}}}}} &
  \raisebox{0.1\height}{\rotatebox[origin=c]{0}{\makecell{~\scalebox{0.8}{\textbf{ES frame (GT)}}}}}
  \end{tabular}
  \caption{3D visualization of motion tracking results using the baseline methods and MulViMotion. Column 1 (blue) shows the ground truth (GT) meshes of ED frame. Columns 2-6 (red) show 3D motion tracking results across the cardiac cycle. These meshes are reconstructed from the warped 3D segmentations (warped from ED frame to different time frames). Column 7 (white) additionally shows the reconstructed meshes of ES frame from the motion tracking results and Column 8 (blue) shows the ground truth meshes of ES frame.}
  \label{comparison_methods_qualitative}
\end{figure*}

We quantitatively compared MulViMotion with baseline methods in Table~\ref{comparison_methods}. With the 3D motion fields generated by different methods, the 3D segmentations of ED frame are warped to ES frame and compared with the ground truth 3D segmentations of ES frame by using metrics introduced in Sec.~\ref{metrics}. From this table, we observe that MulViMotion outperforms all baseline methods for Dice and Hausdorff distance, demonstrating the effectiveness of the proposed method on estimating 3D motion fields. MulViMotion achieves the lowest volume difference, indicating that the proposed method is more capable of preserving the volume of the myocardial wall during cardiac motion tracking. Compared to a diffeomorphic motion tracking method (dDemons~\cite{Vercauteren2007}), the proposed method has a similar number of voxels with a negative Jacobian determinant. This illustrates that the learned motion field is smooth and preserves topology. 

We further qualitatively compared MulViMotion with baseline methods in Fig.~\ref{comparison_methods_qualitative}. A geometric mesh is used to provide 3D visualization of the myocardial wall. Specifically, 3D segmentations of ED frame are warped to any $t$-th frame in the cardiac cycle and geometric meshes are reconstructed from these warped 3D segmentations.
Red meshes in Fig.~\ref{comparison_methods_qualitative} demonstrate that in contrast to all baseline methods which only show motion within SAX plane (\emph{i.e.}, along the $X$ and $Y$ directions), MulViMotion is able to estimation through-plane motion along the longitudinal direction (\emph{i.e.}, the $Z$ direction) in the cardiac cycle, \emph{e.g.}, the reconstructed meshes of $t=20$ frame is deformed in the $X$, $Y$, $Z$ directions compared to $t=0$ and $t=40$ frames. In addition, white meshes in Fig.~\ref{comparison_methods_qualitative} illustrate that compared to all baseline methods, the 3D motion field generated by MulViMotion performs best in warping ED frame to ES frame and obtains the reconstructed mesh of ES frame which is most similar to the ground truth (GT) ES frame mesh (blue meshes). These results demonstrate the effectiveness of MulViMotion for 3D motion tracking, especially for estimating through-plane motion. 

\subsubsection{Runtime}
Table~\ref{comparison_methods} shows runtime results of MulViMotion and baseline methods using Intel Xeon E5-2643 CPU and NVIDIA Tesla T4 GPU. The average inference time for a single subject is reported. FFD~\cite{Rueckert1999} and dDemons~\cite{Vercauteren2007} are only available on CPUs while the 3D-UNet~\cite{ociek2016} and MulViMotion are available on both CPU and GPU. The results show that our method achieves similar runtime to 3D-UNet~\cite{ociek2016} on GPU and at least 5 times faster than baseline methods on CPU.

\subsubsection{Ablation study}
For the proposed method, we explore the effects of using different anatomical views and the importance of the shape regularization module. We use evaluation metrics in Sec.~\ref{metrics} to show quantitative results.

Table~\ref{diffviews} shows the motion tracking results using different anatomical views. In particular, \textit{M1} only uses images and 2D edge maps from SAX view to train the proposed method, \textit{M2} uses those from both SAX and 2CH views and \textit{M3} uses those from both SAX and 4CH views. \textit{M2} and \textit{M3} outperforms \textit{M1}, illustrating the importance of LAX view images. In addition, MulViMotion (\textit{M}) outperforms other variant models. This might be because more LAX views can introduce more high-resolution 3D anatomical information for 3D motion tracking.   

\begin{table}[tb]
\centering
\caption{3D motion tracking \textbf{with different anatomical views}. \textit{M1} and \textit{M2} are variants of the proposed method and \textit{M} refers to MulViMotion. Results are reported the same way as Table~\ref{comparison_methods}. Best results in bold.}
\label{diffviews}
\resizebox{0.5\textwidth}{!}{
\begin{tabular}{c|ccc|ccc}
\toprule[1.2pt]
~~~   &
\multicolumn{3}{c|}{Anatomical views}                   &
\multirow{2}{*}[-0.5em]{Dice $\uparrow$}    &
\multirow{2}{*}[-0.5em]{HD (mm) $\downarrow$}                   &
\multirow{2}{*}[-0.5em]{VD ($\%$) $\downarrow$}             \\
\cmidrule{2-4}
~~~   &
SAX                        &
2CH                                              &
4CH                                              &
~~~   &
~~~   &
~~~   \\
\midrule
M1  &
$\surd$  &
~~~   &
~~~   &
0.7780 (0.0275)   &
18.2564 (3.4031)   &
30.66 (7.73)   \\
M2  &
$\surd$  &
$\surd$   &
~~~   &
0.7964 (0.0273)   &
18.1014 (3.7146)   &
24.05 (5.24)   \\
M3   &
$\surd$  &
~~~   &
$\surd$  &
0.7904 (0.0305)   &
19.2265 (3.2441)   &
17.50 (4.55)   \\
M  &
$\surd$  &
$\surd$   &
$\surd$  &
\textbf{0.8200 (0.0348)}   &
\textbf{14.5937 (4.2449)}   &
\textbf{8.62 (4.85)}   \\
\bottomrule[1.2pt]
\end{tabular}
}
\end{table}

In Table~\ref{diffloss}, the proposed method is trained using all three anatomical views but optimized by different combination of losses. \textit{A1} optimizes the proposed method without shape regularization (\emph{i.e.}, without $\mathcal{L}_{shape}$ in Eq.~\ref{Loss}). \textit{A2} introduces basic shape regularization on top of  \textit{A1}, which adds $\mathcal{L}_0^S$ and $\mathcal{L}_{0\to t}^S$ for $\mathcal{L}_{shape}$. MulViMotion (\textit{M}) outperforms \textit{A1}, illustrating the importance of shape regularization. MulViMotion also outperforms \textit{A2}. This is likely because $\mathcal{L}_0^S$ and $\mathcal{L}_t^S$ are both needed to guarantee the generation of distinct and correct 3D edge maps for all frames in the cardiac cycle. These results show the effectiveness of all proposed components in $\mathcal{L}_{shape}$.

\begin{table}[tb]
\centering
\caption{3D motion tracking \textbf{with different combination of loss functions}. \textit{A1} optimizes the proposed method without shape regularization (without $\mathcal{L}_{shape}$ in Eq.~\ref{Loss}). \textit{A2} adds basic shape regularization on top of \textit{A1}. \textit{M} refers to MulViMotion. All models are trained by three anatomical views. Results are reported the same way as Table~\ref{comparison_methods}. Best results in bold.}
\label{diffloss}
\resizebox{0.5\textwidth}{!}{
\begin{tabular}{c|ccc|ccc}
\toprule[1.2pt]
~~~   &
\multicolumn{3}{c|}{$\mathcal{L}_{shape}$}                   &
\multirow{2}{*}[-0.5em]{Dice $\uparrow$}    &
\multirow{2}{*}[-0.5em]{HD (mm) $\downarrow$}                   &
\multirow{2}{*}[-0.5em]{VD ($\%$) $\downarrow$}             \\
\cmidrule{2-4}
~~~   &
$\mathcal{L}_0^S$                        &
$\mathcal{L}_t^S$                      &
$\mathcal{L}_{0\to t}^S$                       &
~~~   &
~~~   &
~~~   \\
\midrule
A1  &
~~~  &
~~~   &
~~~   &
0.7134 (0.0316)  &
18.9555 (3.1054)  &
33.93 (10.27)  \\
A2   &
$\surd$  &
~~~   &
$\surd$ &
0.7294 (0.0295)   &
17.5047 (3.7485)   &
12.51 (4.28)   \\
M  &
$\surd$  &
$\surd$   &
$\surd$  &
\textbf{0.8200 (0.0348)}   &
\textbf{14.5937 (4.2449)}   &
\textbf{8.62 (4.85)}   \\
\bottomrule[1.2pt]
\end{tabular}
}
\end{table}

\begin{figure}[t]
    \centering
    \begin{tabular}{@{\hspace{-1\tabcolsep}}c@{\hspace{0.5\tabcolsep}}c@{\hspace{-0.8\tabcolsep}}}
         \includegraphics[height=3.2cm, trim=0cm 0.2cm 1cm 0.6cm, clip]{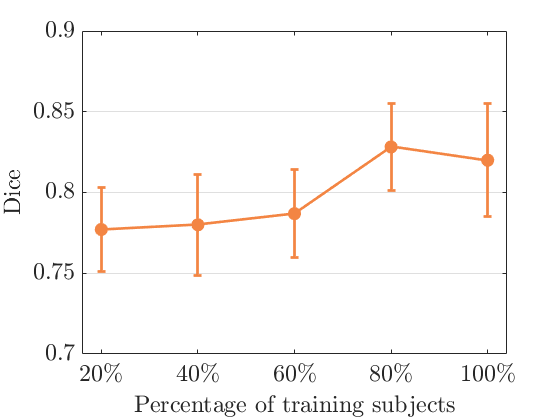} &
         \includegraphics[height=3.2cm, trim=0.2cm 0.2cm 1cm 0.6cm, clip]{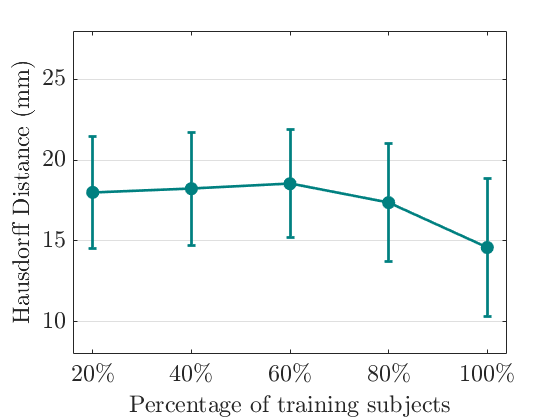}
    \end{tabular}
    \caption{3D motion tracking \textbf{with different strength of shape regularization}, where the shape loss ($\mathcal{L}_{shape}$) is computed by different percentage of training subjects ($20\%, 40\%, 60\%, 80\%, 100\%$). The left column is Dice score and the right column is Hausdorff distance.}
    \label{strength_Lshape}
\end{figure}

Fig.~\ref{strength_Lshape} shows motion estimation performance using different strength of shape regularization. In detail, the proposed method is trained by three anatomical views and all loss components, but the shape loss ($\mathcal{L}_{shape}$) is computed by different percentage of training subjects ($20\%, 40\%, 60\%, 80\%, 100\%$). From Fig.~\ref{strength_Lshape}, we observe that motion estimation performance is improved with an increased percentage of subjects. 

\subsubsection{The influence of hyper-parameters}
\begin{figure}[t]
    \centering
    \setcounter{subfigure}{0}
    \hspace{-0.7cm}
    \subfloat[Various $\lambda$]{
    \begin{tabular}{c}
         \includegraphics[height=3.3cm, trim=0cm 0cm 1.5cm 1cm, clip]{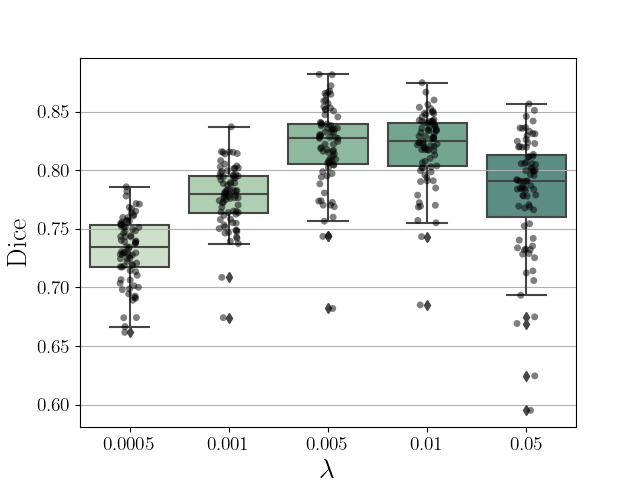} \\
         \includegraphics[height=3.3cm, trim=0cm 0cm 1.5cm 1cm, clip]{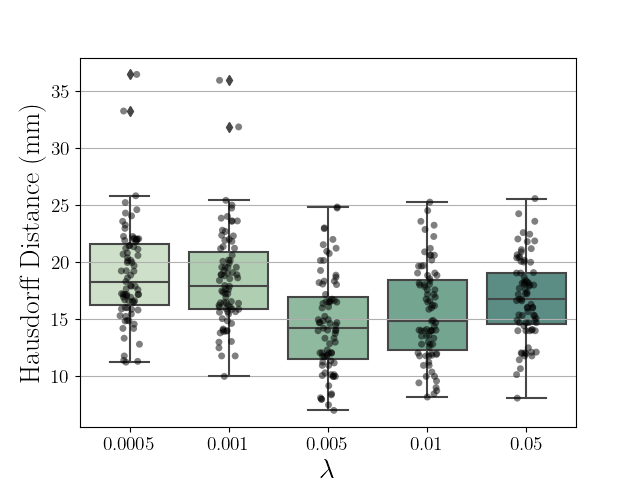}
    \end{tabular}
    }
    \hspace{-0.9cm}
    \setcounter{subfigure}{1}
    \subfloat[Various $\beta$]{
    \begin{tabular}{c}
         \includegraphics[height=3.3cm, trim=0cm 0cm 1.5cm 1cm, clip]{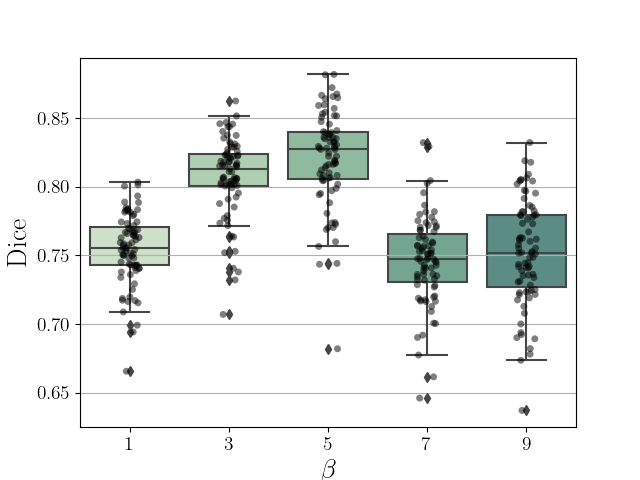}  \\
         \includegraphics[height=3.3cm, trim=0cm 0cm 1.5cm 1cm, clip]{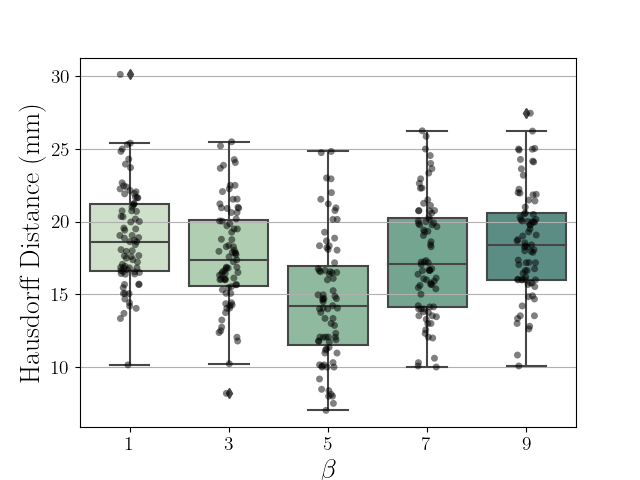} 
    \end{tabular}
    } 
    \caption{Effects of varied hyper-parameters on Dice and Hausdorff distance. (a) shows the results of using various $\lambda$ under $\beta=5$. (b) shows the results of using various $\beta$ under $\lambda =0.005$.}
    \label{hyperpara}
\end{figure}

Fig.~\ref{hyperpara} presents Dice and Hausdorff distance (HD) on the test data for various smoothness loss weight $\lambda$ and shape regularization weight $\beta$ (Eq.~\ref{Loss}). The Dice scores and HDs are computed according to Sec.~\ref{metrics}. We observe that a strong constraint on motion field smoothness may scarify registration accuracy (see Fig.~\ref{hyperpara} (a)).
Moreover, registration performance improves as $\beta$ increases from $1$ to $5$ and then deteriorates with a further increased $\beta$ (from $5$ to $9$).
This might be because a strong shape regularization can enforce motion estimation to focus mainly on the few 2D planes which contain sparse labels.  

\subsubsection{The performance on subjects with slice misalignment}
\begin{figure}[tb]
 \centering
 \subfloat[SAX, 2CH and 4CH views]{
 \begin{tabular}{@{\hspace{-0.5\tabcolsep}}c@{\hspace{0.3\tabcolsep}}c@{\hspace{0.3\tabcolsep}}c@{\hspace{0.3\tabcolsep}}c@{\hspace{0.3\tabcolsep}}c@{\hspace{0.3\tabcolsep}}c}
 \raisebox{0.8\height}{\rotatebox[origin=c]{90}{\makecell{~\scalebox{0.8}{\textbf{ED frame}}}}}
  \includegraphics[height=1.6cm]{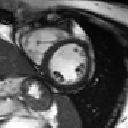} &
  \includegraphics[height=1.6cm]{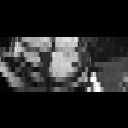} &
  \includegraphics[height=1.6cm]{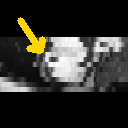} &
  \includegraphics[height=1.6cm]{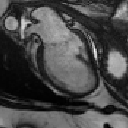} &
  \includegraphics[height=1.6cm]{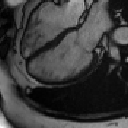} \\
  \end{tabular}
  } \\
 \subfloat[Motion tracking results]{
 \begin{tabular}{@{\hspace{-1.5\tabcolsep}}c@{\hspace{-0.1\tabcolsep}}c@{\hspace{0.1\tabcolsep}}c@{\hspace{0.1\tabcolsep}}c@{\hspace{0.1\tabcolsep}}c@{\hspace{0.1\tabcolsep}}c}
 \multirow{3}{*}[1.7em]{\raisebox{1\height}{\rotatebox[origin=c]{90}{\makecell{~\scalebox{0.7}{\textbf{Motion tracking results on 2CH view}}}}}}
  \raisebox{0.8\height}{\rotatebox[origin=c]{90}{\makecell{~\scalebox{0.6}{3D-UNet~\cite{ociek2016}}}}} &
  \includegraphics[height=1.6cm]{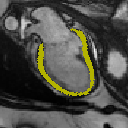} &
  \includegraphics[height=1.6cm]{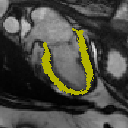} &
  \includegraphics[height=1.6cm]{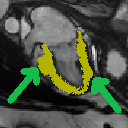} &
  \includegraphics[height=1.6cm]{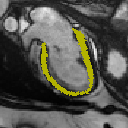} &
  \includegraphics[height=1.6cm]{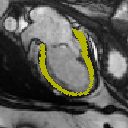} \\
  ~~~
  \raisebox{0.8\height}{\rotatebox[origin=c]{90}{\makecell{~\scalebox{0.6}{MulViMotion}}}} &
  \includegraphics[height=1.6cm]{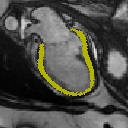} &
  \includegraphics[height=1.6cm]{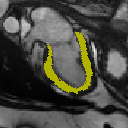} &
  \includegraphics[height=1.6cm]{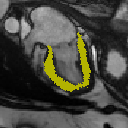} &
  \includegraphics[height=1.6cm]{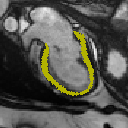} &
  \includegraphics[height=1.6cm]{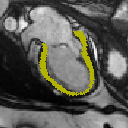} \\
  ~~~
  \raisebox{0.8\height}{\rotatebox[origin=c]{90}{\makecell{~\scalebox{0.6}{\makecell{MulViMotion \\ w/o $\mathbf{\mathcal{L}_{shape}}$}}}}}&
  \includegraphics[height=1.6cm]{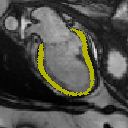} &
  \includegraphics[height=1.6cm]{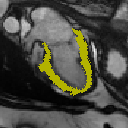} &
  \includegraphics[height=1.6cm]{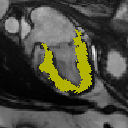} &
  \includegraphics[height=1.6cm]{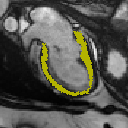} &
  \includegraphics[height=1.6cm]{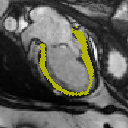} \\
  \multirow{3}{*}[1.7em]{\raisebox{1\height}{\rotatebox[origin=c]{90}{\makecell{~\scalebox{0.7}{\textbf{Motion tracking results on 4CH view}}}}}}
  \raisebox{0.7\height}{\rotatebox[origin=c]{90}{\makecell{~\scalebox{0.6}{3D-UNet~\cite{ociek2016}}}}} &
  \includegraphics[height=1.6cm]{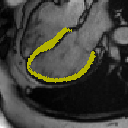} &
  \includegraphics[height=1.6cm]{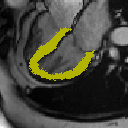} &
  \includegraphics[height=1.6cm]{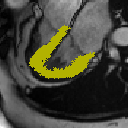} &
  \includegraphics[height=1.6cm]{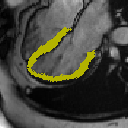} &
  \includegraphics[height=1.6cm]{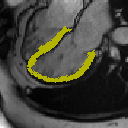} \\
  ~~~
  \raisebox{0.7\height}{\rotatebox[origin=c]{90}{\makecell{~\scalebox{0.6}{MulViMotion}}}} &
  \includegraphics[height=1.6cm]{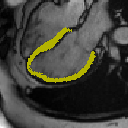} &
  \includegraphics[height=1.6cm]{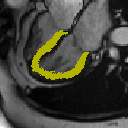} &
  \includegraphics[height=1.6cm]{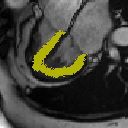} &
  \includegraphics[height=1.6cm]{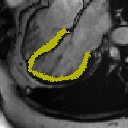} &
  \includegraphics[height=1.6cm]{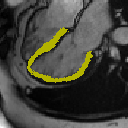} \\
  ~~~ 
  \raisebox{0.7\height}{\rotatebox[origin=c]{90}{\makecell{~\scalebox{0.6}{\makecell{MulViMotion \\ w/o $\mathbf{\mathcal{L}_{shape}}$}}}}}&
  \includegraphics[height=1.6cm]{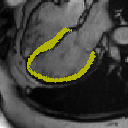} &
  \includegraphics[height=1.6cm]{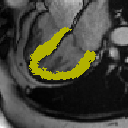} &
  \includegraphics[height=1.6cm]{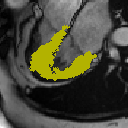} &
  \includegraphics[height=1.6cm]{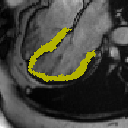} &
  \includegraphics[height=1.6cm]{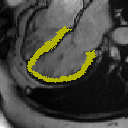} \\
  ~~~ 
  ~~~ &
  \raisebox{0.1\height}{\rotatebox[origin=c]{0}{\makecell{~\scalebox{0.8}{\textbf{t=0}}}}} &
  \raisebox{0.1\height}{\rotatebox[origin=c]{0}{\makecell{~\scalebox{0.8}{\textbf{t=10}}}}} &
  \raisebox{0.1\height}{\rotatebox[origin=c]{0}{\makecell{~\scalebox{0.8}{\textbf{t=17}}}}} &
  \raisebox{0.1\height}{\rotatebox[origin=c]{0}{\makecell{~\scalebox{0.8}{\textbf{t=30}}}}} &
  \raisebox{0.1\height}{\rotatebox[origin=c]{0}{\makecell{~\scalebox{0.8}{\textbf{t=40}}}}}
  \end{tabular}
  }
  \caption{Motion tracking results on the test subject with slice misalignment. The first three columns in (a) are the three orthogonal planes of the SAX stack and the last two columns are 2CH and 4CH view images, respectively. (b) presents examples of motion tracking results using 3D-UNet~\cite{ociek2016}, MulViMotion and MulViMotion without $\mathcal{L}_{shape}$. The yellow arrow shows an example of slice misalignment while green arrows show examples of motion tracking failures using 3D-UNet. Note that we show the results in frame $t=17$ for a more distinct comparison.}
  \label{sliceshift}
\end{figure}

Acquired SAX stacks may contain slice misalignment due to poor compliance with breath holding instructions or the change of position during breath-holding acquisitions~\cite{Biffi2019}. This leads to an incorrect representation of cardiac volume and result in difficulties for accurate 3D motion tracking. Fig.~\ref{sliceshift} compares the motion tracking results of 3D-UNet~\cite{ociek2016}, MulViMotion and MulViMotion without $\mathcal{L}_{shape}$ for the test subject with the severe slice misalignment (\emph{e.g.}, Fig.~\ref{sliceshift} (a) middle column). Fig.~\ref{sliceshift} (b) shows that in contrast to 3D-UNet, the motion fields generated by MulViMotion enables topology preservation of the myocardial wall (\emph{e.g.}, mesh of $t=17$). 
MulViMotion outperforms MulViMotion without $\mathcal{L}_{shape}$, which indicates the importance of the shape regularization module for reducing negative effect of slice misalignment. These results demonstrate the advantage of integrating shape information from multiple views and shows the effectiveness of the proposed method on special cases.

\subsubsection{Wall thickening measurement}
We have computed regional and global myocardial wall thickness at ED frame and ES frame based on ED frame segmentation and warped ES frame segmentation\footnote{Implemented based on https://github.com/baiwenjia/ukbb$\_$cardiac}, respectively. The global wall thickness at ED frame is $6.6\pm0.9mm$, which is consistent with results obtained by~\cite{Bai2020} ($5.5\pm0.8mm$). The wall thickness at the ES frame for American Heart Association 16-segments are shown in Table~\ref{wall_thickness}. In addition, we have computed the fractional wall thickening between ED frame and ES frame by $(ES-ED)/ED*100\%$. The results in Table~\ref{wall_thickness} shows that the regional and global fractional wall thickening are comparable with results reported in literature~\cite{Ubachs2009, Dong1994}. 

\begin{table}[h]
\centering
\caption{Wall thickness at the ES frame and fractional wall thickening between ED and ES frames. Results are reported as ``mean (standard deviation)".}
\label{wall_thickness}
\resizebox{0.5\textwidth}{!}{
\begin{tabular}{cc|cc}
\toprule[1.2pt]
\multicolumn{2}{c|}{Segments}   &
Wall thickness (mm)                  &
Fractional wall thickening ($\%$)    \\
\midrule
\multirow{6}{*}[-0.5em]{Basal}  &
Anterior (1)   &
9.7 (2.7)     &
34.0 (39.5)    \\
~~~ &
Anteroseptal (2)   &
5.7 (2.9)     &
-24.4 (38.7)    \\
~~~ &
Inferoseptal (3)   &
5.5 (2.0)     &
-17.3 (30.2)    \\
~~~ &
Inferior (4)   &
9.0 (1.7)     &
47.8 (28.5)    \\
~~~ &
Inferolateral (5)   &
11.0 (2.0)     &
72.8 (25.9)    \\
~~~ &
Anterolateral (6)   &
10.9 (1.8)     &
62.0 (23.8)   \\
\midrule
\multirow{6}{*}[-0.5em]{Mid-ventricle}  &
Anterior (7)   &
10.9 (1.5)     &
79.9 (21.0)  \\
~~~ &
Anteroseptal (8)   &
11.9 (1.6)     &
76.2 (21.4)    \\
~~~ &
Inferoseptal (9)   &
10.8 (1.4)     &
39.8 (12.3)    \\
~~~ &
Inferior (10)   &
10.9 (1.3)     &
62.5 (15.5)    \\
~~~ &
Inferolateral (11)   &
11.2 (1.5)     &
73.3 (17.1)    \\
~~~ &
Anterolateral (12)   &
10.5 (1.2)     &
63.9 (15.6)    \\
\midrule
\multirow{4}{*}[-0.5em]{Apical}  &
Anterior (13)   &
10.8 (1.1)     &
86.3 (23.2)    \\
~~~ &
Septal (14)  &
10.9 (1.4)     &
76.7 (20.5)    \\
~~~ &
Inferior (15)   &
10.6 (1.4)     &
76.2 (15.1)    \\
~~~ &
Lateral (16)  &
11.1 (1.4)    &
84.3 (18.9)    \\
\midrule
\multicolumn{2}{c|}{Global} &
10.1 (2.5)     &
55.9 (40.6)    \\
\bottomrule[1.2pt]
\end{tabular}
}
\end{table}

\section{Discussion}
In this paper, we propose a deep learning-based method for estimating 3D myocardial motion from 2D multi-view cine CMR images. A na\"ive alternative to our method would be to train a fully unsupervised motion estimation network using high-resolution 3D cine CMR images. However, such 3D images are rarely available because (1) 3D cine imaging requires long breath holds during acquisition and are not commonly used in clinical practice, and (2) recovering high-resolution 3D volumes purely from 2D multi-view images is challenging due to the sparsity of multi-view planes.

Our focus has been on LV myocardial motion tracking because it is important for clinical assessment of cardiac function. Our model can be easily adapted to 3D right ventricular myocardial motion tracking by using the corresponding 2D edge maps in the shape regularization module during training.

In shape regularization, we use edge maps to represent anatomical shape, \emph{i.e.}, we predict 3D edge maps of the myocardial wall and we use 2D edge maps defined in the multi-view images to provide shape information. This is because (1) the contour of the myocardial wall is more representative of anatomical shape than the content, (2) compared to 3D dense segmentation, 3D edge maps with sparse labels are more likely to be estimated by images from sparse multi-view planes, and (3) using edge maps offers the potential of using automatic contour detection algorithms to obtain shape information directly from images. 


An automated algorithm is utilized to obtain 2D edge maps for providing shape information in the shape regularization module. This is because manual data labeling is time-consuming, costly and usually unavailable. The proposed method can be robust to these automatically obtained 2D edge maps since the 2D edge maps only provides constraint to spatially sparse planes for the estimated 3D edge maps.

\begin{figure}[pt]
 \centering
 \begin{tabular}{@{\hspace{-1\tabcolsep}}c@{\hspace{0.3\tabcolsep}}c@{\hspace{0.3\tabcolsep}}c@{\hspace{0.3\tabcolsep}}c@{\hspace{0.3\tabcolsep}}c@{\hspace{0.3\tabcolsep}}c}
 \raisebox{0.8\height}{\rotatebox[origin=c]{90}{\makecell{~\scalebox{0.75}{\textbf{3D-UNet~\cite{ociek2016}}}}}}
  \includegraphics[height=1.6cm, trim=14cm 2cm 15cm 3.5cm, clip]{3dunet_0.png} &
  \includegraphics[height=1.6cm, trim=14cm 2cm 15cm 3.5cm, clip]{3dunet_10.png} &
  \includegraphics[height=1.6cm, trim=14cm 2cm 15cm 3.5cm, clip]{3dunet_20.png} &
  \includegraphics[height=1.6cm, trim=14cm 2cm 15cm 3.5cm, clip]{3dunet_30.png} &
  \includegraphics[height=1.6cm, trim=14cm 2cm 15cm 3.5cm, clip]{3dunet_40.png} \\
  \raisebox{0.6\height}{\rotatebox[origin=c]{90}{\makecell{~\scalebox{0.75}{\textbf{MulViMotion\textbf{*}}}}}}
  \includegraphics[height=1.6cm, trim=14cm 2cm 15cm 3.5cm, clip]{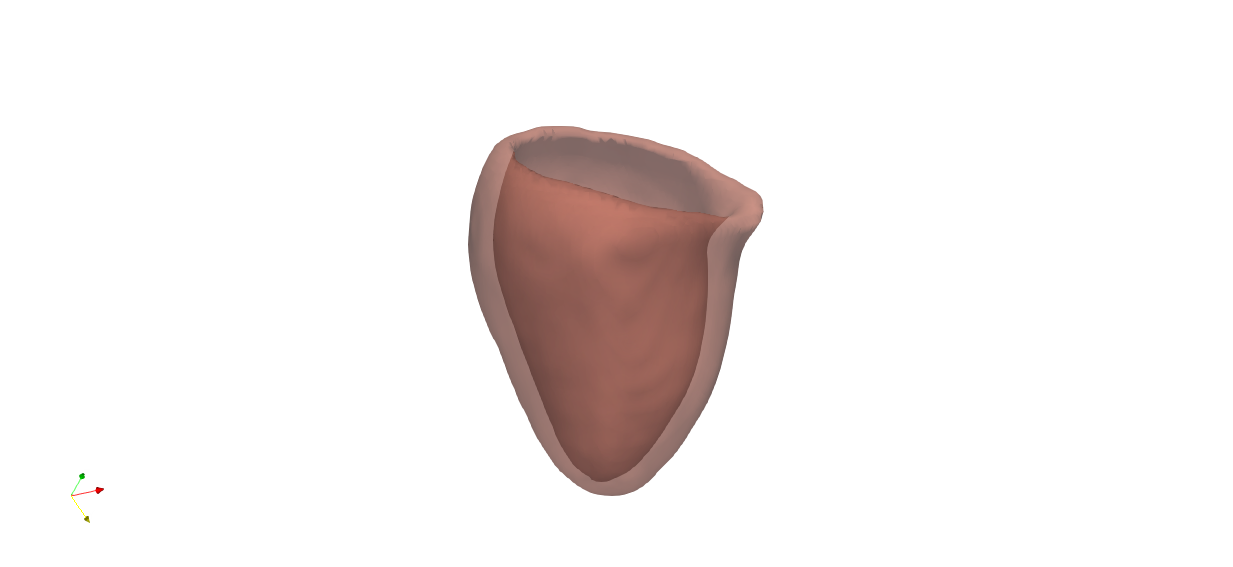} &
  \includegraphics[height=1.6cm, trim=14cm 2cm 15cm 3.5cm, clip]{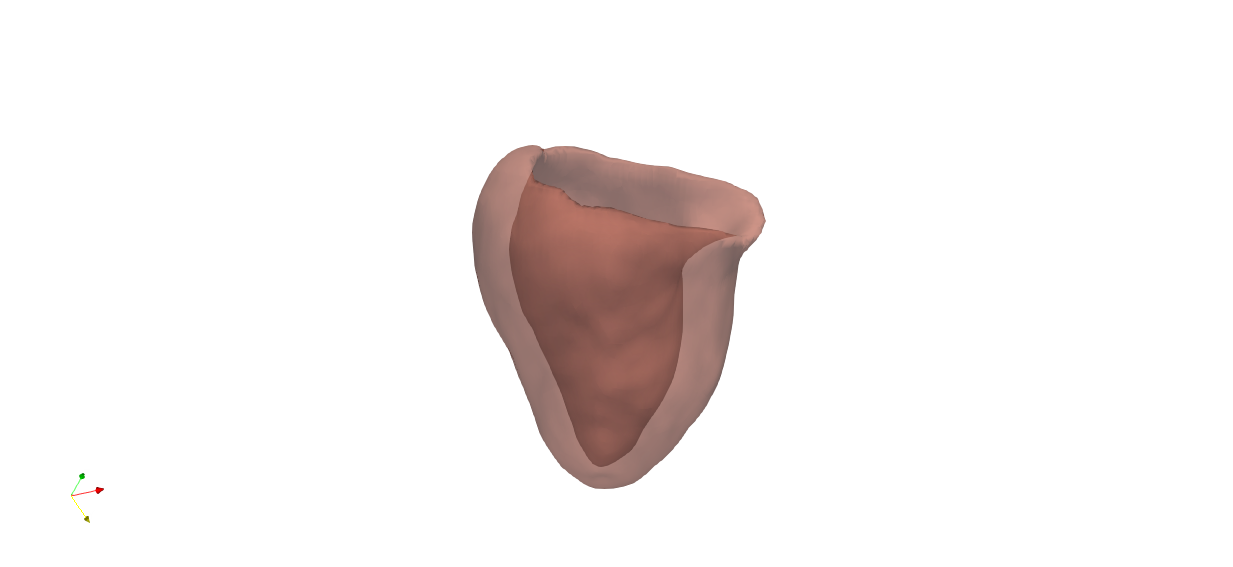} &
  \includegraphics[height=1.6cm, trim=14cm 2cm 15cm 3.5cm, clip]{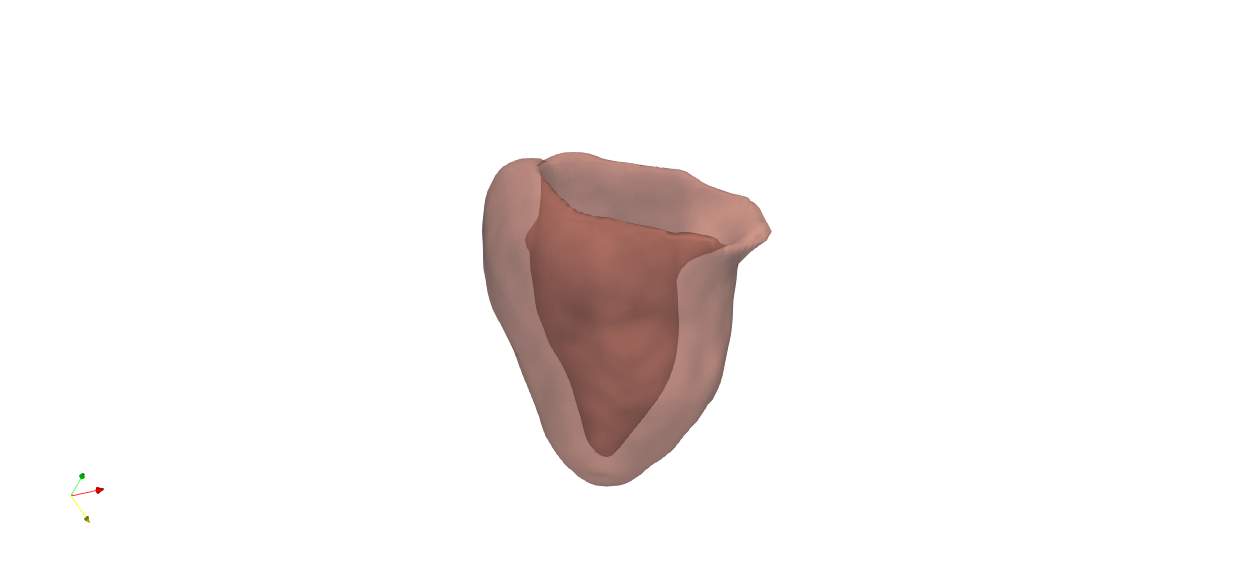} &
  \includegraphics[height=1.6cm, trim=14cm 2cm 15cm 3.5cm, clip]{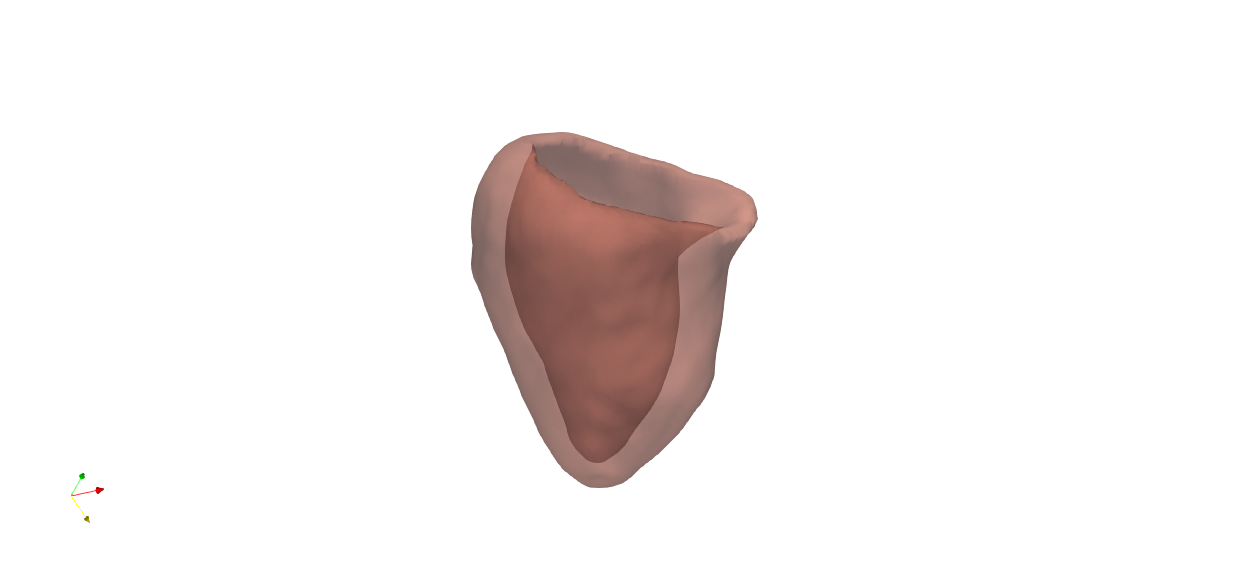} &
  \includegraphics[height=1.6cm, trim=14cm 2cm 15cm 3.5cm, clip]{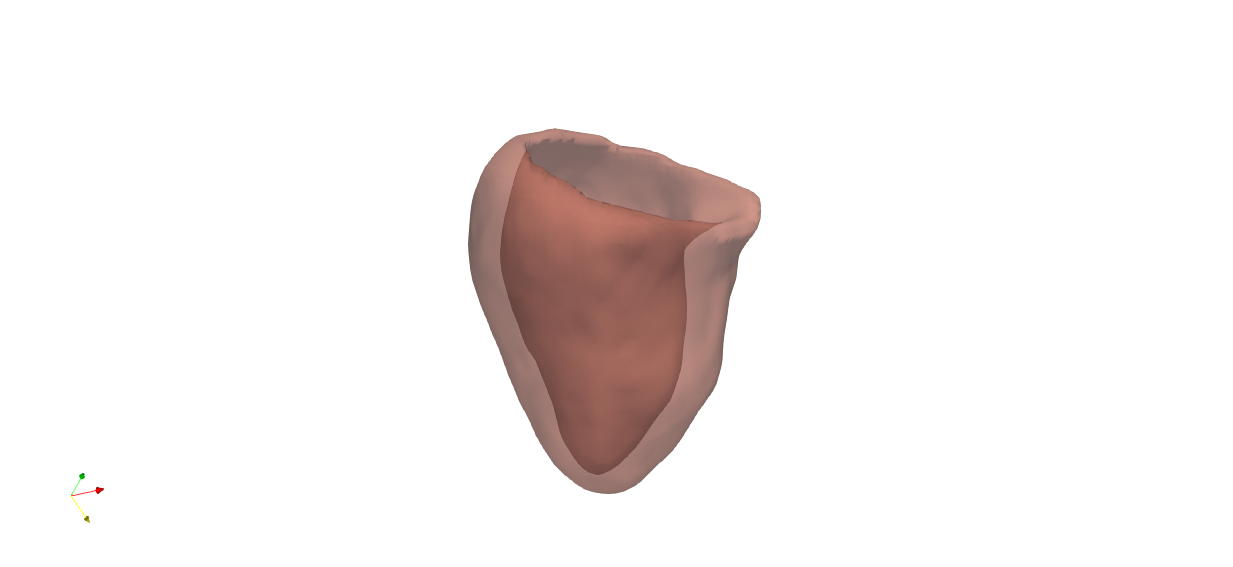} \\
  ~~~
  \raisebox{0.1\height}{\rotatebox[origin=c]{0}{\makecell{~\scalebox{0.8}{\textbf{t=0}}}}} &
  \raisebox{0.1\height}{\rotatebox[origin=c]{0}{\makecell{~\scalebox{0.8}{\textbf{t=10}}}}} &
  \raisebox{0.1\height}{\rotatebox[origin=c]{0}{\makecell{~\scalebox{0.8}{\textbf{t=20}}}}} &
  \raisebox{0.1\height}{\rotatebox[origin=c]{0}{\makecell{~\scalebox{0.8}{\textbf{t=30}}}}} &
  \raisebox{0.1\height}{\rotatebox[origin=c]{0}{\makecell{~\scalebox{0.8}{\textbf{t=40}}}}}
  \end{tabular}
  \caption{3D visualization of motion tracking results using 3D-UNet and MulViMotion\textbf{*}. MulViMotion\textbf{*} uses unaligned SAX ground truth edge maps during training}
  \label{quali_unalign}
\end{figure}

We use the aligned 2D edge maps of SAX stacks to train MulViMotion. This is reasonable because aligned SAX ground truth edge maps can introduce correct shape information of the heart, and thus can explicitly constrain the estimated 3D motion field to reflect the real motion of the heart. Nevertheless, we further test the effectiveness of the proposed method by utilizing unaligned SAX edge maps during training. In specific, MulViMotion\textbf{*} uses the algorithm in~\cite{Bai2018} to predict the 2D segmentation of myocardium for each SAX slice independently without accounting for the inter-slice misalignment. The contour of this 2D segmentation is used as the SAX ground truth edge map during training. LAX ground truth edge maps are still generated based on~\cite{Duan2019}. Table~\ref{quanti_unalign} and Fig.~\ref{quali_unalign} (\emph{e.g.}, $t=20$) show that the proposed method is capable of estimating 3D motion even if it is trained with unaligned SAX edge maps. This indicates that the LAX 2CH and 4CH view images that provides correct longitudinal anatomical shape information can compensate the slice misalignment in the SAX stacks and thus makes a major contribution to the improved estimation accuracy of through-plane motion.

\begin{table}[tb]
\centering
\caption{Quantitative comparison between 3D-UNet and MulViMotion\textbf{*} on test set. MulViMotion\textbf{*} uses unaligned SAX ground truth edge maps during training. Results are reported the same way as Table~\ref{comparison_methods}. Best results in bold.}
\label{quanti_unalign}
\resizebox{0.5\textwidth}{!}{
\begin{tabular}{c|ccc}
\toprule[1.2pt]
Methods   &
Dice $\uparrow$    &
HD (mm) $\downarrow$                  &
VD ($\%$) $\downarrow$             \\
\midrule
3D-UNet~\cite{ociek2016} &
0.7382 (0.0293)  &
17.4785 (3.1030)   &
30.97 (9.89)   \\
MulViMotion\textbf{*}   &
\textbf{0.7856 (0.0295)}  &
\textbf{16.0028 (3.9749)}   &
\textbf{21.35 (5.32)}  \\
\bottomrule[1.2pt]
\end{tabular}
}
\end{table}

In the proposed method, a hybrid 2D/3D network is built to estimate 3D motion fields, where the 2D CNNs combine multi-view features and the 3D CNNs learn 3D representations from the combined features. Such a hybrid network can occupy less GPU memory compared to a pure 3D network. In particular, the number of parameters in this hybrid network is 21.7 millions, much less than 3D-UNet (41.5 millions). Moreover, this hybrid network is able to take full advantage of 2D multi-view images because it enables learning 2D features from each anatomical view before learning 3D representations.

In the experiment, we use 580 subjects for model training and evaluation.
This is mainly because our work tackles 3D data and the number of training subjects is limited by the cost of model training. In specific, we used 500 subjects to train our model for 300 epochs with a NVIDIA Tesla T4 GPU, which requires $\sim 60$ hours of training for each model. In addition, this work focus on developing the methodology for multi-view motion tracking and this sample size align with other previous cardiac analysis work for method development~\cite{Qin2018, ChenC2019, Yu2020, Ye2021}. A population-based clinical study for the whole UK Biobank (currently $\sim50,000$ subjects) still requires future investigation.

With the view planning step in standard cardiac MRI acquisition, the acquired multi-view images are aligned and thus are able to describe a heart from different views~\cite{LuX2011}. In order to preserve such spatial connection between multiple separate anatomical views, data augmentations (\emph{e.g.}, rotation and scaling) that used in some 2D motion estimation works are excluded in this multi-view 3D motion tracking task. 

We use two LAX views (2CH and 4CH) in this work for 3D motion estimation but the number of anatomical views is not a limitation of the proposed method. More LAX views (\emph{e.g.}, 3-chamber view) can be integrated into MulViMotion by adding extra encoders in FeatureNet and extra views in $\mathcal{L}_{shape}$ for shape regularization. However, each additional anatomical view can lead to an increased occupation of GPU memory and extra requirement of image annotation (\emph{i.e.}, 2D edge maps).

The data used in the experiment is acquired by a 1.5 Tesla (1.5T) scanner but the proposed method can be applied on 3T CMR images. The possible dark band artifacts in 3T CMR images may affect the image similarity loss ($\mathcal{L}_{sim}$). However, the high image quality of 3T CMR and utilizing high weights for the regularization terms (\emph{e.g.}, shape regularization and the local smoothness loss) may potentially reduce the negative effect of these artifacts.

We utilize the ED frame and the $t$-th frame ($t={0,1,...,T}$, $T$ is the number of frames) as paired frames to estimate the 3D motion field. This is mainly because the motion estimated from such frame pairing is needed for downstream tasks such as strain estimation~\cite{Sinclair2018, Puyol2018, Ferdian2020}. In the cardiac motion tracking task, the reference frame is commonly chosen as the ED or ES frame~\cite{Yu2020}. Such frame pairing can often be observed in other cardiac motion tracking literature, \emph{e.g.},~\cite{Qin2018, ZhengQ2019, Yu2020}.

In this work, apart from two typical and widely used conventional algorithms, we also compared the proposed method with a learning-based method~\cite{Ronneberger2015} which can represent most of the recent image registration works. In specific, the architecture of~\cite{Ronneberger2015} has been used in many recent works, \emph{e.g.}, ~\cite{XuZ2020, Ta2020, Balakrishnan2019}, and many other recent works, \emph{e.g.},~\cite{deVos2019, Krebs2019, Balakrishnan2019}, are similar to~\cite{Ronneberger2015} where only single view images are utilized for image registration. Nevertheless, we further thoroughly compared the proposed method with another recent and widely used learning-based image registration method~\cite{Balakrishnan2019} (VoxelMorph\footnote{https://github.com/voxelmorph/voxelmorph}). We train VoxelMorph following the optimal architecture and hyper-parameters suggested by the authors (\textit{VM}) and we also train VoxelMorph with a bigger architecture\footnote{\label{VMfootnote}Filters in encoder are $[64,128,256,512]$ while filters in decoder are $[512,512,256,256, 128, 64,64]$. The weight of the smoothness loss is chosen with grid search ($\lambda \in\{0.5,0.6,0.7,0.8,0.9,1\}$) and we select the value with the best result on validation data $\lambda=0.7$. The weight for auxiliary segmentation is chosen from $\gamma \in\{0.1,0.3,0.5,0.7,1\}$ and we select $\gamma=0.5$.} (\textit{VM$^\dag$}). For fair comparison, 2D ground truth edge maps ($E_0^{sa}$, $E_t^{sa}$ in Eq. 8) are used to generate the segmentation of SAX stacks for adding auxiliary information. Table VI shows that the proposed method outperforms VoxelMorph for 3D cardiac motion tracking. This is expected because SAX segmentation used in VoxelMorph has low through-plane resolution and thus can hardly help improve 3D motion estimation. Moreover, VoxelMorph only uses single view images while the proposed method utilizes information from multiple views.

\begin{table}[tb]
\centering
\caption{Quantitative comparison between VoxelMorph (VM)~\cite{Balakrishnan2019} and MulViMotion on test set. VM follows the optimal architecture and hyper-parameters suggested by the authors. VM$^\dag$ uses a bigger architecture$^{\ref{VMfootnote}}$. Results are reported the same way as Table~\ref{comparison_methods}. Best results in bold.}
\label{VMcompare}
\resizebox{0.5\textwidth}{!}{
\begin{tabular}{c|ccc}
\toprule[1.2pt]
Methods   &
Dice $\uparrow$    &
HD (mm) $\downarrow$                  &
VD ($\%$) $\downarrow$             \\
\midrule
VM~\cite{ociek2016} &
0.7115 (0.0339)  &
15.3277 (2.7690)   &
34.71 (11.84)   \\
VM$^\dag$~\cite{ociek2016} &
0.7147 (0.0307)  &
17.6747 (4.3181)   &
31.75 (10.80)   \\
MulViMotion   &
\textbf{0.8200 (0.0348)}   &
\textbf{14.5937 (4.2449)}   &
\textbf{8.62 (4.85)}   \\
\bottomrule[1.2pt]
\end{tabular}
}
\end{table}

\section{Conclusion}
In this paper, we propose multi-view motion estimation network (MulViMotion) for 3D myocardial motion tracking. The proposed method takes full advantage of routinely acquired multi-view 2D cine CMR images to accurately estimate 3D motion fields. Experiments on the UK Biobank dataset demonstrate the effectiveness and practical applicability of our method compared with other competing methods. 


\bibliographystyle{abbrv}
\balance

\newpage

\clearpage

\section*{Appendix}

\subsection{Examples of 3D masks}
\begin{figure}[h]
    \centering
    \setcounter{subfigure}{0}
    \subfloat[SAX view]{
    \begin{tabular}{cc}
         \includegraphics[height=2.5cm, trim=7cm 4.5cm 7.5cm 7.5cm, clip]{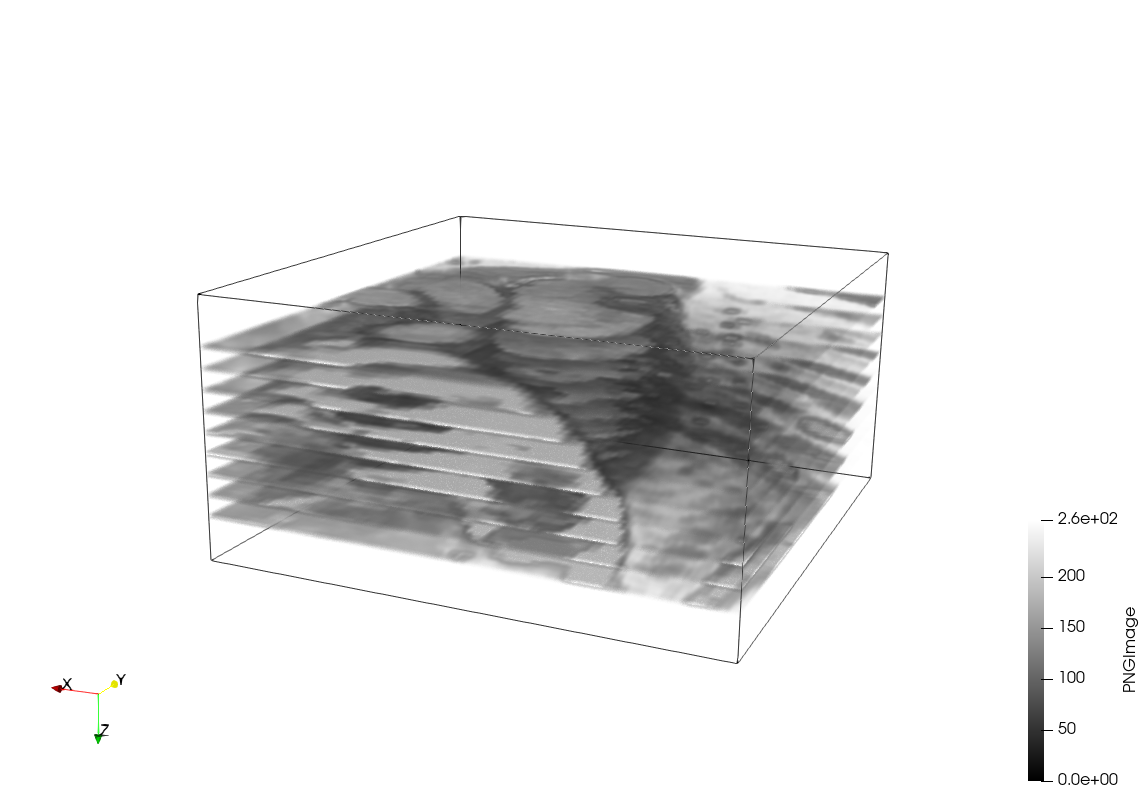} &
         \includegraphics[height=2.5cm, trim=7cm 4.5cm 7.5cm 7.5cm, clip]{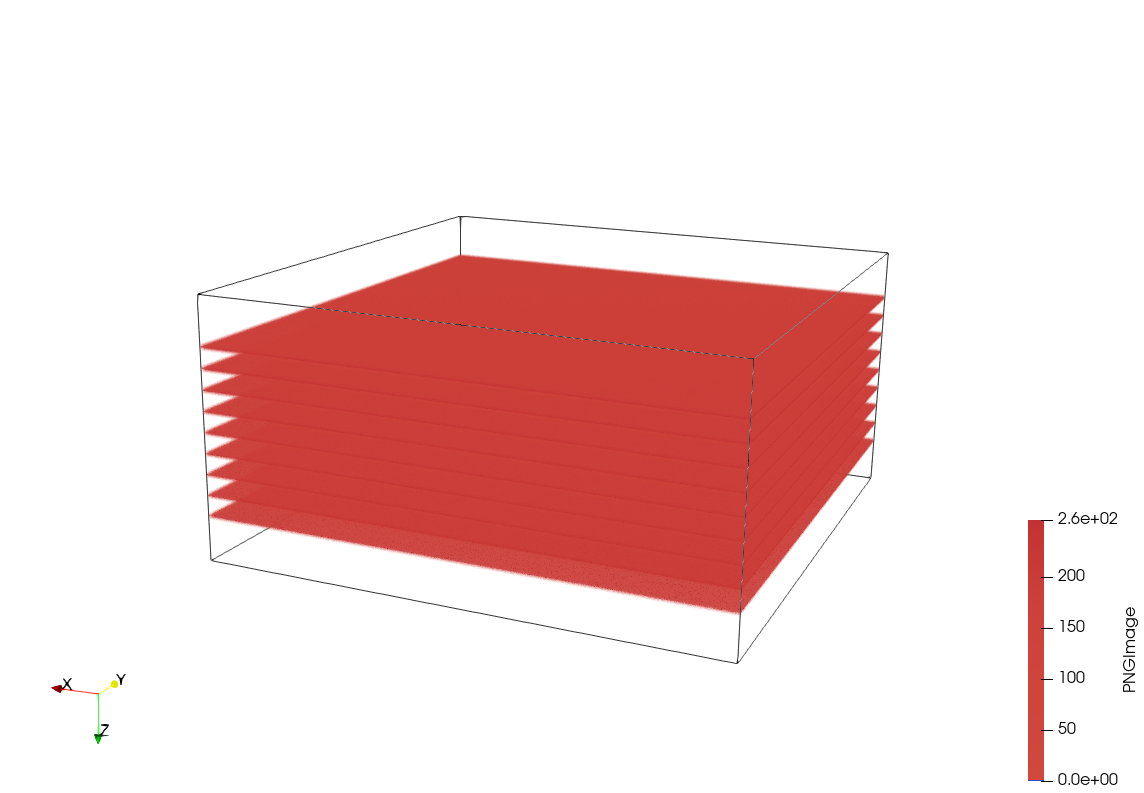}
    \end{tabular}
    }
    \\
    \setcounter{subfigure}{1}
    \subfloat[2CH view]{
    \begin{tabular}{cc}
         \includegraphics[height=2.5cm, trim=7cm 4.5cm 7.5cm 7.5cm, clip]{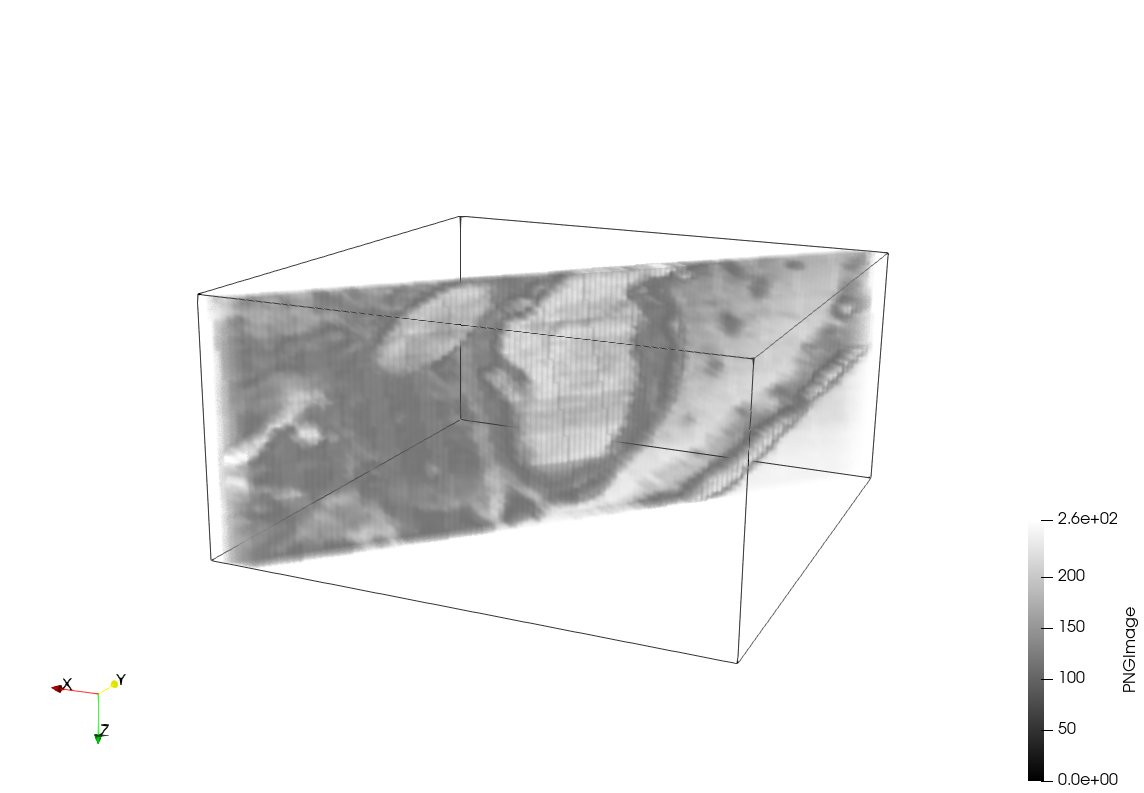}  &
         \includegraphics[height=2.5cm, trim=7cm 4.5cm 7.5cm 7.5cm, clip]{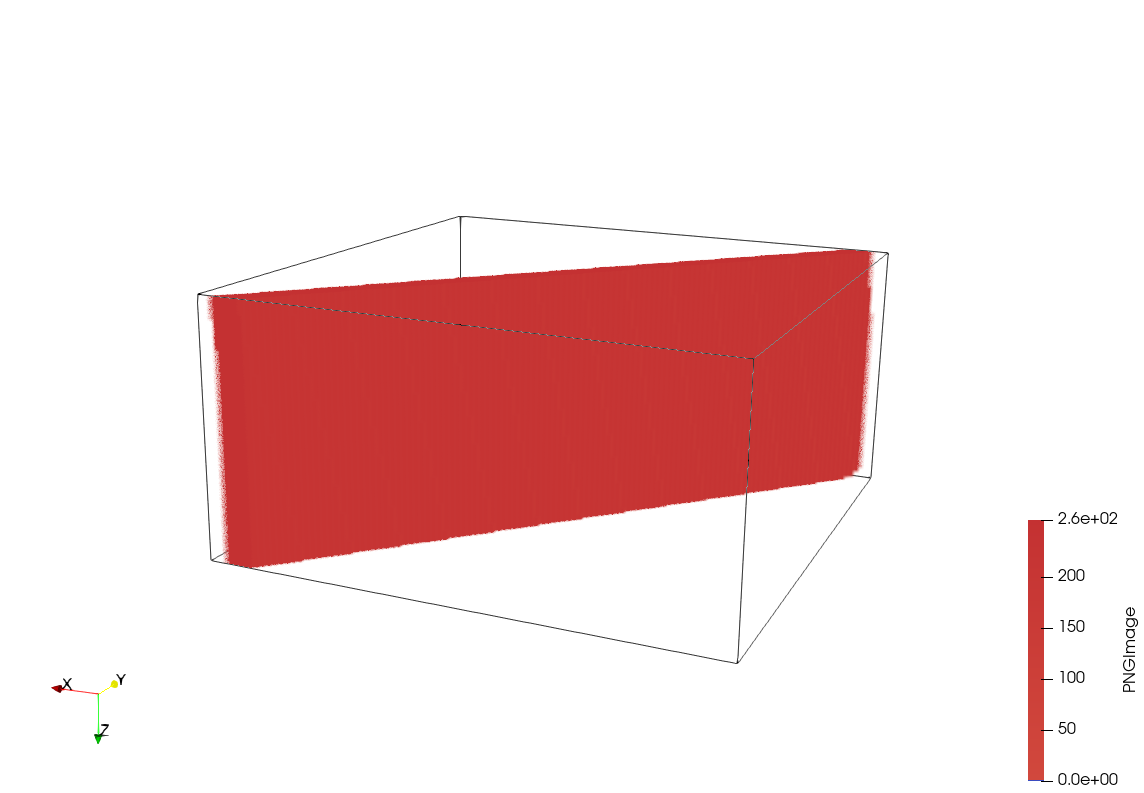} 
    \end{tabular}
    } 
    \\
    \setcounter{subfigure}{2}
    \subfloat[4CH view]{
    \begin{tabular}{cc}
         \includegraphics[height=2.5cm, trim=7cm 4.5cm 7.5cm 7.5cm, clip]{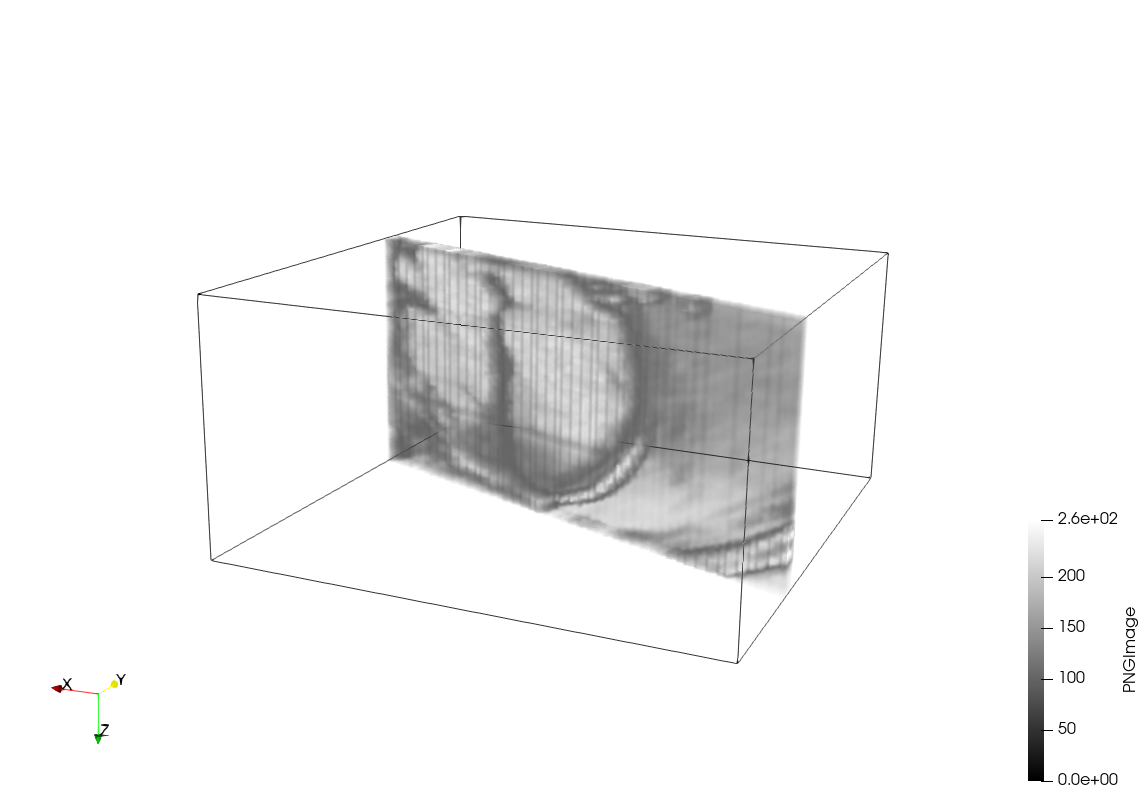}  &
         \includegraphics[height=2.5cm, trim=7cm 4.5cm 7.5cm 7.5cm, clip]{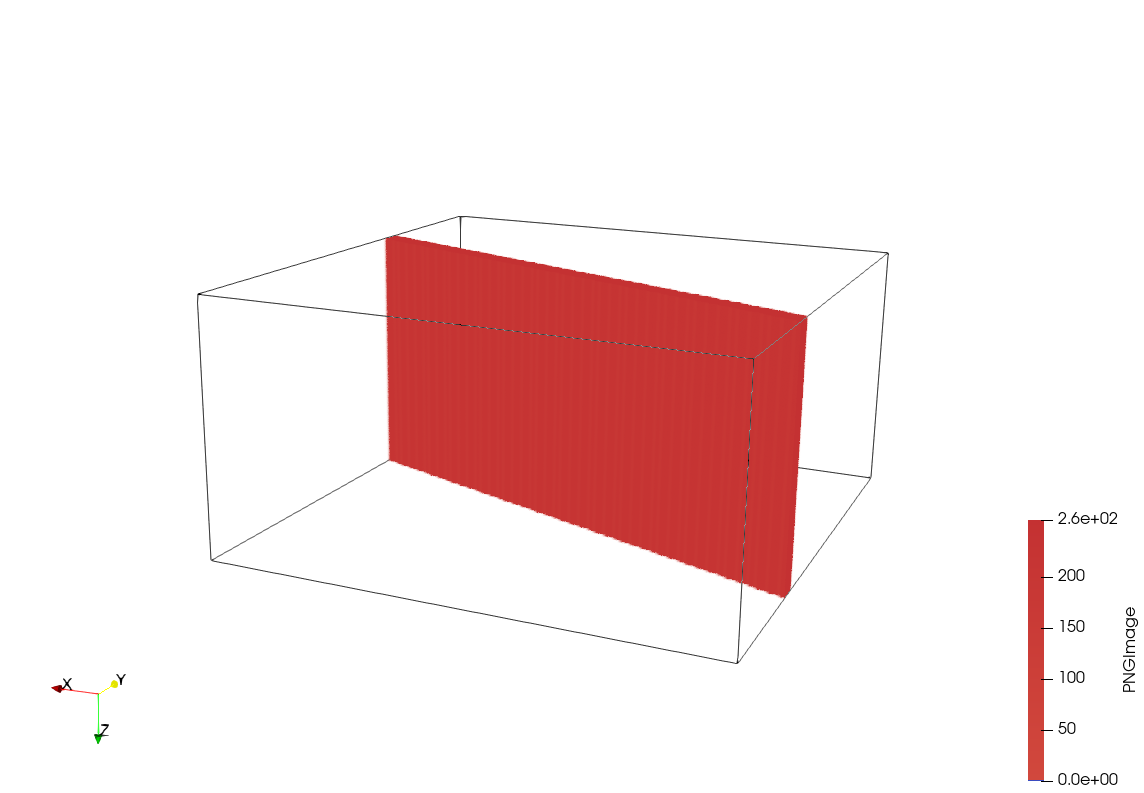} 
    \end{tabular}
    } 
    \caption{Examples of 3D masks used in the shape regularization module of MulViMotion. The top row show the 2D images from different anatomical views in the space of the SAX stack. The bottom row show the 3D masks which represent the locations of these 2D images in the SAX stack. (a) The 2D images from SAX view (9 slices). (b) The single 2D image from 2CH view. (c) The single 2D image from 4CH view.}
    \label{3dmask}
\end{figure}

Fig.~\ref{3dmask} shows the examples of 3D masks used in the shape regularization module of MulViMotion. These 3D masks identify the locations of multi-view images in the SAX stack. We generate these 3D masks in image preprocessing step by a coordinate transformation using DICOM image header information.

\subsection{The dynamic videos of motion tracking results}
The dynamic videos of motion tracking results of different motion estimation methods have been attached as ``Dynamic$\_$videos.zip" in the supplementary material. This file contains four MPEG-4 movies where ``FFD.mp4", ``dDemons.mp4", ``3D-UNet.mp4" are the results of the corresponding baseline methods and ``MulViMotion.mp4" is the result of the proposed method. All methods are applied on the same test subject. The Codecs of these videos is H.264. We have opened the uploaded videos in computers with (1) Win10 operating system, Movies$\&$TV player, (2) Linux Ubuntu 20.04 operating system, Videos player, and (3) Mac OS, QuickTime Player. However, if there is any difficulty to open the attached videos, the same dynamic videos can be found in \textit{https://github.com/qmeng99/dynamic$\_$videos/blob/main/README.md}

\subsection{Additional 3D motion tracking results}
\begin{figure}[h]
 \centering
 \setcounter{subfigure}{0}
 \subfloat[Warped 3D segmentation overlaid on SAX view]{
    \begin{tabular}{@{\hspace{-0.1\tabcolsep}}c@{\hspace{0.1\tabcolsep}}c@{\hspace{0.1\tabcolsep}}c@{\hspace{0.1\tabcolsep}}c@{\hspace{0.1\tabcolsep}}c@{\hspace{0.1\tabcolsep}}c}
    \raisebox{0.8\height}{\rotatebox[origin=c]{90}{\makecell{~\scalebox{0.75}{\textbf{3D-UNet~\cite{ociek2016}}}}}} &
    \raisebox{0.5\height}{\includegraphics[height=0.8cm]{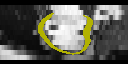}} &
    \raisebox{0.5\height}{\includegraphics[height=0.8cm]{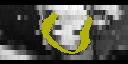}} &
    \raisebox{0.5\height}{\includegraphics[height=0.8cm]{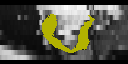}} &
    \raisebox{0.5\height}{\includegraphics[height=0.8cm]{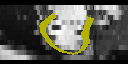}} &
    \raisebox{0.5\height}{\includegraphics[height=0.8cm]{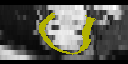}}
     \\
    \raisebox{0.8\height}{\rotatebox[origin=c]{90}{\makecell{~\scalebox{0.75}{\textbf{MulViMotion}}}}} &
   \raisebox{0.5\height}{\includegraphics[height=0.8cm]{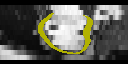}} &
   \raisebox{0.5\height}{\includegraphics[height=0.8cm]{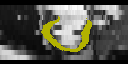}} &
   \raisebox{0.5\height}{\includegraphics[height=0.8cm]{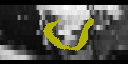}} &
   \raisebox{0.5\height}{\includegraphics[height=0.8cm]{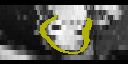}} &
   \raisebox{0.5\height}{\includegraphics[height=0.8cm]{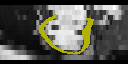}} \\
  \raisebox{0.8\height}{\rotatebox[origin=c]{90}{\makecell{~\scalebox{0.75}{\textbf{\makecell{MulViMotion\\w/o $\mathbf{\mathcal{L}_{shape}}$}}}}}} &
  \raisebox{0.5\height}{\includegraphics[height=0.8cm]{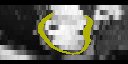}} &
  \raisebox{0.5\height}{\includegraphics[height=0.8cm]{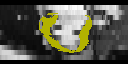}} &
  \raisebox{0.5\height}{\includegraphics[height=0.8cm]{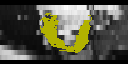}} &
  \raisebox{0.5\height}{\includegraphics[height=0.8cm]{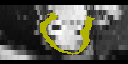}} &
  \raisebox{0.5\height}{\includegraphics[height=0.8cm]{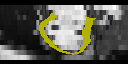}} \\
  ~~~ &
  \raisebox{0.1\height}{\rotatebox[origin=c]{0}{\makecell{~\scalebox{0.8}{\textbf{t=0}}}}} &
  \raisebox{0.1\height}{\rotatebox[origin=c]{0}{\makecell{~\scalebox{0.8}{\textbf{t=10}}}}} &
  \raisebox{0.1\height}{\rotatebox[origin=c]{0}{\makecell{~\scalebox{0.8}{\textbf{t=17}}}}} &
  \raisebox{0.1\height}{\rotatebox[origin=c]{0}{\makecell{~\scalebox{0.8}{\textbf{t=30}}}}} &
  \raisebox{0.1\height}{\rotatebox[origin=c]{0}{\makecell{~\scalebox{0.8}{\textbf{t=40}}}}} 
  \end{tabular}
  }

  \setcounter{subfigure}{1}
    \subfloat[3D visualization]{
    \begin{tabular}{@{\hspace{-0.1\tabcolsep}}c@{\hspace{0.1\tabcolsep}}c@{\hspace{0.1\tabcolsep}}c@{\hspace{0.1\tabcolsep}}c@{\hspace{0.1\tabcolsep}}c@{\hspace{0.1\tabcolsep}}c}
    \raisebox{0.8\height}{\rotatebox[origin=c]{90}{\makecell{~\scalebox{0.75}{\textbf{3D-UNet~\cite{ociek2016}}}}}} &
    \includegraphics[height=1.6cm, trim=14cm 2cm 15cm 3.5cm, clip]{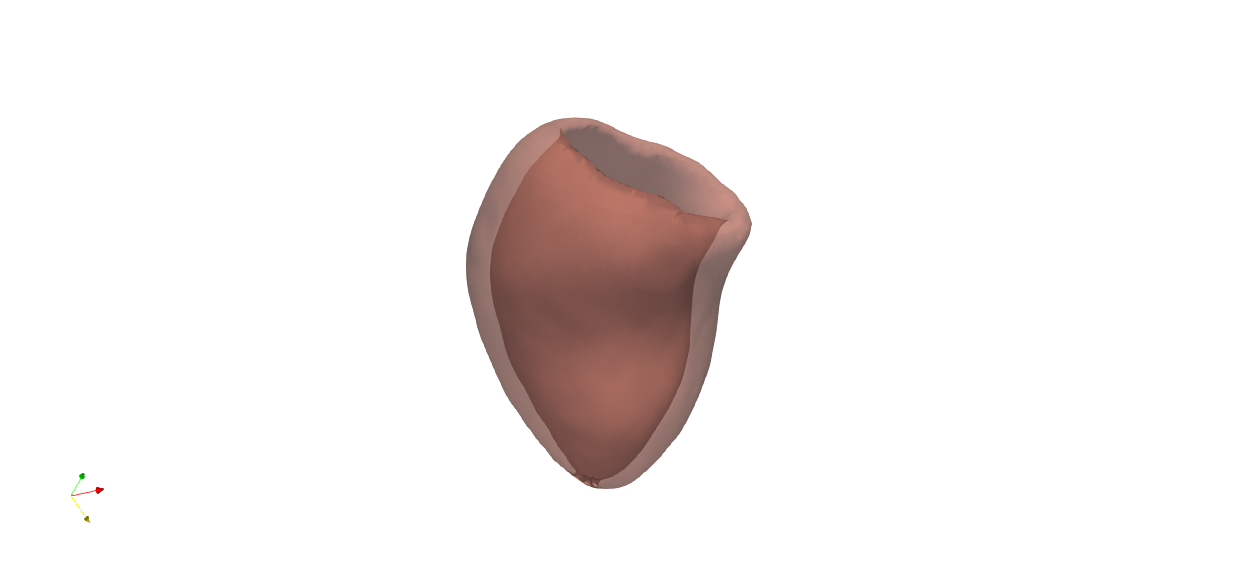} &
    \includegraphics[height=1.6cm, trim=14cm 2cm 15cm 3.5cm, clip]{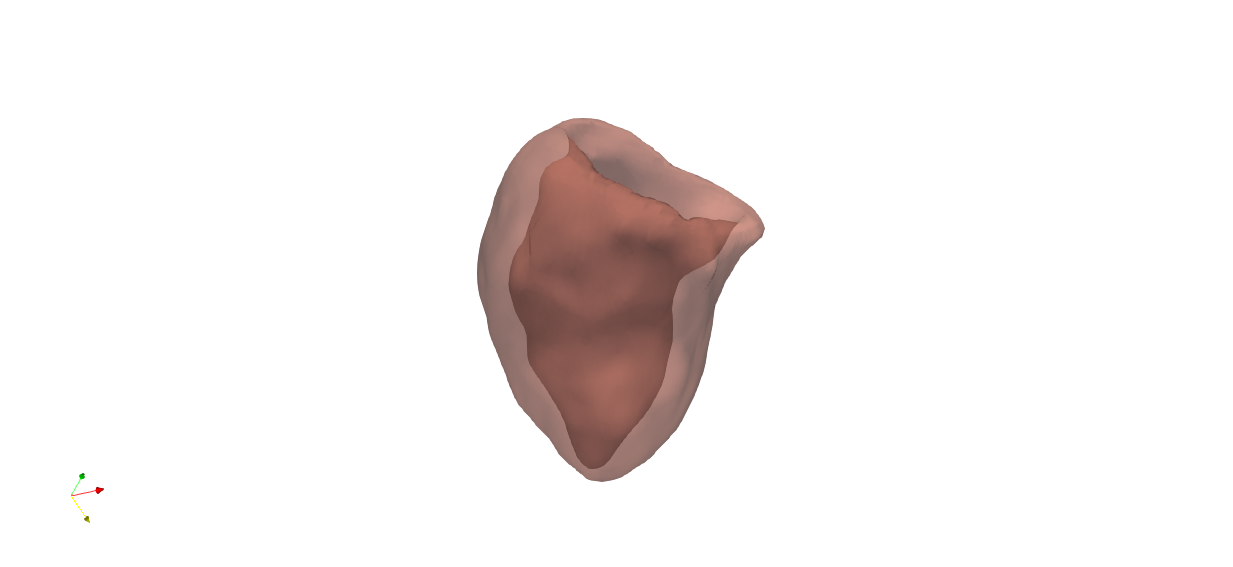} &
    \includegraphics[height=1.6cm, trim=14cm 2cm 15cm 3.5cm, clip]{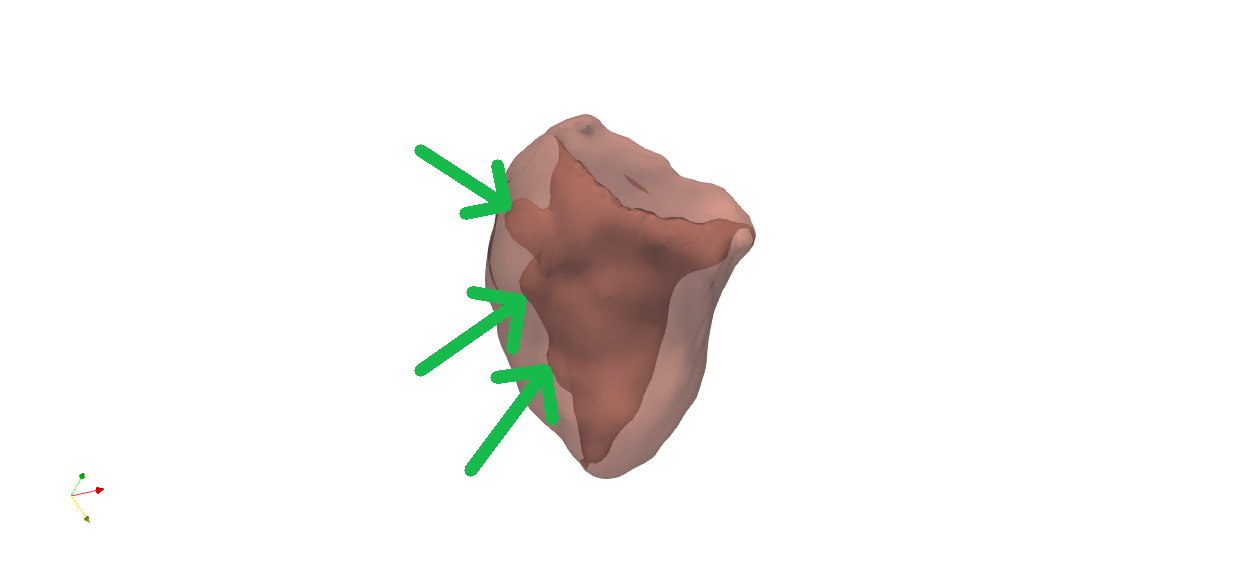} &
    \includegraphics[height=1.6cm, trim=14cm 2cm 15cm 3.5cm, clip]{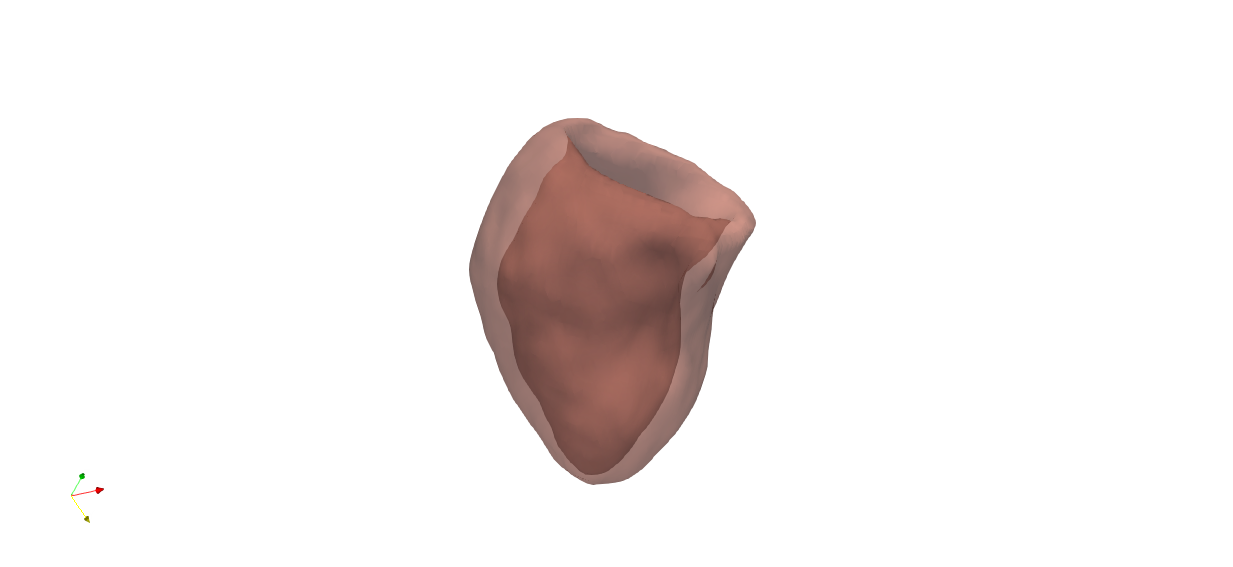} &
    \includegraphics[height=1.6cm, trim=14cm 2cm 15cm 3.5cm, clip]{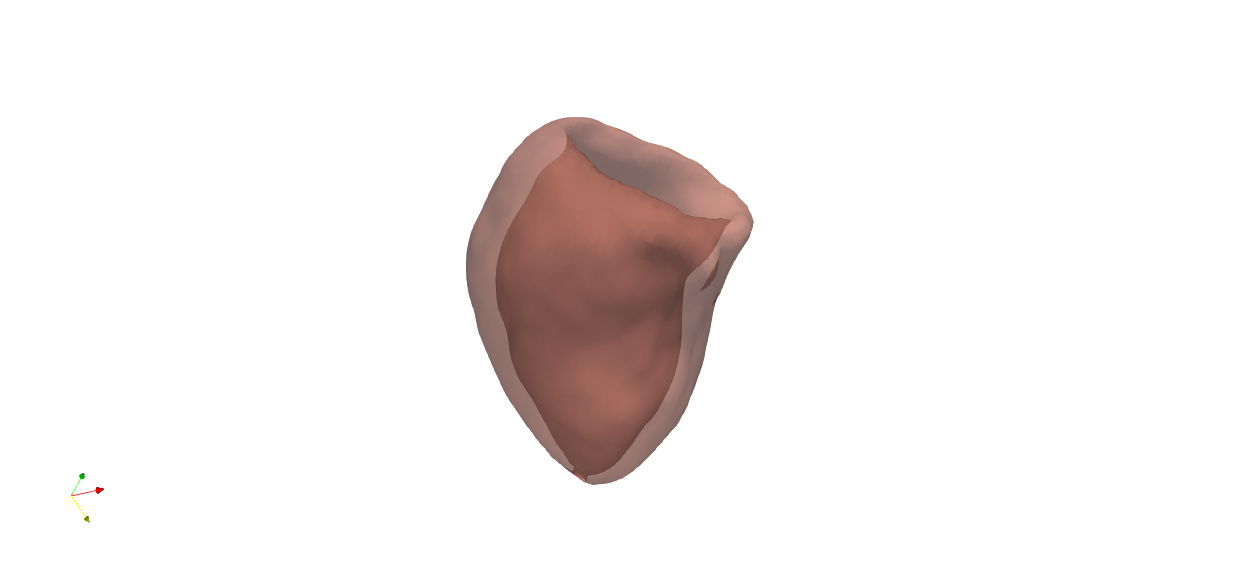} \\
    \raisebox{0.8\height}{\rotatebox[origin=c]{90}{\makecell{~\scalebox{0.75}{\textbf{MulViMotion}}}}} &
    \includegraphics[height=1.6cm, trim=14cm 2cm 15cm 3.5cm, clip]{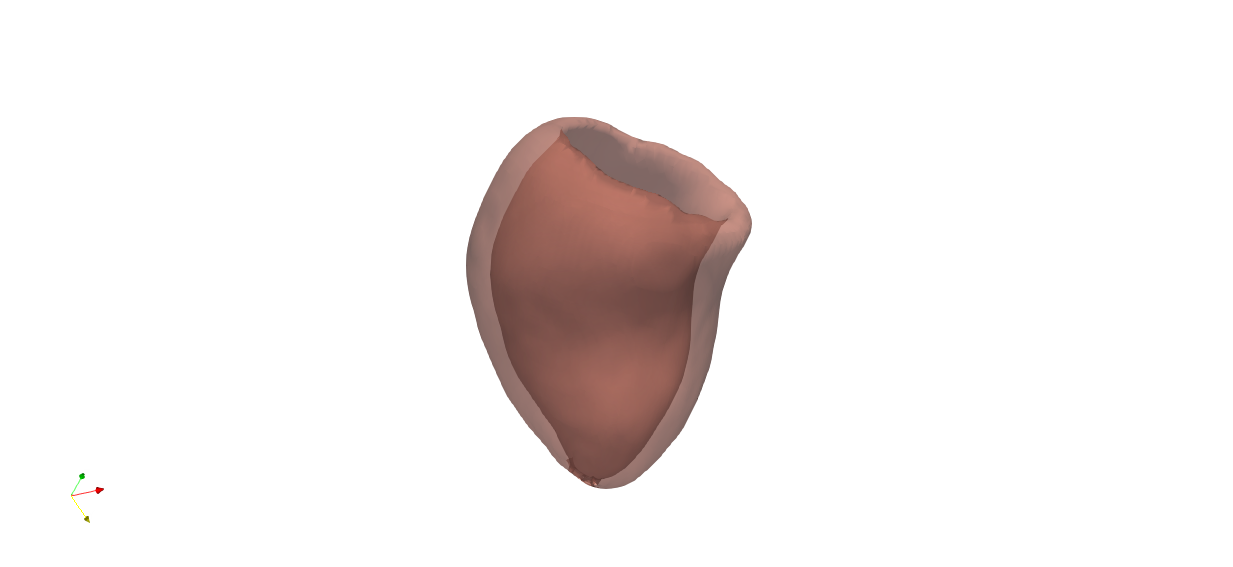} &
    \includegraphics[height=1.6cm, trim=14cm 2cm 15cm 3.5cm, clip]{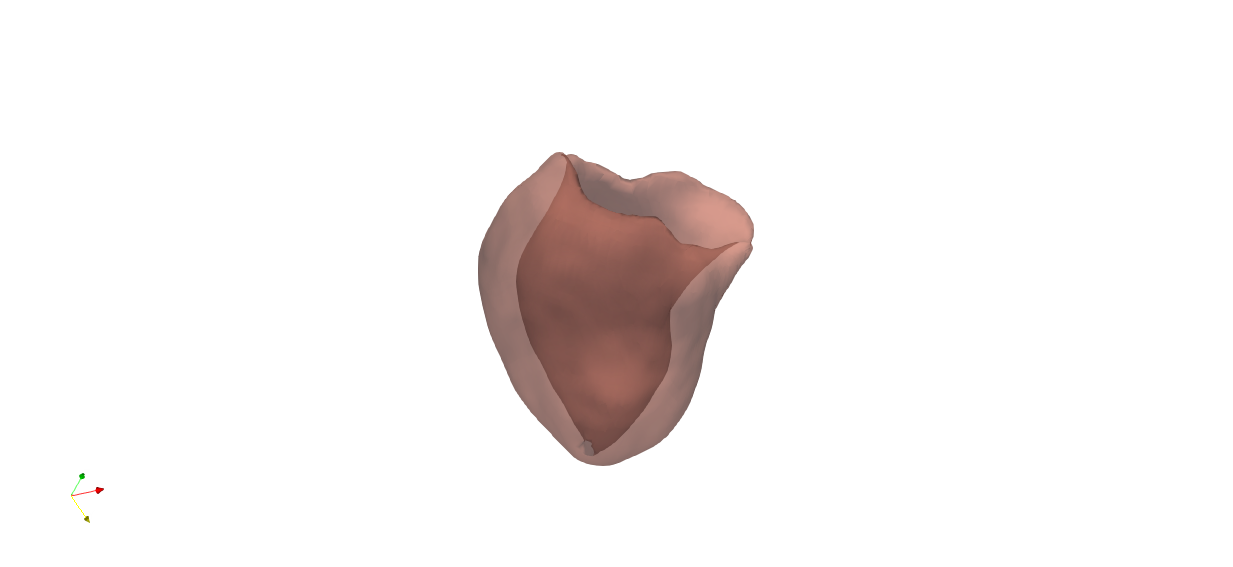} &
    \includegraphics[height=1.6cm, trim=14cm 2cm 15cm 3.5cm, clip]{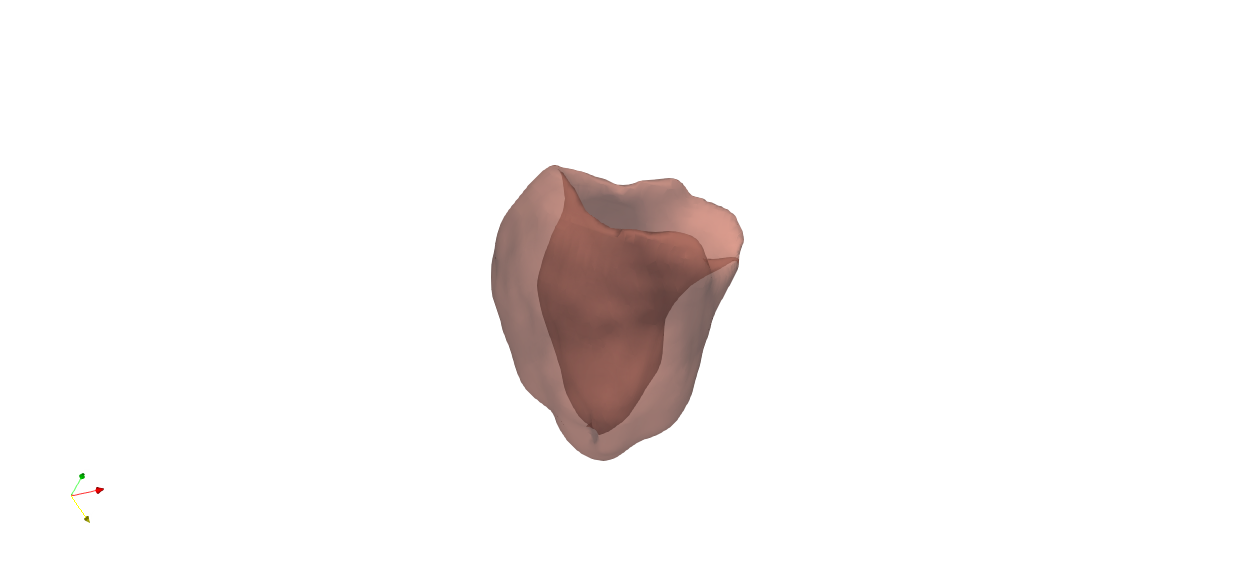} &
    \includegraphics[height=1.6cm, trim=14cm 2cm 15cm 3.5cm, clip]{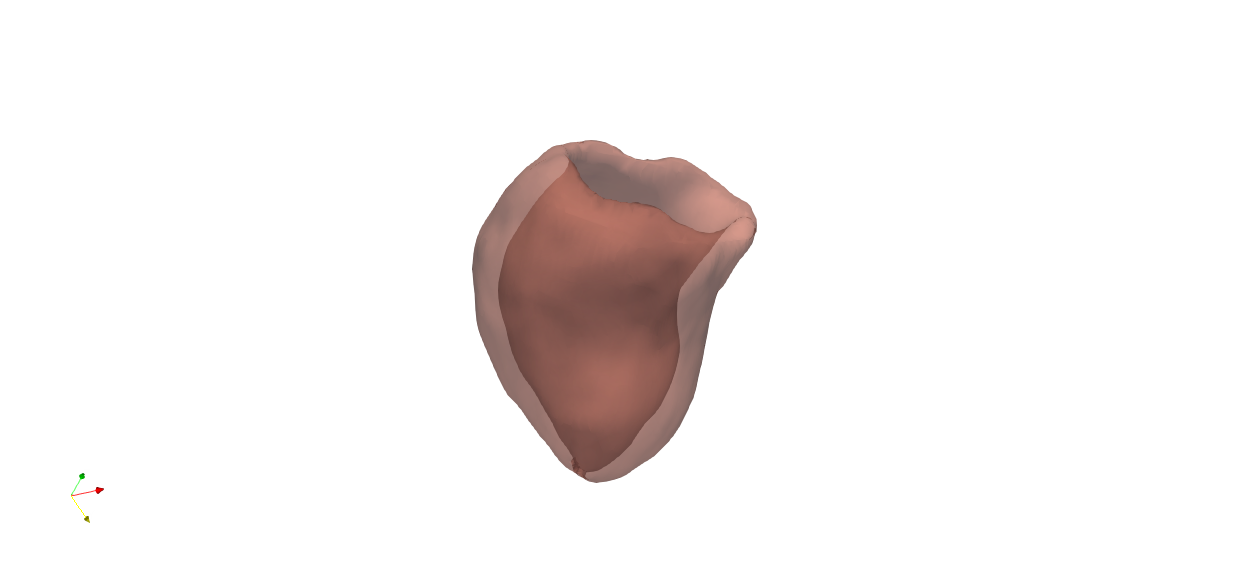} &
    \includegraphics[height=1.6cm, trim=14cm 2cm 15cm 3.5cm, clip]{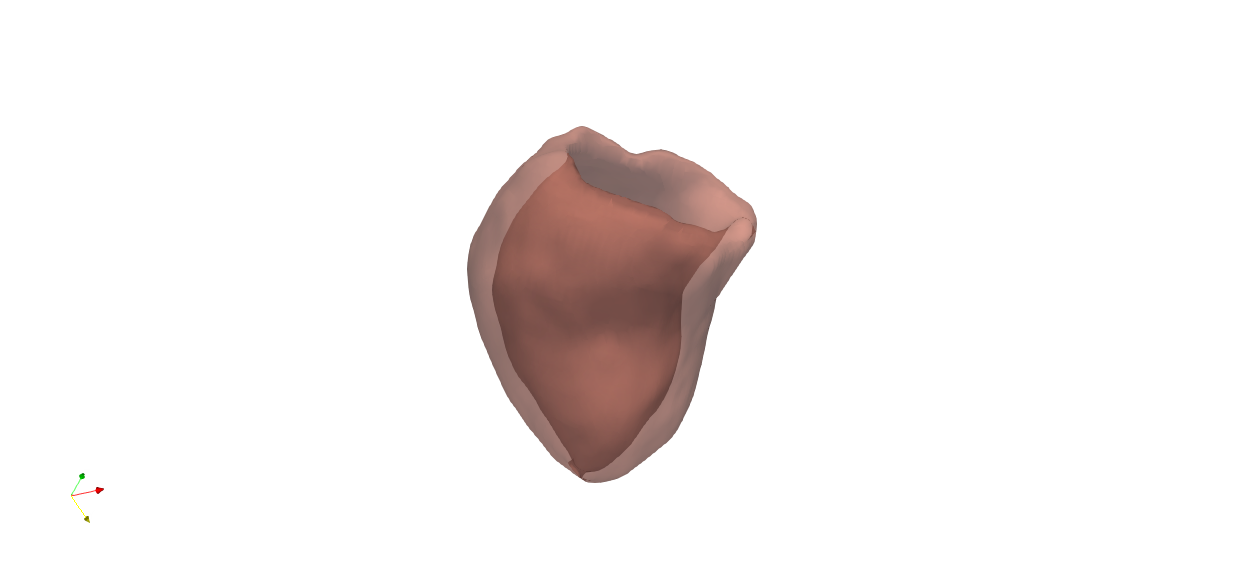} \\
    \raisebox{0.8\height}{\rotatebox[origin=c]{90}{\makecell{~\scalebox{0.75}{\textbf{\makecell{MulViMotion\\W/O $\mathbf{\mathcal{L}_{shape}}$}}}}}} &
    \includegraphics[height=1.6cm, trim=14cm 2cm 15cm 3.5cm, clip]{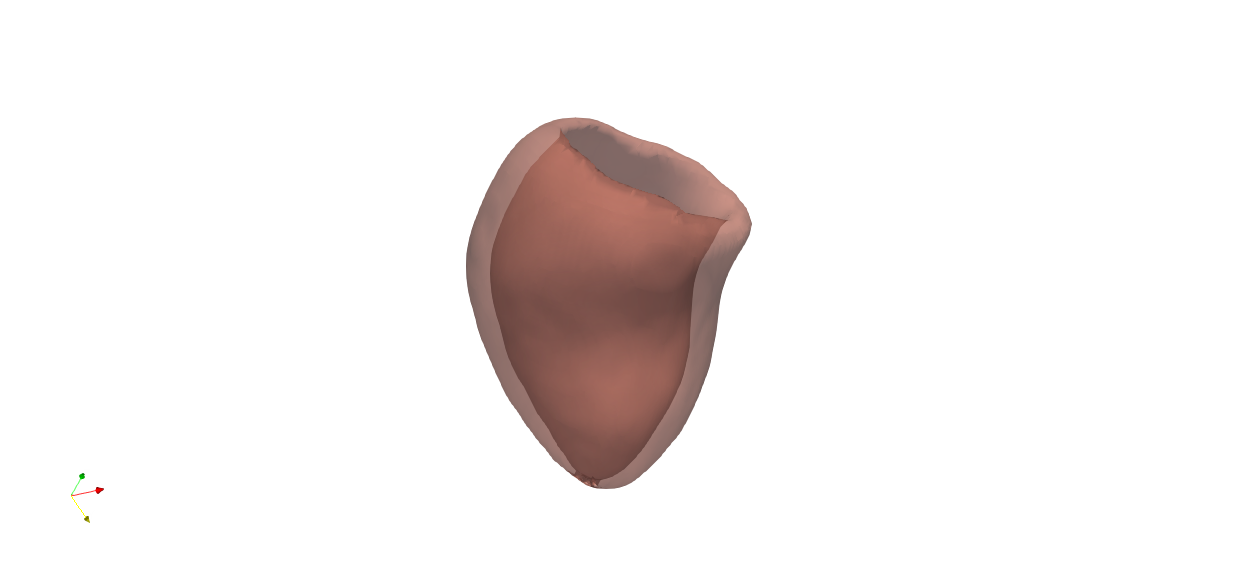} &
    \includegraphics[height=1.6cm, trim=14cm 2cm 15cm 3.5cm, clip]{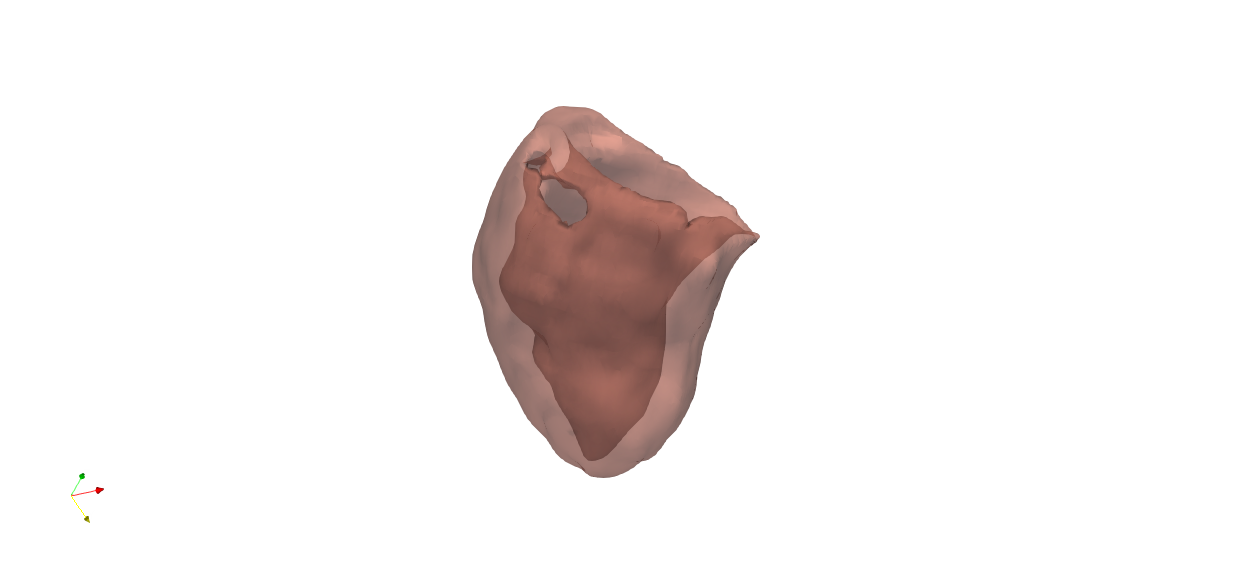} &
    \includegraphics[height=1.6cm, trim=14cm 2cm 15cm 3.5cm, clip]{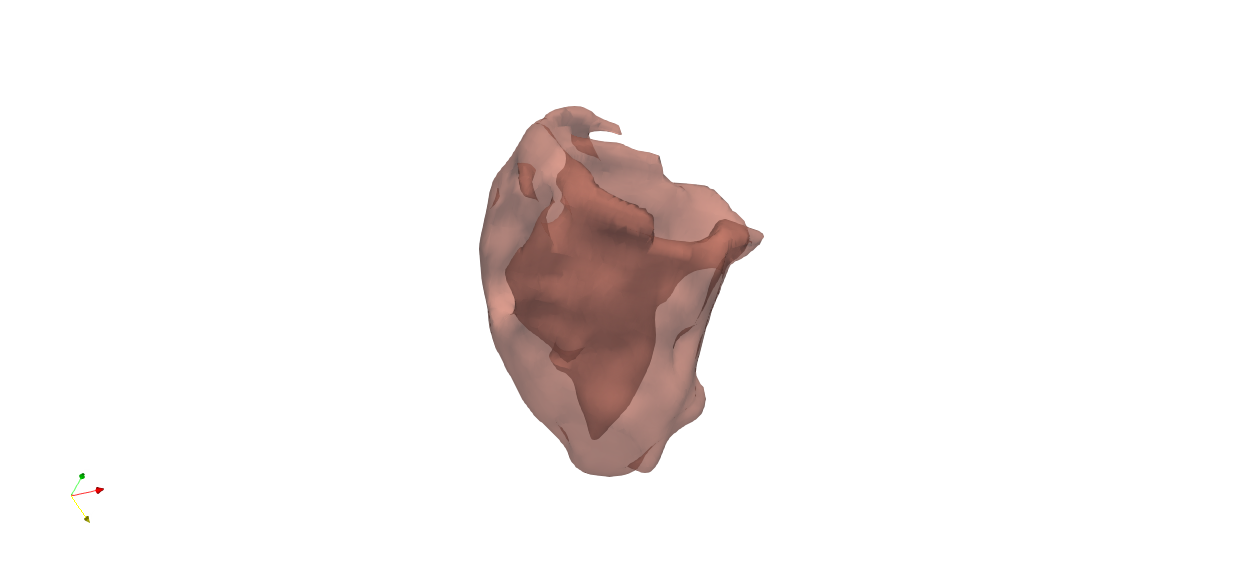} &
    \includegraphics[height=1.6cm, trim=14cm 2cm 15cm 3.5cm, clip]{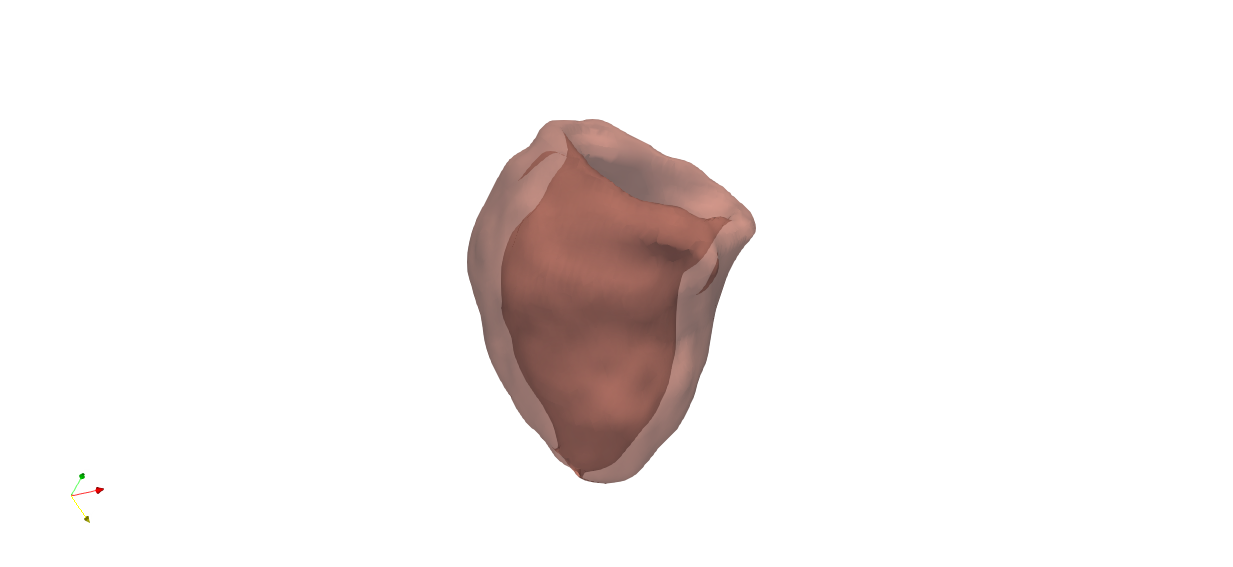} &
    \includegraphics[height=1.6cm, trim=14cm 2cm 15cm 3.5cm, clip]{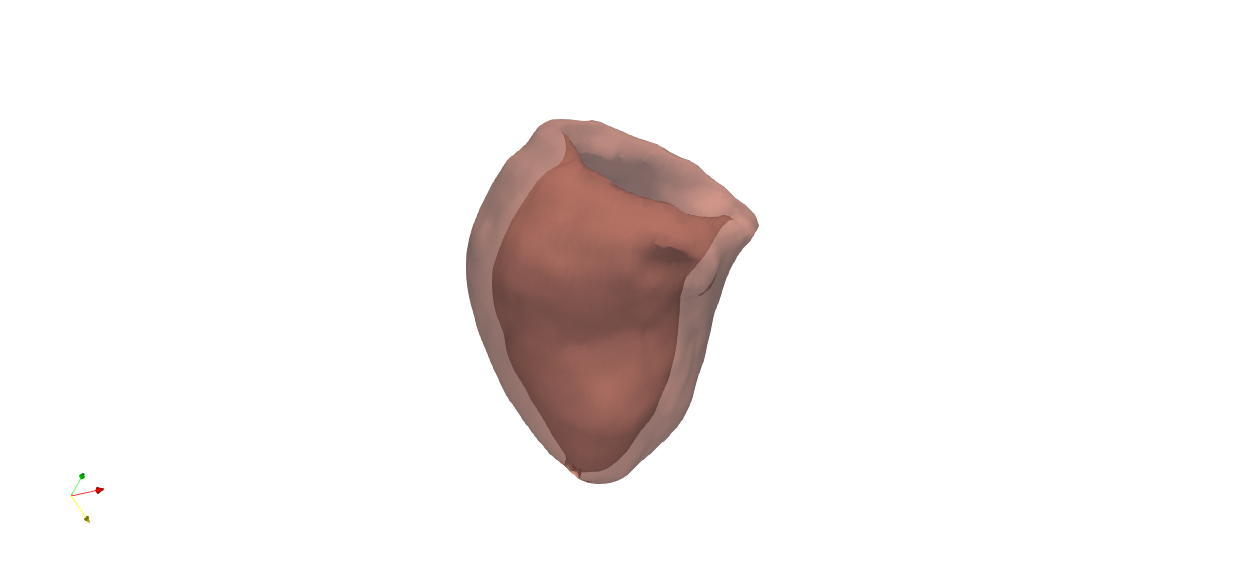} \\
   ~~~ &
   \raisebox{0.1\height}{\rotatebox[origin=c]{0}{\makecell{~\scalebox{0.8}{\textbf{t=0}}}}} &
   \raisebox{0.1\height}{\rotatebox[origin=c]{0}{\makecell{~\scalebox{0.8}{\textbf{t=10}}}}} &
   \raisebox{0.1\height}{\rotatebox[origin=c]{0}{\makecell{~\scalebox{0.8}{\textbf{t=17}}}}} &
   \raisebox{0.1\height}{\rotatebox[origin=c]{0}{\makecell{~\scalebox{0.8}{\textbf{t=30}}}}} &
   \raisebox{0.1\height}{\rotatebox[origin=c]{0}{\makecell{~\scalebox{0.8}{\textbf{t=40}}}}}
   \end{tabular}
   }
  \caption{Motion tracking results on the test subject with slice misalignment. using 3D-UNet~\cite{ociek2016}, MulViMotion, and MulViMotion without $\mathcal{L}_{shape}$. (a) The warped 3D segmentation overlaid on SAX view. (b) The 3D visualization of the motion tracking results. The green arrows show examples of motion tracking failures using 3D-UNet. Note that we show results in frame $t=17$ for a more distinct comparison.}
  \label{sliceshift_more}
\end{figure}

\begin{figure}[h]
    \centering
    \begin{tabular}{@{\hspace{-1\tabcolsep}}c@{\hspace{0.5\tabcolsep}}c@{\hspace{-0.8\tabcolsep}}}
         \subfloat[A single test subject]{\includegraphics[height=3.2cm, trim=0.6cm 0.5cm 0.5cm 0.5cm, clip]{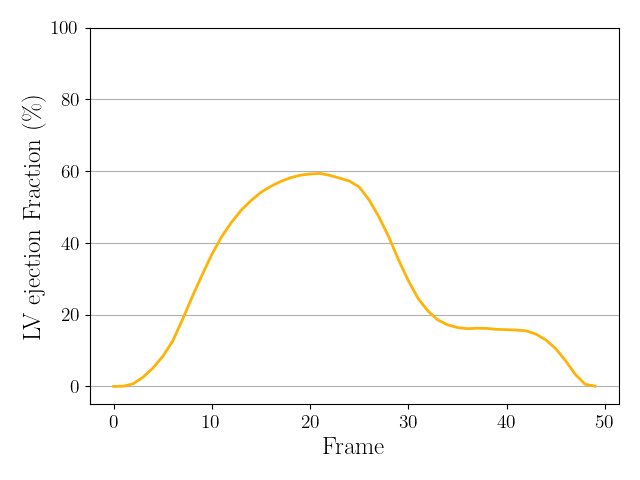}} &
         \subfloat[All test subjects]{\includegraphics[height=3.2cm, trim=0.6cm 0.5cm 0.5cm 0.5cm, clip]{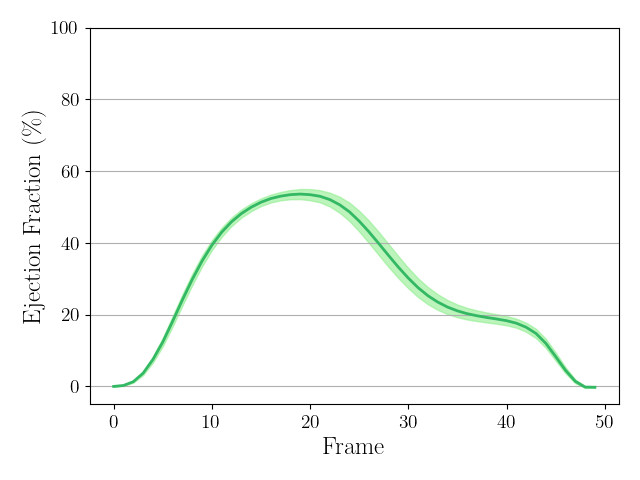}
         }
    \end{tabular}
    \caption{The results of temporal ejection fraction across the cardiac cycle. (a) Results on a randomly selected test subject. (b) Results on all test subjects (mean values and confidence interval are presented).}
    \label{lv_ef}
\end{figure}

Fig.~\ref{sliceshift_more} shows the additional 3D motion tracking results on a test subject with slice misalignment. This test subject is the same subject used in Fig.~\ref{sliceshift} in the main paper. These more results further demonstrate that the proposed method is able to reduce the negative effect of slice misalignment on 3D motion tracking. In addition, we have computed more established clinical biomarkers. Fig.~\ref{lv_ef} shows the temporal ejection fraction across the cardiac cycle. 

\subsection{Applications}

\subsubsection{Strain estimation}

Myocardial strain provide a quantitative evaluation for the total deformation of a region of tissue during the heartbeat. It is typically evaluated along three orthogonal directions, namely radial, circumferential and longitudinal. Here, we evaluate the performance of the proposed method by estimating the three strains based on the estimated 3D motion field $\Phi_t$. The myocardial mesh at the ED frame is warped to the $t$-th frame using a numeric method and vertex-wise strain is calculated using the Lagrangian strain tensor formula~\cite{Petitjean2005} (implemented by https://github.com/Marjola89/3Dstrain$\_$analysis). Subsequently, global strain is computed by averaging across all the vertices of the myocardial wall. 


\begin{figure}[thb]
    \centering
    \begin{tabular}{@{\hspace{-1\tabcolsep}}c@{\hspace{0.5\tabcolsep}}c@{\hspace{-0.8\tabcolsep}}}
         \subfloat[Single subject]{\includegraphics[height=3.2cm, trim=0.6cm 0.5cm 0.5cm 0.5cm, clip]{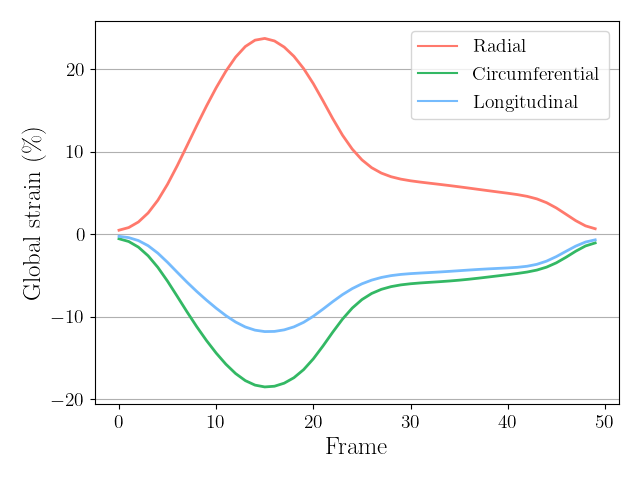}} &
         \subfloat[All test subject]{\includegraphics[height=3.2cm, trim=0.6cm 0.5cm 0.5cm 0.5cm, clip]{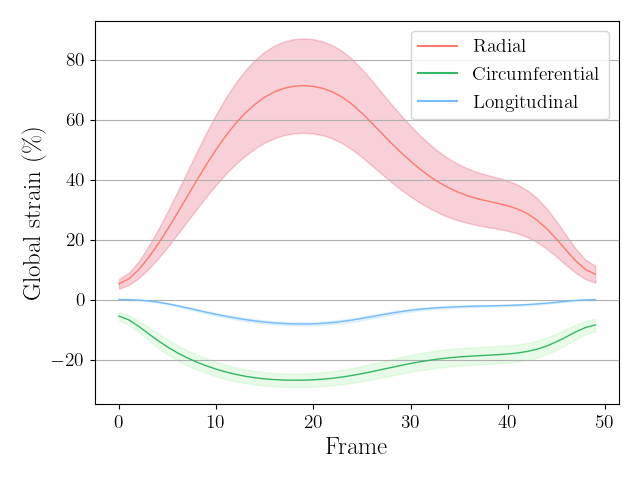}
         }
    \end{tabular}
    \caption{Global strains across the cardiac cycle which are estimated base on MulViMotion. (a) Results on a randomly selected test subject. (b) Results on all test subjects (mean values and confidence interval are presented).}
    \label{strain}
\end{figure}

Fig.~\ref{strain} shows the estimated global strain curves on test subjects. 
Both the shapes of the curves and the value ranges of peak strains are consistent with reported results in the literature~\cite{KawelBoehm2015, Cao2018, Ferdian2020}, \emph{i.e.}, radial peak strain is $\sim 20\%$ to $\sim 70\%$, circumferential peak strain is $\sim -15\%$ to $\sim -22\%$ and longitudinal peak strain is $\sim -8\%$ to $\sim -20\%$. 

To get more reference strains, we have separately computed global longitudinal and circumferential strains on the 2D LAX and SAX slices according to the algorithm in~\cite{Bai2020}. On the test set, global longitudinal peak strain is $-18.55\%\pm2.74\%$ (ours is $-9.72\%\pm2.49\%$) while global circumferential peak strain is $-22.76\%\pm3.31\%$ (ours is $-27.38\%\pm9.63\%$). It is possible that our strains are different from these strains. This is because these strains in ~\cite{Bai2020} are computed only on sparse 2D slices by 2D motion field estimation, and in contrast, we compute global strains by considering the whole myocardium wall with 3D motion fields.



Compared to echocardiograpy, another widely used imaging modality for strain estimation, the average circumferential peak strain reported in our work ($-27.38\%$) is consistent with those typically reported in echocardiograpy ($\sim-22\%$ to $\sim-32\%$~\cite{Amzulescu2019}). The average longitudinal peak strain in our study ($-9.72\%$) is lower than that reported in echocardiograpy ($\sim-20\%$ to $\sim-25\%$~\cite{Amzulescu2019}). This difference is likely due to the higher spatial and temporal resolution of echocardiography (\emph{e.g.}, $0.2-0.3mm$ for spatial resolution and $40-60$ frames/s for temporal resolution) compared to CMR (\emph{e.g.}, our data has $\sim1.8mm$ in-plane resolution, $\sim10mm$ through-plane resolution and $50$ frames/heart-beat temporal resolution)~\cite{Amzulescu2019, Petersen2015}.

For strain estimation, our results are in general consistent with the value ranges reported in~\cite{KawelBoehm2015, Cao2018,Ferdian2020}. However, it has to be noted that we calculate the strain based on 3D motion fields, whereas most existing strain analysis methods or software packages are based on 2D motion fields, i.e. only accounting for in-plane motion within SAX or LAX views. This may result in difference between our estimated strain values and the reported strain values in literature. In addition, there is still a lack of agreement of strain value ranges (in particular for radial strains) even among mainstream commercial software packages~\cite{Cao2018}. This is because strain value ranges can vary depending on the vendors, imaging modalities, image quality and motion estimation techniques~\cite{Cao2018, Amzulescu2019}. It still requires further investigations to set up a reference standard for strain evaluation and to carry out clinical association studies using the reported strain values. Moreover, when manual segmentation is available, it could be used to provide more perfect and accurate shape constraint, which may further improve 3D motion estimation and thus strain estimation.

\end{document}